\documentclass[longauth]{aa}  

\usepackage{graphicx}
\usepackage{txfonts}
\usepackage[colorlinks=true,
            linkcolor=black,
            urlcolor=blue,
            citecolor=blue]{hyperref}
\usepackage{placeins}

\usepackage{booktabs}
\usepackage{xspace}

\newcommand{\pyttv}{\texttt{PyTTV}\xspace}
\newcommand{\pytransit}{\texttt{PyTransit}\xspace}
\newcommand{\emcee}{\texttt{emcee}\xspace}
\newcommand{\rebound}{\texttt{REBOUND}\xspace}
\newcommand{\pipe}{\texttt{PIPE}\xspace}
\newcommand{\tess}{TESS\xspace}
\newcommand{\NP}{\mathcal{N}}
\newcommand{\UP}{\mathcal{U}}

\begin{document}

   \title{A close outer companion to the ultra-hot Jupiter TOI-2109\,b?
   \thanks{The raw and detrended photometric time-series data from CHEOPS are available in electronic from at the CDS via anonymous ftp to cdsarc.u-strasbg.fr (130.79.128.5) or via \href{http://cdsweb.u-strasbg.fr/cgi-bin/qcat?/A+A/}{http://cdsweb.u-strasbg.fr/cgi-bin/qcat?/A+A/}.}$^\mathrm{,}$\thanks{This study uses CHEOPS data observed as part of the Guaranteed Time Observation (GTO) programmes CH\_PR100012 and CH\_PR140063.}}

    \titlerunning{A close outer companion to the ultra-hot Jupiter TOI-2109\,b?}

   \author{J.-V. Harre\inst{1}\thanks{E-mail: jan-vincent.harre@dlr.de} $^{\href{https://orcid.org/0000-0001-8935-2472}{\includegraphics[scale=0.04]{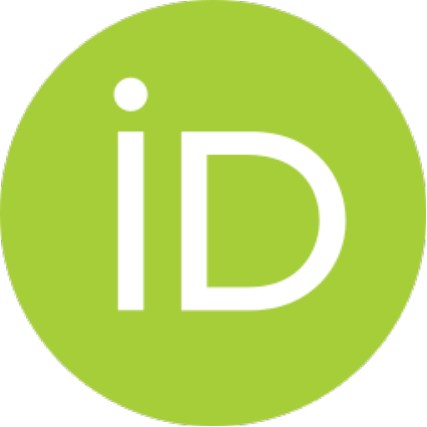}}}$\and 
A. M. S. Smith\inst{1} $^{\href{https://orcid.org/0000-0002-2386-4341}{\includegraphics[scale=0.04]{figures/ORCID_iD.pdf}}}$\and 
S. C. C. Barros\inst{2,3} $^{\href{https://orcid.org/0000-0003-2434-3625}{\includegraphics[scale=0.04]{figures/ORCID_iD.pdf}}}$\and 
V. Singh\inst{4} $^{\href{https://orcid.org/0000-0002-7485-6309}{\includegraphics[scale=0.04]{figures/ORCID_iD.pdf}}}$\and 
J. Korth\inst{5} $^{\href{https://orcid.org/0000-0002-0076-6239}{\includegraphics[scale=0.04]{figures/ORCID_iD.pdf}}}$\and 
A. Brandeker\inst{6} $^{\href{https://orcid.org/0000-0002-7201-7536}{\includegraphics[scale=0.04]{figures/ORCID_iD.pdf}}}$\and 
A. Collier Cameron\inst{7} $^{\href{https://orcid.org/0000-0002-8863-7828}{\includegraphics[scale=0.04]{figures/ORCID_iD.pdf}}}$\and 
M. Lendl\inst{8} $^{\href{https://orcid.org/0000-0001-9699-1459}{\includegraphics[scale=0.04]{figures/ORCID_iD.pdf}}}$\and 
T. G. Wilson\inst{9} $^{\href{https://orcid.org/0000-0001-8749-1962}{\includegraphics[scale=0.04]{figures/ORCID_iD.pdf}}}$\and 
L. Borsato\inst{10} $^{\href{https://orcid.org/0000-0003-0066-9268}{\includegraphics[scale=0.04]{figures/ORCID_iD.pdf}}}$\and 
Sz. Csizmadia\inst{1} $^{\href{https://orcid.org/0000-0001-6803-9698}{\includegraphics[scale=0.04]{figures/ORCID_iD.pdf}}}$\and 
J. Cabrera\inst{1} $^{\href{https://orcid.org/0000-0001-6653-5487}{\includegraphics[scale=0.04]{figures/ORCID_iD.pdf}}}$\and 
H. Parviainen\inst{15} $^{\href{https://orcid.org/0000-0001-5519-1391}{\includegraphics[scale=0.04]{figures/ORCID_iD.pdf}}}$\and
A. C. M. Correia\inst{11}\and 
B. Akinsanmi\inst{8} $^{\href{https://orcid.org/0000-0001-6519-1598}{\includegraphics[scale=0.04]{figures/ORCID_iD.pdf}}}$\and 
N. Rosario\inst{2}\and 
P. Leonardi\inst{12}\and 
L. M. Serrano\inst{33} $^{\href{https://orcid.org/0000-0001-9211-3691}{\includegraphics[scale=0.04]{figures/ORCID_iD.pdf}}}$\and 
Y. Alibert\inst{13,14} $^{\href{https://orcid.org/0000-0002-4644-8818}{\includegraphics[scale=0.04]{figures/ORCID_iD.pdf}}}$\and 
R. Alonso\inst{15,16} $^{\href{https://orcid.org/0000-0001-8462-8126}{\includegraphics[scale=0.04]{figures/ORCID_iD.pdf}}}$\and 
J. Asquier\inst{17}\and 
T. Bárczy\inst{18} $^{\href{https://orcid.org/0000-0002-7822-4413}{\includegraphics[scale=0.04]{figures/ORCID_iD.pdf}}}$\and 
D. Barrado Navascues\inst{19} $^{\href{https://orcid.org/0000-0002-5971-9242}{\includegraphics[scale=0.04]{figures/ORCID_iD.pdf}}}$\and 
W. Baumjohann\inst{20} $^{\href{https://orcid.org/0000-0001-6271-0110}{\includegraphics[scale=0.04]{figures/ORCID_iD.pdf}}}$\and 
W. Benz\inst{14,13} $^{\href{https://orcid.org/0000-0001-7896-6479}{\includegraphics[scale=0.04]{figures/ORCID_iD.pdf}}}$\and 
N. Billot\inst{8} $^{\href{https://orcid.org/0000-0003-3429-3836}{\includegraphics[scale=0.04]{figures/ORCID_iD.pdf}}}$\and 
C. Broeg\inst{14,13} $^{\href{https://orcid.org/0000-0001-5132-2614}{\includegraphics[scale=0.04]{figures/ORCID_iD.pdf}}}$\and 
M.-D. Busch\inst{21}\and 
P. E. Cubillos\inst{22,20}\and 
M. B. Davies\inst{23} $^{\href{https://orcid.org/0000-0001-6080-1190}{\includegraphics[scale=0.04]{figures/ORCID_iD.pdf}}}$\and 
M. Deleuil\inst{24} $^{\href{https://orcid.org/0000-0001-6036-0225}{\includegraphics[scale=0.04]{figures/ORCID_iD.pdf}}}$\and 
A. Deline\inst{8}\and 
L. Delrez\inst{25,26,27} $^{\href{https://orcid.org/0000-0001-6108-4808}{\includegraphics[scale=0.04]{figures/ORCID_iD.pdf}}}$\and 
O. D. S. Demangeon\inst{2,3} $^{\href{https://orcid.org/0000-0001-7918-0355}{\includegraphics[scale=0.04]{figures/ORCID_iD.pdf}}}$\and 
B.-O. Demory\inst{13,14} $^{\href{https://orcid.org/0000-0002-9355-5165}{\includegraphics[scale=0.04]{figures/ORCID_iD.pdf}}}$\and 
A. Derekas\inst{28}\and 
B. Edwards\inst{29}\and 
D. Ehrenreich\inst{8,30} $^{\href{https://orcid.org/0000-0001-9704-5405}{\includegraphics[scale=0.04]{figures/ORCID_iD.pdf}}}$\and 
A. Erikson\inst{1}\and 
A. Fortier\inst{14,13} $^{\href{https://orcid.org/0000-0001-8450-3374}{\includegraphics[scale=0.04]{figures/ORCID_iD.pdf}}}$\and 
L. Fossati\inst{20} $^{\href{https://orcid.org/0000-0003-4426-9530}{\includegraphics[scale=0.04]{figures/ORCID_iD.pdf}}}$\and 
M. Fridlund\inst{31,32} $^{\href{https://orcid.org/0000-0002-0855-8426}{\includegraphics[scale=0.04]{figures/ORCID_iD.pdf}}}$\and 
D. Gandolfi\inst{33} $^{\href{https://orcid.org/0000-0001-8627-9628}{\includegraphics[scale=0.04]{figures/ORCID_iD.pdf}}}$\and 
K. Gazeas\inst{34}\and 
M. Gillon\inst{25} $^{\href{https://orcid.org/0000-0003-1462-7739}{\includegraphics[scale=0.04]{figures/ORCID_iD.pdf}}}$\and 
M. Güdel\inst{35}\and 
M. N. Günther\inst{36} $^{\href{https://orcid.org/0000-0002-3164-9086}{\includegraphics[scale=0.04]{figures/ORCID_iD.pdf}}}$\and 
A. Heitzmann\inst{8} $^{\href{https://orcid.org/0000-0002-8091-7526}{\includegraphics[scale=0.04]{figures/ORCID_iD.pdf}}}$\and 
Ch. Helling\inst{20,37}\and 
K. G. Isaak\inst{36} $^{\href{https://orcid.org/0000-0001-8585-1717}{\includegraphics[scale=0.04]{figures/ORCID_iD.pdf}}}$\and 
L. L. Kiss\inst{38,39}\and 
K. W. F. Lam\inst{1} $^{\href{https://orcid.org/0000-0002-9910-6088}{\includegraphics[scale=0.04]{figures/ORCID_iD.pdf}}}$\and 
J. Laskar\inst{40} $^{\href{https://orcid.org/0000-0003-2634-789X}{\includegraphics[scale=0.04]{figures/ORCID_iD.pdf}}}$\and 
A. Lecavelier des Etangs\inst{41} $^{\href{https://orcid.org/0000-0002-5637-5253}{\includegraphics[scale=0.04]{figures/ORCID_iD.pdf}}}$\and 
D. Magrin\inst{10} $^{\href{https://orcid.org/0000-0003-0312-313X}{\includegraphics[scale=0.04]{figures/ORCID_iD.pdf}}}$\and 
P. F. L. Maxted\inst{42} $^{\href{https://orcid.org/0000-0003-3794-1317}{\includegraphics[scale=0.04]{figures/ORCID_iD.pdf}}}$\and 
B. Merín\inst{43} $^{\href{https://orcid.org/0000-0002-8555-3012}{\includegraphics[scale=0.04]{figures/ORCID_iD.pdf}}}$\and 
C. Mordasini\inst{14,13}\and 
V. Nascimbeni\inst{10} $^{\href{https://orcid.org/0000-0001-9770-1214}{\includegraphics[scale=0.04]{figures/ORCID_iD.pdf}}}$\and 
G. Olofsson\inst{6} $^{\href{https://orcid.org/0000-0003-3747-7120}{\includegraphics[scale=0.04]{figures/ORCID_iD.pdf}}}$\and 
R. Ottensamer\inst{35}\and 
I. Pagano\inst{4} $^{\href{https://orcid.org/0000-0001-9573-4928}{\includegraphics[scale=0.04]{figures/ORCID_iD.pdf}}}$\and 
E. Pallé\inst{15,16} $^{\href{https://orcid.org/0000-0003-0987-1593}{\includegraphics[scale=0.04]{figures/ORCID_iD.pdf}}}$\and 
G. Peter\inst{44} $^{\href{https://orcid.org/0000-0001-6101-2513}{\includegraphics[scale=0.04]{figures/ORCID_iD.pdf}}}$\and 
D. Piazza\inst{14}\and 
G. Piotto\inst{10,45} $^{\href{https://orcid.org/0000-0002-9937-6387}{\includegraphics[scale=0.04]{figures/ORCID_iD.pdf}}}$\and 
D. Pollacco\inst{9}\and 
D. Queloz\inst{46,47} $^{\href{https://orcid.org/0000-0002-3012-0316}{\includegraphics[scale=0.04]{figures/ORCID_iD.pdf}}}$\and 
R. Ragazzoni\inst{10,45} $^{\href{https://orcid.org/0000-0002-7697-5555}{\includegraphics[scale=0.04]{figures/ORCID_iD.pdf}}}$\and 
N. Rando\inst{36}\and 
H. Rauer\inst{1,48} $^{\href{https://orcid.org/0000-0002-6510-1828}{\includegraphics[scale=0.04]{figures/ORCID_iD.pdf}}}$\and 
I. Ribas\inst{49,50} $^{\href{https://orcid.org/0000-0002-6689-0312}{\includegraphics[scale=0.04]{figures/ORCID_iD.pdf}}}$\and 
N. C. Santos\inst{2,3} $^{\href{https://orcid.org/0000-0003-4422-2919}{\includegraphics[scale=0.04]{figures/ORCID_iD.pdf}}}$\and 
G. Scandariato\inst{4} $^{\href{https://orcid.org/0000-0003-2029-0626}{\includegraphics[scale=0.04]{figures/ORCID_iD.pdf}}}$\and 
D. Ségransan\inst{8} $^{\href{https://orcid.org/0000-0003-2355-8034}{\includegraphics[scale=0.04]{figures/ORCID_iD.pdf}}}$\and 
A. E. Simon\inst{14,13} $^{\href{https://orcid.org/0000-0001-9773-2600}{\includegraphics[scale=0.04]{figures/ORCID_iD.pdf}}}$\and 
S. G. Sousa\inst{2} $^{\href{https://orcid.org/0000-0001-9047-2965}{\includegraphics[scale=0.04]{figures/ORCID_iD.pdf}}}$\and 
M. Stalport\inst{26,25}\and 
S. Sulis\inst{24} $^{\href{https://orcid.org/0000-0001-8783-526X}{\includegraphics[scale=0.04]{figures/ORCID_iD.pdf}}}$\and 
Gy. M. Szabó\inst{28,51} $^{\href{https://orcid.org/0000-0002-0606-7930}{\includegraphics[scale=0.04]{figures/ORCID_iD.pdf}}}$\and 
S. Udry\inst{8} $^{\href{https://orcid.org/0000-0001-7576-6236}{\includegraphics[scale=0.04]{figures/ORCID_iD.pdf}}}$\and 
B. Ulmer\inst{44}\and 
V. Van Grootel\inst{26} $^{\href{https://orcid.org/0000-0003-2144-4316}{\includegraphics[scale=0.04]{figures/ORCID_iD.pdf}}}$\and 
J. Venturini\inst{8} $^{\href{https://orcid.org/0000-0001-9527-2903}{\includegraphics[scale=0.04]{figures/ORCID_iD.pdf}}}$\and 
E. Villaver\inst{15,16}\and 
V. Viotto\inst{10} $^{\href{https://orcid.org/0000-0001-5700-9565}{\includegraphics[scale=0.04]{figures/ORCID_iD.pdf}}}$\and 
N. A. Walton\inst{52} $^{\href{https://orcid.org/0000-0003-3983-8778}{\includegraphics[scale=0.04]{figures/ORCID_iD.pdf}}}$\and 
R. West\inst{53,54} $^{\href{https://orcid.org/0000-0001-6604-5533}{\includegraphics[scale=0.04]{figures/ORCID_iD.pdf}}}$\and 
K. Westerdorff\inst{44}
}

\institute{
\label{inst:1} Institute of Planetary Research, German Aerospace Center (DLR), Rutherfordstrasse 2, 12489 Berlin, Germany \and
\label{inst:2} Instituto de Astrofisica e Ciencias do Espaco, Universidade do Porto, CAUP, Rua das Estrelas, 4150-762 Porto, Portugal \and
\label{inst:3} Departamento de Fisica e Astronomia, Faculdade de Ciencias, Universidade do Porto, Rua do Campo Alegre, 4169-007 Porto, Portugal \and
\label{inst:4} INAF, Osservatorio Astrofisico di Catania, Via S. Sofia 78, 95123 Catania, Italy \and
\label{inst:5} Lund Observatory, Division of Astrophysics, Department of Physics, Lund University, Box 118, 22100 Lund, Sweden \and
\label{inst:6} Department of Astronomy, Stockholm University, AlbaNova University Center, 10691 Stockholm, Sweden \and
\label{inst:7} Centre for Exoplanet Science, SUPA School of Physics and Astronomy, University of St Andrews, North Haugh, St Andrews KY16 9SS, UK \and
\label{inst:8} Observatoire astronomique de l'Université de Genève, Chemin Pegasi 51, 1290 Versoix, Switzerland \and
\label{inst:9} Department of Physics, University of Warwick, Gibbet Hill Road, Coventry CV4 7AL, United Kingdom \and
\label{inst:10} INAF, Osservatorio Astronomico di Padova, Vicolo dell'Osservatorio 5, 35122 Padova, Italy \and
\label{inst:11} CFisUC, Department of Physics, University of Coimbra, 3004-516 Coimbra, Portugal \and
\label{inst:12} Dipartimento di Fisica e Astronomia, Università degli Studi di Padova, Vicolo dell’Osservatorio 3, 35122 Padova, Italy \and
\label{inst:13} Center for Space and Habitability, University of Bern, Gesellschaftsstrasse 6, 3012 Bern, Switzerland \and
\label{inst:14} Weltraumforschung und Planetologie, Physikalisches Institut, University of Bern, Gesellschaftsstrasse 6, 3012 Bern, Switzerland \and
\label{inst:15} Instituto de Astrofísica de Canarias, Vía Láctea s/n, 38200 La Laguna, Tenerife, Spain \and
\label{inst:16} Departamento de Astrofísica, Universidad de La Laguna, Astrofísico Francisco Sanchez s/n, 38206 La Laguna, Tenerife, Spain \and
\label{inst:17} European Space Agency (ESA), ESTEC, Keplerlaan 1, 2201 AZ Noordwijk, The Netherlands \and
\label{inst:18} Admatis, 5. Kandó Kálmán Street, 3534 Miskolc, Hungary \and
\label{inst:19} Depto. de Astrofísica, Centro de Astrobiología (CSIC-INTA), ESAC campus, 28692 Villanueva de la Cañada (Madrid), Spain \and
\label{inst:20} Space Research Institute, Austrian Academy of Sciences, Schmiedlstrasse 6, A-8042 Graz, Austria \and
\label{inst:21} Physikalisches Institut, University of Bern, Gesellschaftsstrasse 6, 3012 Bern, Switzerland \and
\label{inst:22} INAF, Osservatorio Astrofisico di Torino, Via Osservatorio, 20, I-10025 Pino Torinese To, Italy \and
\label{inst:23} Centre for Mathematical Sciences, Lund University, Box 118, 221 00 Lund, Sweden \and
\label{inst:24} Aix Marseille Univ, CNRS, CNES, LAM, 38 rue Frédéric Joliot-Curie, 13388 Marseille, France \and
\label{inst:25} Astrobiology Research Unit, Université de Liège, Allée du 6 Août 19C, B-4000 Liège, Belgium \and
\label{inst:26} Space sciences, Technologies and Astrophysics Research (STAR) Institute, Université de Liège, Allée du 6 Août 19C, 4000 Liège, Belgium \and
\label{inst:27} Institute of Astronomy, KU Leuven, Celestijnenlaan 200D, 3001 Leuven, Belgium \and
\label{inst:28} ELTE Gothard Astrophysical Observatory, 9700 Szombathely, Szent Imre h. u. 112, Hungary \and
\label{inst:29} SRON Netherlands Institute for Space Research, Niels Bohrweg 4, 2333 CA Leiden, Netherlands \and
\label{inst:30} Centre Vie dans l’Univers, Faculté des sciences, Université de Genève, Quai Ernest-Ansermet 30, 1211 Genève 4, Switzerland \and
\label{inst:31} Leiden Observatory, University of Leiden, PO Box 9513, 2300 RA Leiden, The Netherlands \and
\label{inst:32} Department of Space, Earth and Environment, Chalmers University of Technology, Onsala Space Observatory, 439 92 Onsala, Sweden \and
\label{inst:33} Dipartimento di Fisica, Università degli Studi di Torino, via Pietro Giuria 1, I-10125, Torino, Italy \and
\label{inst:34} National and Kapodistrian University of Athens, Department of Physics, University Campus, Zografos GR-157 84, Athens, Greece \and
\label{inst:35} Department of Astrophysics, University of Vienna, Türkenschanzstrasse 17, 1180 Vienna, Austria \and
\label{inst:36} European Space Agency (ESA), European Space Research and Technology Centre (ESTEC), Keplerlaan 1, 2201 AZ Noordwijk, The Netherlands \and
\label{inst:37} Institute for Theoretical Physics and Computational Physics, Graz University of Technology, Petersgasse 16, 8010 Graz, Austria \and
\label{inst:38} Konkoly Observatory, Research Centre for Astronomy and Earth Sciences, 1121 Budapest, Konkoly Thege Miklós út 15-17, Hungary \and
\label{inst:39} ELTE E\"otv\"os Lor\'and University, Institute of Physics, P\'azm\'any P\'eter s\'et\'any 1/A, 1117 Budapest, Hungary \and
\label{inst:40} IMCCE, UMR8028 CNRS, Observatoire de Paris, PSL Univ., Sorbonne Univ., 77 av. Denfert-Rochereau, 75014 Paris, France \and
\label{inst:41} Institut d'astrophysique de Paris, UMR7095 CNRS, Université Pierre \& Marie Curie, 98bis blvd. Arago, 75014 Paris, France \and
\label{inst:42} Astrophysics Group, Lennard Jones Building, Keele University, Staffordshire, ST5 5BG, United Kingdom \and
\label{inst:43} European Space Agency, ESA - European Space Astronomy Centre, Camino Bajo del Castillo s/n, 28692 Villanueva de la Cañada, Madrid, Spain \and
\label{inst:44} Institute of Optical Sensor Systems, German Aerospace Center (DLR), Rutherfordstrasse 2, 12489 Berlin, Germany \and
\label{inst:45} Dipartimento di Fisica e Astronomia "Galileo Galilei", Università degli Studi di Padova, Vicolo dell'Osservatorio 3, 35122 Padova, Italy \and
\label{inst:46} ETH Zurich, Department of Physics, Wolfgang-Pauli-Strasse 2, CH-8093 Zurich, Switzerland \and
\label{inst:47} Cavendish Laboratory, JJ Thomson Avenue, Cambridge CB3 0HE, UK \and
\label{inst:48} Institut fuer Geologische Wissenschaften, Freie Universitaet Berlin, Maltheserstrasse 74-100,12249 Berlin, Germany \and
\label{inst:49} Institut de Ciencies de l'Espai (ICE, CSIC), Campus UAB, Can Magrans s/n, 08193 Bellaterra, Spain \and
\label{inst:50} Institut d'Estudis Espacials de Catalunya (IEEC), 08860 Castelldefels (Barcelona), Spain \and
\label{inst:51} HUN-REN-ELTE Exoplanet Research Group, Szent Imre h. u. 112., Szombathely, H-9700, Hungary \and
\label{inst:52} Institute of Astronomy, University of Cambridge, Madingley Road, Cambridge, CB3 0HA, United Kingdom \and
\label{inst:53} Centre for Exoplanets and Habitability, University of Warwick, Gibbet Hill Road, Coventry CV4 7AL, UK \and
\label{inst:54} Department of Physics, University of Warwick, Gibbet Hill Road, Coventry CV4 7AL, UK
}

   \date{Received June 06, 2024; accepted XX XX, 202X}

 
  \abstract
   {Hot Jupiters with close-by planetary companions are rare, with only a handful of them having been discovered so far. This could be due to their suggested dynamical histories, leading to the possible ejection of other planets. TOI-2109\,b is special in this regard because it is the hot Jupiter with the closest relative separation from its host star, being separated by less than 2.3 stellar radii. Unexpectedly, transit timing measurements from recently obtained CHEOPS observations show low amplitude transit-timing variations (TTVs).}
   {We aim to search for signs of orbital decay and to characterise the apparent TTVs, trying to gain information about a possible companion.}
   {We fit the newly obtained CHEOPS light curves using \texttt{TLCM} and extract the resulting mid-transit timings. Successively, we use these measurements in combination with TESS and archival photometric data and radial velocity data to estimate the rate of tidal orbital decay of TOI-2109\,b, as well as characterise the TTVs using the N-body code \texttt{TRADES} and the photodynamical approach of \texttt{PyTTV}.}
   {
   We find tentative evidence at $3\sigma$ for orbital decay in the TOI-2109 system, when we correct the mid-transit timings using the best-fitting sinusoidal model of the TTVs.
   We do not detect additional transits in the available photometric data, but find evidence towards the authenticity of the apparent TTVs, indicating
   a close-by, outer companion with $P_\mathrm{c} > 1.125\,$d. Due to the fast rotation of the star, the new planetary candidate cannot be detected in the available radial velocity (RV) measurements, and its parameters can only be loosely constrained by our joint TTV and RV modelling. }
   {TOI-2109 could join a small group of rare hot Jupiter systems that host close-by planetary companions, only one of which (WASP-47\,b) has an outer companion.
   More high-precision photometric measurements are necessary to confirm the existence of this planetary companion.}

   \keywords{Planet-star interactions -- Planets and satellites: dynamical evolution and stability -- Techniques: photometric}

   \maketitle
%
\section{Introduction}

\begin{figure}
    \centering
    \includegraphics[width=0.9\linewidth]{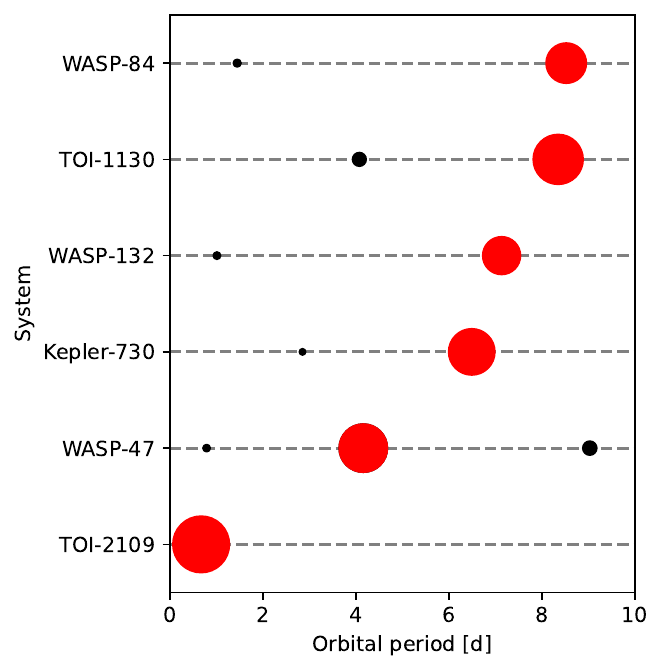}
    \caption{Overview of hot Jupiters (red) with close-by companions (black), including TOI-2109\,b. Sizes given to scale. We note that the WASP-47 system also contains a long-period companion that is not shown here. System parameters taken from \citet{2019ApJ...870L..17C, 2021AJ....162..256W, 2022AJ....164...13H, 2023A&A...675A.115K, 2023MNRAS.525L..43M, 2023A&A...673A..42N}.}
    \label{fig:HJ_companion_overview}
\end{figure}

There are more than 500\footnote{According to the \href{https://exoplanetarchive.ipac.caltech.edu/}{NASA Exoplanet Archive} as of 26.02.2024.} hot Jupiter (HJ) systems known, only a few of which are known to host close-by planetary companions. These are WASP-47 \citep{2012MNRAS.426..739H, 2015ApJ...812L..18B, 2016A&A...586A..93N, 2023A&A...673A..42N}, Kepler-730 \citep{2018ApJS..235...38T, 2018RNAAS...2..160Z, 2019ApJ...870L..17C}, TOI-1130 \citep{2020ApJ...892L...7H, 2023A&A...675A.115K}, TOI-1408 \citep{2023MNRAS.526L.111G, 2024ApJ...971L..28K}, WASP-132 \citep{2017MNRAS.465.3693H, 2022AJ....164...13H}, and WASP-84 \citep{2014MNRAS.445.1114A, 2023MNRAS.525L..43M}. The HJs in these systems all host only inner companions in the Earth- to Neptune-regime, except for WASP-47, which also contains a close outer sub-Neptune companion to the HJ, and a long-period giant planet. An overview of these systems is given in Fig.~\ref{fig:HJ_companion_overview}.
Other systems worth mentioning are TOI-2000 \citep{2023MNRAS.524.1113S}, containing a hot Saturn with an inner companion, TOI-5398 \citep{2024A&A...682A.129M}, hosting a warm Saturn with an inner sub-Neptune, and WASP-148 \citep{2020A&A...640A..32H}, consisting of a hot Saturn and a non-transiting misaligned outer Saturn, near a 4:1 mean-motion resonance (MMR).
Furthermore, a recent study by \citet{2023AJ....165..171W} finds that at least $12\%\pm6\%$ of hot Jupiters should have close-by companions, according to their analysis of the full Kepler data set. This indicates that there should be more of these hitherto rare systems within the sample of known hot Jupiter systems.

Theories for the origins of hot Jupiters include in situ formation, disk migration and high-eccentricity tidal migration (for a review see e.g. \citealt{2018ARA&A..56..175D}). 
As found by \citet{2022ApJ...926L..17R}, the distributions of obliquities and eccentricities that have been observed, are consistent with high-eccentricity migration and tidal damping, and do not strictly require in situ formation or disk migration.
Due to strong tidal interactions with their host star, misalignments and eccentricities can be damped relatively quickly, effectively erasing the imprints of formation and migration. However, this only applies to hot Jupiters orbiting cool stars, since tidal interactions are less efficient for hot host stars, as has been observed by \citet{2021ApJ...916L...1A}. A recent review on stellar obliquities is given by \citet{2022PASP..134h2001A}.
Violent migration scenarios, like planet-planet scattering, make it hard to explain close planetary companions to hot Jupiters. A study on in situ formation using N-body simulations, focussed on hot Jupiters with companion super-Earths, found that they can produce such systems, obtaining similar occurrence rates for hot Jupiters with inner companions, but according to their simulations, hot Jupiters should not be detected as single transiting planets as often. A fraction of their simulations also led to two close-in gas giants \citep{2021MNRAS.505.2500P}.

Close-by planetary companions, especially when in resonance, can induce strong transit timing variations (TTVs) on one another, see e.g. \citet{2023A&A...673A..42N} for a dynamical modeling of the WASP-47 system, making use of high-precision CHEOPS photometry. A few other prominent examples of the effect of TTVs are the Kepler-9 system \citep{2010Sci...330...51H}, Kepler-80 \citep{2016AJ....152..105M}, TRAPPIST-1 \citep{2017Natur.542..456G, 2018A&A...613A..68G}, and TOI-178 \citep{2021A&A...649A..26L}. Examining TTVs has the potential to deliver more information about the observed system, like constraining planetary masses or orbital periods of non-transiting companions \citep{2012ApJ...761..122L, 2013ApJ...772...74W, 2014ApJ...787...80H, 2017AJ....154....5H}.

Tidal orbital decay is caused by gravitational interactions of a planet with its host star. These interactions raise a bulge on the surface of the star, which can lead to the transfer of angular momentum between the two bodies. Orbital decay only happens if the star is rotating more slowly than the planet is orbiting around the star in the case of spin-orbit alignment because the tidal bulge will be dragged behind the moving sub-planetary point on the stellar surface. Consequently, the planetary orbit shrinks, while the star spins up \citep{1973ApJ...180..307C, 1996ApJ...470.1187R}. There is always an offset between the position of the stellar bulge and the sub-planetary point due to the viscosity of the plasma, except if the orbital period of the planet and the rotation of the star are synchronised (see e.g. \citealt{1981A&A....99..126H}). 
Even outward migration of the planet is possible, if the star is rotating faster than the planet is orbiting. A measure of the efficiency of the dissipation of orbital kinetic energy due to friction within the star, in the case of tidal orbital decay, is given by the modified stellar tidal quality factor $Q'_\star$ \citep{1966Icar....5..375G}. Theory suggests values ranging from 10$^5$ to 10$^{9}$ for stars, with smaller values indicating a higher dissipation efficiency (see e.g., \citealt{2005ApJ...620..970M, 2008ApJ...678.1396J, 2007ApJ...661.1180O, 2011ApJ...731...67P, 2012MNRAS.422.3151H, 2012ApJ...751...96P, 2014ARA&A..52..171O, 2018MNRAS.476.2542C}).
\citet{2018AJ....155..165P} treated $Q'_\star$ as being dependent on the tidal forcing period rather than being a constant for different stellar types, and found $Q'_\star$ in the range from $10^5$ to $10^7$ for orbital periods between 0.5 days and 2 days.
Measurements of $Q'_\star$ from WASP-12\,b and Kepler-1658\,b give values of $Q'_\star = 1.8 \times 10^5$ \citep{2020ApJ...888L...5Y} and $Q'_\star = 2.5 \times 10^4$ \citep{2022ApJ...941L..31V}, respectively, where the host star of the latter is a subgiant. Additionally, several lower limits for $Q'_\star$ were established in systems where tidal orbital decay could not be confirmed yet (see e.g., \citealt{2020AcA....70....1M, 2020AJ....159..150P, 2022A&A...657A..52B, 2022AJ....163..120G, 2022ApJS..259...62I, 2022A&A...668A.114R, 2023A&A...669A.124H, 2024ApJS..270...14W}).

TOI-2109 is a fast-rotating ($v\sin i = (81.2\pm1.6)$\,km\,s$^{-1}$) F-type star, with a mass of $M_\star = 1.447^{+0.075}_{-0.078}\,M_\odot$ and a radius of $R_\star = 1.698^{+0.062}_{0.057}$ \citep{2021AJ....162..256W}. It hosts an ultra-hot Jupiter, TOI-2109\,b, in a tight $P=0.672\,$d orbit, making it the shortest orbital period hot Jupiter orbiting a Sun-like star, with a separation of less than 2.3 stellar radii from the center of its star \citep{2021AJ....162..256W}.
Its predicted orbital decay rate is one of the highest amongst all hot Jupiters (see Fig.~14 of \citealt{2021AJ....162..256W}), hence making this planet a very interesting target for orbital decay studies.
However, the fast rotation of the stellar host impacts the expected magnitude of the orbital decay signature due to the small difference of rotation period and orbital period, which, in combination with the shallower convective zones of these stars, might lead to less efficient tidal interactions (see e.g. \citealt{2019CeMDA.131...30B, 2023A&A...669A.124H}).
Its equilibrium temperature is the second highest of all known exoplanets with $T_\mathrm{eq} = 3646\pm88\,$K, being slightly cooler than KELT-9\,b with $T_\mathrm{eq} = 3921^{+182}_{-174}$ \citep{2017Natur.546..514G, 2022A&A...666A.118J}.
Contrary to many other hot Jupiters orbiting hot stars, TOI-2109\,b is aligned ($\lambda = 1.7^\circ \pm 1.7^\circ$). Yet, the stellar effective temperature is not far off the Kraft break \citep{1967ApJ...150..551K} at $T_\mathrm{eff} = 6530^{+130}_{-120}\,$K.

CHEOPS (CHaracterising ExOPlanet Satellite) is an S-class mission of the European Space Agency (ESA), dedicated for photometric follow-up of transiting planets of bright stars \citep{2021ExA....51..109B}. With its defocused 32\,cm telescope, it delivers high-precision photometry, which is able to precisely constrain planetary radii (e.g. \citealt{2021ExA....51..109B, 2022A&A...659A..74D}). We observed TOI-2109 under the ``Tidal decay (ID 0012)'' and the ``TIDES (ID 0063)'' programmes, the first of which is dedicated to observing hot Jupiters for measurements of tidal orbital decay as part of the Guaranteed Time Observing (GTO) programme of CHEOPS \citep{2022A&A...657A..52B}, with the latter being part of the first extension of the CHEOPS mission.
CHEOPS' performance and its potential for the characterisation of ultra-hot Jupiters have been demonstrated in \citep{2020A&A...643A..94L}.

In this paper, we present new transit and phase curve observation of TOI-2109\,b, taken with the CHEOPS space telescope, and analyse the apparent transit timing variations. In Sect.~\ref{sec:obs}, we describe the observations that were used consecutively in our light curve analysis (Sect.~\ref{sec:lcs}), orbital decay analysis (Sect.~\ref{sec:decay}), and TTV analysis (Sect.~\ref{sec:TTVs}). The conclusions are presented in Sect.~\ref{sec:conclusion}.

\section{Observations\label{sec:obs}}

TOI-2109 was observed by CHEOPS in 31 visits, eight of which were phase curve observations, with the rest having been transit observations. Details, including file keys for data access, can be found in Table\,\ref{table:obs_log}.
These data can be accessed via the DACE (Data \& Analysis Center for Exoplanets) website\footnote{\href{https://dace.unige.ch/cheopsDatabase/?}{https://dace.unige.ch/cheopsDatabase/?}} of the University of Geneva, via the CHEOPS archive\footnote{\href{https://cheops-archive.astro.unige.ch/archive_browser/}{https://cheops-archive.astro.unige.ch/archive\_browser/}}, via the \texttt{PYCHEOPS} Python package \citep{2022MNRAS.514...77M} using the provided file keys, or at the CDS. We fully process this data ourselves.

In addition to observations by CHEOPS, TOI-2109 has also been observed in two TESS Sectors, 25 and 52 (Cycle 2 and 4). However, only the latter is available at a cadence of 120\,s, with the first being extracted from the full frame images FFIs), resulting in a cadence of 30\,min\footnote{These data are freely available via the \href{https://mast.stsci.edu/portal/Mashup/Clients/Mast/Portal.html}{MAST Portal}.}. We fit these light curves to derive the mid-transit times.

Furthermore, \citet{2021AJ....162..256W} took 20 (partial) transit observations with ground-based telescopes\footnote{These, and more, are freely available on \href{https://exofop.ipac.caltech.edu/tess/target.php?id=392476080}{ExoFOP}.}. These include observations from the Fred Lawrence Whipple Observatory with KeplerCam, the University of Louisville Manner Telescope, the Maury Lewin Astronomical Observatory, MuSCAT2 with TCS at Teide Observatory, MuSCAT3 with FTN at Haleakala Observatory, the Wild Boar Remote Observatory, the Grand-Pra Observatory with the RCO, and the Las Cumbres Observatory Global Telescope network using the telescopes of the McDonald observatory, Siding Spring Observatory, South African Astronomical Observatory and the Cerro Tololo Interamerican Observatory.
Aside from photometric observations, \citet{2021AJ....162..256W} also published a total of 58 measured radial velocities (RVs) from TRES and FIES, contributing 47 (including 28 from transit spectroscopy observations) and 11, respectively.
For these data, we use the derived timings and RVs as provided by \citet{2021AJ....162..256W}.

Besides these data, TOI-2109 was also observed by several cameras of SuperWASP-N  \citep{2006PASP..118.1407P} from 2006 to 2011, giving us early photometry for our orbital decay analysis. After phase-folding, the transit is clearly visible.
We fully process this data ourselves.

\begin{table*}
\caption{Log of CHEOPS observations of TOI-2109. }
\label{table:obs_log}
\renewcommand{\arraystretch}{1.1} 
\centering
\begin{tabular}{c c c c c c}
\hline\hline
Visit no. & Start date & Duration & No. of data & Eff. & File key \\ 
          & [UTC]      & [h]       & points     & [$\%$] & \\\hline
    1 & 2021-05-14 22:35:37 & 5.29 & 249 & 78.3 & CH\_PR100012\_TG002001\_V0300 \\ 
    2 & 2021-05-18 23:00:37 & 5.62 & 275 & 81.3 & CH\_PR100012\_TG002002\_V0300 \\ 
    3 & 2021-05-21 15:55:17 & 5.44 & 260 & 79.5 & CH\_PR100012\_TG002003\_V0300 \\ 
    4 & 2021-05-27 17:00:17 & 5.42 & 255 & 78.2 & CH\_PR100012\_TG002004\_V0300 \\ 
    5 & 2021-05-31 01:40:37 & 5.64 & 301 & 88.7 & CH\_PR100012\_TG002101\_V0300 \\ 
    6 & 2021-06-04 02:36:18 & 5.64 & 305 & 89.9 & CH\_PR100012\_TG002102\_V0300 \\ 
    7 & 2021-06-10 19:44:37 & 5.64 & 277 & 81.7 & CH\_PR100012\_TG002103\_V0300 \\ 
    8 & 2021-06-12 04:36:17 & 5.95 & 276 & 77.1 & CH\_PR100012\_TG002104\_V0300 \\ 
    9 & 2021-06-13 12:33:36 & 5.60 & 302 & 89.6 & CH\_PR100012\_TG002105\_V0300 \\ 
    10 & 2021-06-16 04:50:17 & 5.64 & 264 & 77.8 & CH\_PR100012\_TG002106\_V0300 \\ 
    11 & 2022-04-10 19:21:38 & 5.54 & 231 & 69.3 & CH\_PR100012\_TG002601\_V0300 \\ 
    12 & 2022-04-16 04:03:18 & 5.64 & 206 & 60.7 & CH\_PR100012\_TG002602\_V0300 \\ 
    13 & 2022-04-25 14:24:38 & 6.44 & 263 & 67.9 & CH\_PR100012\_TG002603\_V0300 \\ 
    14 & 2022-05-30 13:29:38 & 5.59 & 294 & 87.5 & CH\_PR100012\_TG003301\_V0300 \\ 
    15 & 2022-05-31 05:32:18 & 5.64 & 273 & 80.5 & CH\_PR100012\_TG003302\_V0300 \\ 
    16 & 2022-05-31 21:41:38 & 5.69 & 303 & 88.6 & CH\_PR100012\_TG003303\_V0300 \\ 
    17 & 2022-06-02 06:13:18 & 5.45 & 260 & 79.2 & CH\_PR100012\_TG003304\_V0300 \\ 
    18 & 2022-06-02 22:18:38 & 5.59 & 301 & 89.5 & CH\_PR100012\_TG003305\_V0300 \\ 
    19 & 2022-06-06 06:37:38 & 5.64 & 276 & 81.4 & CH\_PR100012\_TG003306\_V0300 \\ 
    20 & 2022-06-12 07:53:18 & 5.64 & 268 & 79.0 & CH\_PR100012\_TG003307\_V0300 \\ 
    21 & 2023-04-15 00:10:19 & 5.62 & 235 & 69.5 & CH\_PR140063\_TG000101\_V0300 \\ 
    22 & 2023-04-21 01:29:19 & 5.34 & 220 & 68.5 & CH\_PR140063\_TG000102\_V0300 \\ 
    23 & 2023-04-27 02:37:19 & 5.25 & 210 & 66.4 & CH\_PR140063\_TG000103\_V0300 \\ 
    24 & 2023-05-13 07:50:19 & 23.74 & 1042 & 73.1 & CH\_PR100012\_TG003901\_V0300 \\ 
    25 & 2023-05-17 21:41:59 & 26.00 & 1165 & 74.6 & CH\_PR100012\_TG003902\_V0300 \\ 
    26 & 2023-06-01 06:26:59 & 25.35 & 1297 & 85.2 & CH\_PR100012\_TG003903\_V0300 \\ 
    27 & 2023-06-02 08:23:18 & 23.48 & 1209 & 85.8 & CH\_PR100012\_TG003904\_V0300 \\ 
    28 & 2023-06-07 09:13:19 & 24.68 & 1240 & 83.7 & CH\_PR100012\_TG004401\_V0300 \\ 
    29 & 2023-06-08 18:59:38 & 23.56 & 1177 & 83.2 & CH\_PR100012\_TG004501\_V0300 \\ 
    30 & 2023-06-12 19:15:39 & 23.74 & 1132 & 79.4 & CH\_PR100012\_TG004801\_V0300 \\ 
    31 & 2023-06-16 03:29:19 & 23.74 & 1106 & 77.6 & CH\_PR100012\_TG005001\_V0300 \\ 
\hline
\end{tabular}
\tablefoot{The efficiency (``Eff.'') describes the percentage of the time on target that is spent collecting data. The file key is a unique identifier that can be used to access the data on the CHEOPS archive.}
\end{table*}

\section{Light curves\label{sec:lcs}}

\subsection{Preparation and reduction}
We make use of both the standard CHEOPS Data Reduction Pipeline (\texttt{DRP} Version 14.1.2, \citealt{2020A&A...635A..24H}) and \texttt{PIPE}\footnote{PIPE can be obtained at \href{https://github.com/alphapsa/PIPE}{GitHub}.} (PSF Imagette Photometric Extraction, Brandeker et al., in prep.; \citealt{2021A&A...651L..12M, 2021A&A...654A.159S, 2022A&A...659L...4B}), which is complementary to the \texttt{DRP}. 
In contrast to the \texttt{DRP}'s aperture photometry on the subarrays, \texttt{PIPE} makes use of the point-spread function (PSF) photometry from the imagettes, coming from the satellite. 
There are trade-offs to be had between the two pipelines. \texttt{PIPE} can, amongst other things, deliver shorter cadence photometry (although not in the case of TOI-2109), provide a better cosmic ray correction, can deal better with hot pixels and faint targets, and doesn't require the roll-angle correction that is necessary in the \texttt{DRP} photometry, originating in CHEOPS' nadir-locked orbit and its rotating field of view. 
Some drawbacks of using \texttt{PIPE} can be caused by inaccuracies in the PSF modelling, which is dependent on pointing jitter, the location on the detector, the temperature of the telescope and the spectral energy distribution of the target star.

Reduction of the \texttt{DRP} and \texttt{PIPE} data follow a similar recipe. In the case of the \texttt{DRP} data, we use \texttt{PYCHEOPS} for the data handling. We use the ``DEFAULT'' aperture data and choose the ``decontaminate'' option when extracting the light curve, which performs a subtraction of the contamination of nearby stars ($> 2000$\,ppm for TOI-2109\,b). After this, we clip outliers that are deviating by $5\sigma$ or more from the mean absolute deviation of the median-smoothed light curve. The next step is removing data points with high background values. We correct for the ramp effect (\citealt{2021A&A...653A.173M}; Fortier et al., submitted) via linear decorrelation against the ``thermfront2'' sensor data and clip any remaining outliers at $5\sigma$ again (this is mostly done to clip remaining outliers in the observed phase curves, the transit observations remain largely unaffected).
For the \texttt{PIPE} data, we start by removing the data points flagged by the pipeline and those that show high sky background values. In the next step, we clip the remaining outliers as we did with the \texttt{DRP} data.

For the reduction of the WASP data, we visually inspect all available fields and mask out all fields except 1, 3, 5, 6, and 8 due to bad transit coverage and/or high scatter. Subsequently, we filter out all data points with values of ``SIGMA\_XS'' $> 0.03$, which is a measure of the root-mean-square (RMS) scatter of the magnitude of all stars in the frame relative to their mean values. This leaves us with 2110 datapoints that were acquired over a range of 121 days in 2006.

\subsection{Light curve modelling}

For modelling of the light curves, we make use of the Transit and Light Curve Modeller (\texttt{TLCM}, \citealt{2020MNRAS.496.4442C}). A short description of the setup and \texttt{TLCM} is given in \citet{2023A&A...669A.124H}. 
We do not fit the gravity darkening parameters, since the planet's orbit is aligned and thus does not show a modulation of the light curve during transit.
For the phase curves we are additionally fitting the geometric albedo $A_\mathrm{g}$ of the planet, mass ratio of star and planet from the ellipsoidal effect $q$, and the shift of the brightest point on the planetary surface from the substellar point $\varepsilon$, using \texttt{TLCM}'s Lambertian phase curve model, akin to the process in \citet{2024arXiv240319468K}.
This means that only the reflected light component is regarded for the phase curve fit, which is sufficient for our analysis, since a detailed atmospheric analysis is not within the scope of this paper and will be published in (Singh et al., in prep.).

In the case of the CHEOPS and TESS data, we firstly complete a combined fit of all available light curves each, yielding a phase curve fit each
to the CHEOPS \texttt{DRP} data set, the CHEOPS \texttt{PIPE} data set and the 2\,min cadence TESS data from Sector 52.
The roll-angle effect of the CHEOPS photometry is corrected in the case of the \texttt{DRP} data using the following equation, while fitting the roll-angle parameters $RA_1$ to $RA_6$:
\begin{align}
    f_\phi =& \,\, RA_1 \sin(\phi) + RA_2\sin(2\cdot\phi) + RA_3\sin(3\cdot\phi)\\
    &+ RA_4\cos(\phi) + RA_5 \cos(2\cdot\phi) + RA_6\cos(3\cdot\phi),\notag
\end{align}
with $f_\phi$ being the roll-angle component of the fitted model used
to decorrelate against the roll-angle of the satellite, and $\phi$ being the roll-angle value itself.

\begin{figure}
    \centering
    \includegraphics[width=\linewidth]{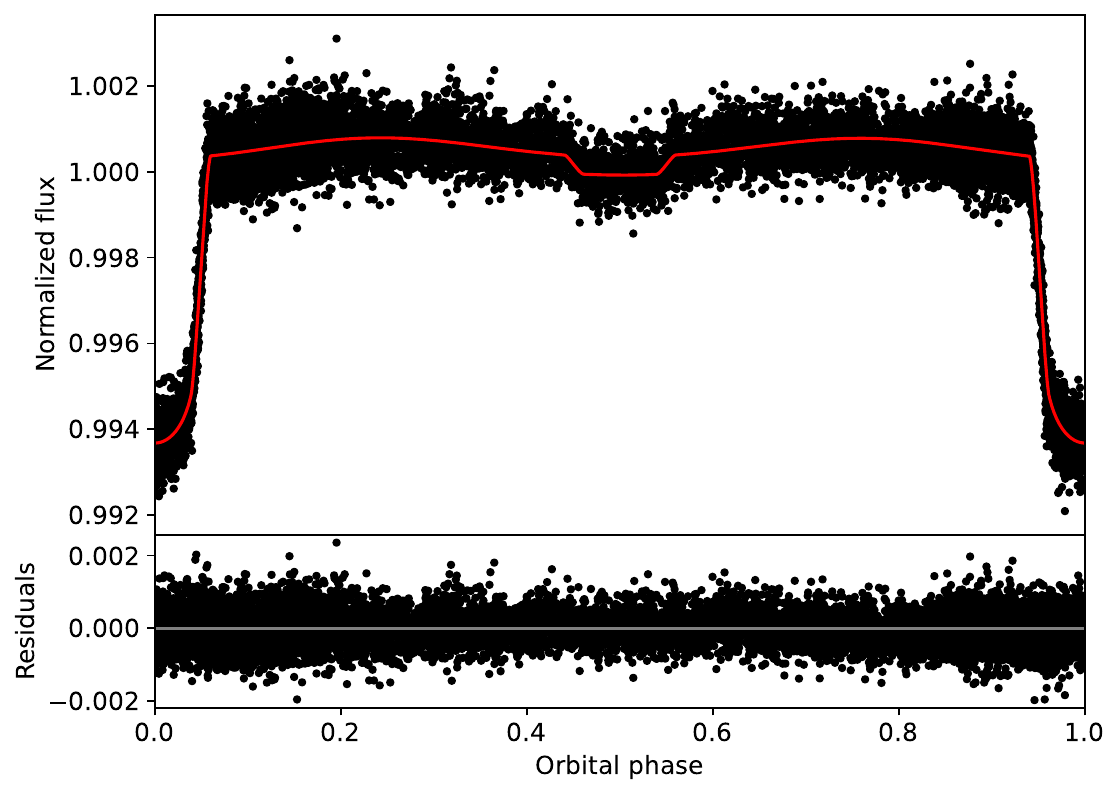}
    \caption{Phase curve fit (top) of the \texttt{PIPE} data (black) shown with \texttt{TLCM}'s median solution phase curve model (red). The residuals of the fit are shown in the bottom plot.}
    \label{fig:pc_pipe_norn}
\end{figure}

\texttt{TLCM} uses the following reparameterisation for the limb darkening parameters:
\begin{align}
    A &= \frac{1}{4}\left(u_1\left(\frac{1}{\alpha} - \frac{1}{\beta}\right) + u_2-\left(\frac{1}{\alpha} + \frac{1}{\beta}\right)\right), \\
    B &= \frac{1}{4}\left(u_1\left(\frac{1}{\alpha} + \frac{1}{\beta}\right) + u_2\left(\frac{1}{\alpha} - \frac{1}{\beta}\right)\right),
\end{align}
as implemented in \citet{2024arXiv240319468K} with $\alpha = \frac{1}{2}\cos(77^\circ)$, $\beta = \frac{1}{2}\sin(77^\circ)$, and $u_1$ and $u_2$ the quadratic limb darkening parameters from \citet{2017A&A...600A..30C} for TESS and \citet{2021RNAAS...5...13C} for CHEOPS.
The priors of the phase curve fits are shown in Table~\ref{tab:priors} and the results are shown in Table~\ref{tab:pc1}. 
An example of a \texttt{TLCM} fit to the phase curve from the \texttt{PIPE} data reduction is shown in Fig.~\ref{fig:pc_pipe_norn}.
In addition, we also tried fitting eccentric orbits in all cases, but found no significant improvements in the fits, which is why we decided to only use the solutions of the circular cases for the following fits to the individual transits.

\begin{table}[]
    \centering
    \renewcommand{\arraystretch}{1.1} 
    \caption{Priors for the \texttt{TLCM} phase curve fits. All data sets use the same uniform priors ($\mathcal{U}$) for the respective parameters, except for the limb darkening parameters, where we differentiate between the CHEOPS and TESS data, and apply Gaussian priors ($\mathcal{N}$). The $\sigma=0.46$ in the Gaussian priors on $A$ and $B$ originates in the conversion from the standard quadratic limb darkening parameters $u_a$ and $u_b$ to \texttt{TLCM}'s parameterisation and reflects a $\sigma=0.1$ on the standard parameters.}
    \begin{tabular}{l c}  \hline\hline
        Parameter & Prior \\ \hline
        $T_{0,\mathrm{CHEOPS}}$ [BJD$_\mathrm{TDB}$] & $\mathcal{U}(10049.387, 10049.787)$ \\
        $T_{0,\mathrm{TESS}}$ [BJD$_\mathrm{TDB}$] & $\mathcal{U}(9729.970, 9730.370)$ \\
        $P$ [d] & $\mathcal{U}(0.671474, 0.673474)$ \\
        $a/R_\star$ & $\mathcal{U}(2.17, 2.37)$ \\
        $R_\mathrm{p}/R_\star$ & $\mathcal{U}(0.0716, 0.0916)$ \\
        $b$ & $\mathcal{U}(0.2, 1.2)$ \\
        $A_\mathrm{g}$ & $\mathcal{U}(-1.0, 3.0)$ \\
        $\varepsilon$ [$^\circ$]& $\mathcal{U}(-100.0, 100.0)$\\
        $q$ & $\mathcal{U}(0.0013, 0.0053)$\\
        $A_\mathrm{CHEOPS}$ & $\mathcal{N}(1.92, 0.46)$ \\
        $B_\mathrm{CHEOPS}$ & $\mathcal{N}(2.33, 0.46)$ \\
        $A_\mathrm{TESS}$ & $\mathcal{N}(1.12, 0.46)$ \\
        $B_\mathrm{TESS}$ & $\mathcal{N}(1.68, 0.46)$ \\
        $RA_1$ & $\mathcal{U}(-0.01, 0.01)$ \\
        $RA_2$ & $\mathcal{U}(-0.01, 0.01)$ \\
        $RA_3$ & $\mathcal{U}(-0.01, 0.01)$ \\
        $RA_4$ & $\mathcal{U}(-0.01, 0.01)$ \\
        $RA_5$ & $\mathcal{U}(-0.01, 0.01)$ \\
        $RA_6$ & $\mathcal{U}(-0.01, 0.01)$ \\\hline
    \end{tabular}
    \tablefoot{The values in the parantheses describe the lower and upper boundaries of the allowed interval for the uniform priors, in the case of Gaussian priors, they describe the mean and standard deviation. $b$ denotes the impact parameter, and $A$ and $B$ \texttt{TLCM}'s limb darkening parameterisation. The roll-angle parameters $RA_1$ to $RA_6$ are only fitted for \texttt{DRP} light curves.}
    \label{tab:priors}
\end{table}

\begin{table*}[]
    \centering
    \renewcommand{\arraystretch}{1.1} 
    \caption{Results of the phase curve fits of TOI-2109\,b, using the three data sets (CHEOPS (\texttt{DRP}), CHEOPS (\texttt{PIPE}) and TESS).
    }
    \begin{tabular}{l c c c}\hline\hline
        Parameter & CHEOPS (\texttt{DRP}) & CHEOPS (\texttt{PIPE}) & TESS \\\hline
        $T_0$ [BJD$_\mathrm{TDB}$] & $10049.588137 \pm 0.000062$ & $10049.588259\pm0.000052$ & $9730.163124\pm0.000100$ \\
        $P$ [d] & $ 0.67247397\pm0.00000009$ & $0.67247417\pm0.00000008 $ & $0.67248057\pm0.00000890$ \\
        $a/R_\star$ & $2.202\pm0.018$ & $2.208\pm0.013$ & $ 2.257\pm0.053$\\
        $R_\mathrm{p}/R_\star$ & $0.08042\pm0.00085$ & $0.08121\pm0.00071$ & $0.07957\pm0.00080$\\
        $b$ & $0.766\pm0.017$ & $0.769\pm0.014$ & $0.745\pm0.054$ \\
        $i$\tablefootmark{a} [$^\circ$] & $69.64\pm0.25$ & $69.61\pm0.19$ & $70.71\pm0.73$ \\
        $A_\mathrm{g}$ & $1.07\pm1.35$ & $ 1.08\pm1.35$ & $ 0.97\pm1.35$\\
        $\varepsilon [^\circ]$ & $-1.16\pm68.35$ & $0.59\pm68.46$ & $ -1.86\pm68.61$\\
        $q$ & $0.0015\pm0.0001$ & $ 0.0014 \pm 0.0001$ & $ 0.0021 \pm 0.0002$\\
        $A$ & $1.753 \pm 0.449$ & $1.313\pm0.315$ & $1.883\pm0.474$\\
        $B$ & $0.932\pm0.187$ & $1.166\pm0.145$ & $0.585\pm0.302$\\
        $RA_1$ & $-0.0000527\pm0.0000069$ & - & -\\
        $RA_2$ & $-0.0000564\pm0.0000069$ &- &- \\
        $RA_3$ & $-0.0000269\pm0.0000069$ &- &- \\
        $RA_4$ & $-0.0002591\pm0.0000078$ &- &- \\
        $RA_5$ & $ 0.0003812\pm0.0000077$ &- &- \\
        $RA_6$ & $-0.0001243\pm0.0000076$ &- &- \\\hline
    \end{tabular}
    \tablefoot{\tablefoottext{a}{Derived quantity.}}
    \label{tab:pc1}
\end{table*}

The mid-transit times obtained from the \texttt{TLCM} fits to the individual transit observations from CHEOPS can be found in Table~\ref{tab:tts} and for TESS, WASP and the data published in \citet{2021AJ....162..256W} in Table~\ref{tab:tts_rest}. For these, the transit shape was fixed to the shape determined from the phase curve fits. Even though these fits showed good consistency for the derived parameters, we derived the individual transit timings for both CHEOPS data reductions for comparison.
Some of the mid-transit timings were biased due to either missing ingress or egress in the observations, which is why they are filtered out and not used in the following analysis steps. This was determined by visually inspecting the light curves and by comparing the timings retrieved from the \texttt{DRP} and \texttt{PIPE} data -- if they agreed well within $1\sigma$, we generally accepted them, except if the timing uncertainties were large ($> 30\,$s). All individual transits that were modelled are shown in Fig.~\ref{fig:transits}.

\subsection{Discussion}

The phase curve fits to the CHEOPS and TESS data show a general agreement between each other for most of the parameters (differences < $1\sigma$ or $2\sigma$ for most of the fitted parameters between the CHEOPS data sets, except for the orbital period at just over $2\sigma$).
Larger deviations between CHEOPS and TESS might be explained by their different wavelength ranges, especially for parameters like for the recovered $R_\mathrm{p}/R_\star$ values. The impact of CHEOPS' increased precision over TESS can be mainly seen in the recovered transit parameters. Both of them reach comparable precision in the measurements of the phase curve parameters, due to the increased phase coverage and number of full phase curves that are available during one sector with TESS.
Assuming only circular orbits is justified by computing the eccentricity damping time scale via \citep{1966Icar....5..375G, 2017AJ....154....4P}:
\begin{equation}
    \tau_e = \frac{e}{\mathrm{d}e/\mathrm{d}t} = \frac{2\,Q_\mathrm{p}}{63\,\pi}\left(\frac{a}{R_\mathrm{p}}\right)^5\,P,
\end{equation}
with $Q_\mathrm{p} \sim 10^6$ the planetary tidal quality factor of Jupiter, we obtain $\tau_e \approx 1.0\,$Myr. Since the system has an age estimate of $1.77^{+0.88}_{-0.68}\,$Gyr, any primordial eccentricity should essentially be reduced to zero.
Nevertheless, a small eccentricity might be induced by the companion \citep{2003ApJ...592.1201L, 2007MNRAS.382.1768M, 2010MNRAS.407.1048M, 2012A&A...538A.105L}, and obtaining a full phase curve observation of this planet with e.g. the Hubble Space Telescope (HST) or the James-Webb Space Telescope (JWST, \citealt{2006SSRv..123..485G}) would allow the measurement of even small eccentricities at a high precision, which would have the added benefit of allowing a precise atmospheric characterisation. Finding a small eccentricity would provide strong evidence towards the existence of the companion.
For a comparison, see the phase curves obtained with JWST for WASP-121 \citep{2023ApJ...943L..17M}, a star that is relatively similar to TOI-2109, also in magnitude. Such a phase curve should allow very precise characterisation of the orbital characteristics.
In particular, the close proximity of this hot Jupiter to its host star and its high equilibrium temperature, in combination with the short orbital period make it a relatively cheap and interesting target to observe. Apart from just determining the atmospheric composition, signs of an escaping atmosphere could be searched for (see e.g. \citealt{2023AJ....165..244D}).

Converting the limb darkening parameters back from $A$ and $B$ (for the values see Table~\ref{tab:pc1}) to the standard quadratic limb darkening parameters yields $u_{1,\mathrm{DRP}} = 0.28\pm0.17$ and $u_{2,\mathrm{DRP}} = -0.40\pm0.17$, $u_{1,\mathrm{PIPE}} = 0.28\pm0.12$ and $u_{2,\mathrm{PIPE}} = -0.07\pm0.12$, and $u_{1,\mathrm{TESS}} = 0.28\pm0.20$ and $u_{2,\mathrm{TESS}} = -0.63\pm0.20$. 
The two solutions coming from the CHEOPS data reductions are in good agreement with each other. However, in all cases, we retrieve negative values for the $u_2$ component.
This and the difference of these limb darkening parameters to the theoretical ones from \citet{2021RNAAS...5...13C}, which are $u_{1,\mathrm{PHX}}=0.48$ and $u_{2,\mathrm{PHX}}=0.20$ for the stellar atmosphere models from PHOENIX \citep{2013A&A...553A...6H}, and $u_{1,\mathrm{ATL}}=0.31$ and $u_{1,\mathrm{ATL}}=0.31$ for those of ATLAS \citep{1970SAOSR.309.....K} might be explained by the fast stellar rotation, leading to a deformation of the star, and hence to gravity darkening. This should impact the obtained limb-darkening parameters. Nevertheless, the obtained mid-transit timings should be relatively unaffected by this.
Comparing our retrieved limb darkening parameters from the TESS data, we find that they agree with the results of \citet{2021AJ....162..256W} in terms of the resulting radial intensity profile again within $2\sigma$, except for the cross-over point and the outermost range, where we only find a $3\sigma$ agreement.

\section{Tidal orbital decay\label{sec:decay}}

To search for signs of orbital decay in the measured mid-transit timings from the individual fits to the observed transits, we make use of the standard three models -- a linear model describing a constant and circular orbit, a quadratic orbital decay model describing the decreasing orbital period, and a sinusoidal model that represents the effect of apsidal precession on the mid-transit timings (see e.g. \citealt{2017AJ....154....4P, 2023A&A...669A.124H}).

The model for the mid-transit times of the constant, circular orbit is given in the following:
\begin{equation}
    t_\mathrm{tra}(N) = T_0 + N\,P,
\end{equation}
where $N$ is the individual transit number or epoch, $T_0$ is the reference timing, and $P$ is the orbital period.

The quadratic orbital decay model describes a change in the orbital period of the planet and is given by:
\begin{equation}
    t_\mathrm{tra}(N) = T_0 + N\,P + \frac{1}{2}\,\frac{\mathrm{d}P}{\mathrm{d}N}\,N^2,
\end{equation}
where the decay rate is introduced into the timing model as $\frac{\mathrm{d}P}{\mathrm{d}N}$.
This can be converted to the period derivative $\dot{P}$ via:
\begin{equation}
    \dot{P} = \frac{\mathrm{d}P}{\mathrm{d}t} = \frac{1}{P} \frac{\mathrm{d}P}{\mathrm{d}N}.
\end{equation}
This value can subsequently be plugged into the formulation of \citet{1966Icar....5..375G} to obtain $Q'_\star$:
\begin{equation}
    \dot{P} = \frac{f\pi}{Q'_\star} \frac{M_\mathrm{p}}{M_\star} \left(\frac{R_\star}{a}\right)^5,
\end{equation}
with the tidal factor $f=-\frac{27}{2}$, $M_\mathrm{p}$ and $M_\star$ the planetary and stellar masses, respectively, $R_\star$ the stellar radius and $a$ the semi-major axis of the planetary orbit. The given tidal factor is only valid in the case where the planetary orbital period $P$ is shorter than the stellar rotation period $P_\star$. For other configurations, $f$ also depends on the true orbital obliquity $\psi$ of the planet, as is shown in Table~2 of \citet{2023A&A...669A.124H}. For TOI-2109\,b we choose the nominal value, although its fast rotation will likely reduce the effect of the tidal decay, depending on which of the rotation periods that \citet{2021AJ....162..256W} found in the photometry data is the true one.

The sinusoidal model assumes a slightly eccentric orbit, leading to an apsidal precession motion of the planetary orbit. Using the formulations of \citet{1995Ap&SS.226...99G}, we get:
\begin{equation}
    t_\mathrm{tra}(N) = T_0 + N\,P_\mathrm{s} - \frac{e\,P_\mathrm{a}}{\pi}\,\cos \omega(N),
\end{equation}
where $P_\mathrm{s}$ is the sidereal period, $P_\mathrm{a}$ is the anomalistic period, and $\omega$ is the argument of pericenter. The sidereal and anomalistic period are related via:
\begin{equation}
    P_\mathrm{s} = P_\mathrm{a} \left(1 - \frac{1}{2\pi}\frac{\mathrm{d}\omega}{\mathrm{d}N}\right).
\end{equation}
The argument of pericenter is linearly related to the transit number, and hence with time, via the following equation:
\begin{equation}
    \omega(N) = \omega_0 + \frac{\mathrm{d}\omega}{\mathrm{d}N} N,
\end{equation}
with $\omega_0$ the value of the argument of pericenter at the reference time $T_0$.

From these fits, using a Markov-chain Monte Carlo algorithm via \texttt{emcee} \citep{2013PASP..125..306F}, we find no significant orbital decay, nor apsidal precession, and that a linear ephemeris fits the measured mid-transit timings best (see Table~\ref{tab:decay_results}). Nevertheless, assuming tidal orbital decay to be occuring at the measured rate, and converting the decay rate into the period derivative, we obtain $\dot{P}_\mathrm{PIPE} = (-1.32\pm1.63)$\,ms\,yr$^{-1}$ and $\dot{P}_\mathrm{DRP} = (-1.10\pm1.69)$\,ms\,yr$^{-1}$. This leads to $95\%$ confidence lower limits of $Q'_{\star,\mathrm{PIPE}} > 1.6 \times 10^7$ ($Q'_{\star,\mathrm{DRP}} > 1.7 \times 10^7$), and for the decay timescale $\tau = \frac{P}{|\dot{P}|}$, to $\tau_\mathrm{PIPE} > 12.7$\,Myr ($\tau_\mathrm{DRP} > 13.1$\,Myr).

\begin{table*}[]
    \centering
    \renewcommand{\arraystretch}{1.1} 
    \caption{Priors and median results of our orbital decay fits for the \texttt{DRP} and \texttt{PIPE} data sets.}
    \begin{tabular}{l c c c c}\hline\hline
        Model & Parameter & Prior & CHEOPS (\texttt{DRP}) & CHEOPS (\texttt{PIPE}) \\\hline
        Circular orbit & $T_0$ [BJD$_\mathrm{TDB}$] & $\mathcal{U}(0,11000)$ & $8984.389096\pm0.000081$ & $8984.389120\pm0.000075$\\ \vspace{0.2cm}
                    & $P$ [d]                    & $\mathcal{U}(0.6, 0.75)$ & $0.67247425\pm0.00000007$ & $0.67247424\pm0.00000007$ \\
        Orbital decay & $T_0$ [BJD$_\mathrm{TDB}$] & $\mathcal{U}(0, 11000)$ & $8984.389120\pm0.000089$ & $8984.389143\pm0.000080$ \\
                    & $P$ [d]                    & $\mathcal{U}(0.6, 0.75)$ & $0.67247424\pm0.00000007$ & $0.67247423\pm0.00000007$ \\\vspace{0.2cm}
                    & $\mathrm{d}P/\mathrm{d}N$    & $\mathcal{U}(-10^{-7}, 10^{-7})$ & $(-2.33\pm3.57)\times 10^{-11}$ & $(-2.80\pm3.46)\times 10^{-11}$ \\
        Apsidal precession & $T_0$ [BJD$_\mathrm{TDB}$] & $\mathcal{U}(0,11000)$ & $8984.389132\pm0.005254$ & $ 8984.389092\pm0.005443 $ \\
                    & $P$ [d]                    & $\mathcal{U}(0.6, 0.75)$ & $ 0.67247431\pm0.00000070 $ & $ 0.67247429\pm0.00000070 $ \\
                    & $\mathrm{d}\omega/\mathrm{d}N$ [rad$/$orbit] & $\mathcal{U}(-0.003, 0.003)$ & $ (0.02\pm7.04)\times10^{-4} $ & $ (0.00\pm6.51)\times10^{-4} $ \\
                    & $e$                        & $\mathcal{U}(0.0, 0.1)$ & $ 0.01619\pm0.02926 $ & $ 0.01896\pm0.02971 $ \\
                    & $\omega_0$ [rad] & $\mathcal{U}(0, 2\pi)$ & $ 3.162\pm1.678 $ & $ 3.165\pm1.633 $ \\
        \hline
    \end{tabular}
    \label{tab:decay_results}
\end{table*}

Even though this target is one of the best candidates for orbital decay in theory, and the early WASP timing that we obtained gives us a long observation baseline, we only measure a decay rate that is in agreement with a constant orbital period. 
Comparing with the predictions from \citet{2021AJ....162..256W} that are in the range of $\sim 10$ -- $740\,$ms\,yr$^{-1}$ for the orbital decay rate, we can clearly rule out the upper end of these predictions.
This implies that there is another effect which makes the tidal processes less efficient.
One possible explanation could be the fast stellar rotation ($v\sin i = (81.2\pm1.6)$\,km\,s$^{-1}$). Even if it is greater ($P_\mathrm{rot}/\sin i_\star \approx 1.0\,$d, see \citealt{2021AJ....162..256W}) than the orbital period of the hot Jupiter, it could still have a negative effect on the efficiency of angular momentum transfer from the planet to the star. 
The presence of a close-by companion could also affect this.
Still, our measured decay rate gives us a constraint on the lower limit of the $Q'_\star$ value of TOI-2109 that is not too far off from the predicted $Q'_\star$ values for F-stars of e.g. \citet{2007ApJ...661.1180O, 2011A&A...529A..50L} that are in the range from $10^5$ to $10^7$.
We also investigated the impact of atmospheric mass loss on the decay rate of the orbit of the hot Jupiter due to its tight orbit. Taking the measurement of the mass loss rate of KELT-9\,b of $\dot{M} \sim 10^{12.8\pm0.3}$\,g\,s$^{-1}$ \citep{2020A&A...638A..87W} as a first estimate, we find that we would still need two to three orders of magnitude more mass loss for it to balance a decay rate of just $\dot{P} \sim -1$\,ms\,yr$^{-1}$, which is unlikely.

One further complication are the apparent sinusoidal TTVs that we observed. 
Correcting the mid-transit timings for the best fitting super-period of about $117\,$d that we found by fitting the sinusoidal model to the data in Section~\ref{sec:TTVs}, we find tentative evidence for orbital decay at a rate of $\dot{P} = (-5.56\pm1.62)\,$ms\,yr$^{-1}$ at $3\sigma$. This would correspond to $95\%$ confidence lower limits of $Q'_\star > 8.4\times 10^6$ and $\tau > 6.6\,$Myr.
We note that this result is sensitive to the applied parameters, because, using the second best fitting super-period of about $88\,$d, the result is less significant with a decreased rate of $\dot{P} = (-3.53\pm1.63)\,$ms\,yr$^{-1}$ at $2\sigma$. This would lead to lower limits of $Q'_\star > 1.1 \times 10^7$ and $\tau > 8.6\,$Myr at $95\%$ confidence.

Furthermore, the gravitational interactions between the two planets, should the second planet exist, might also have an impact on the orbital decay of the hot Jupiter. In addition to that, should this excite a slight eccentricity in the orbit of TOI-2109\,b, it would also introduce apsidal precession. Disentangling the two effects is only possible with very long baselines or very precise mid-occultation timings, which we were not able to obtain with sufficient precision for this task from neither CHEOPS nor TESS. In the future, if a significant orbital decay or apsidal precession signature will be detected from continued monitoring with e.g. CHEOPS, JWST should be able to provide the necessary precision with a few occultation observations. These would also be relatively cheap to obtain due to their short duration, and could give us information about the atmosphere.

However, as shown in \citet{2023Univ....9..506H}, a distant massive planetary companion would also influence the measured mid-transit timings due to the introduction of a light-time effect. Astrometric observations of Gaia's \citep{2016A&A...595A...1G} fourth data release will be able to detect such companions, if they are in the right mass and distance range.

\section{An outer companion?}\label{sec:TTVs} 
\begin{figure}
    \centering
    \includegraphics[width=\linewidth]{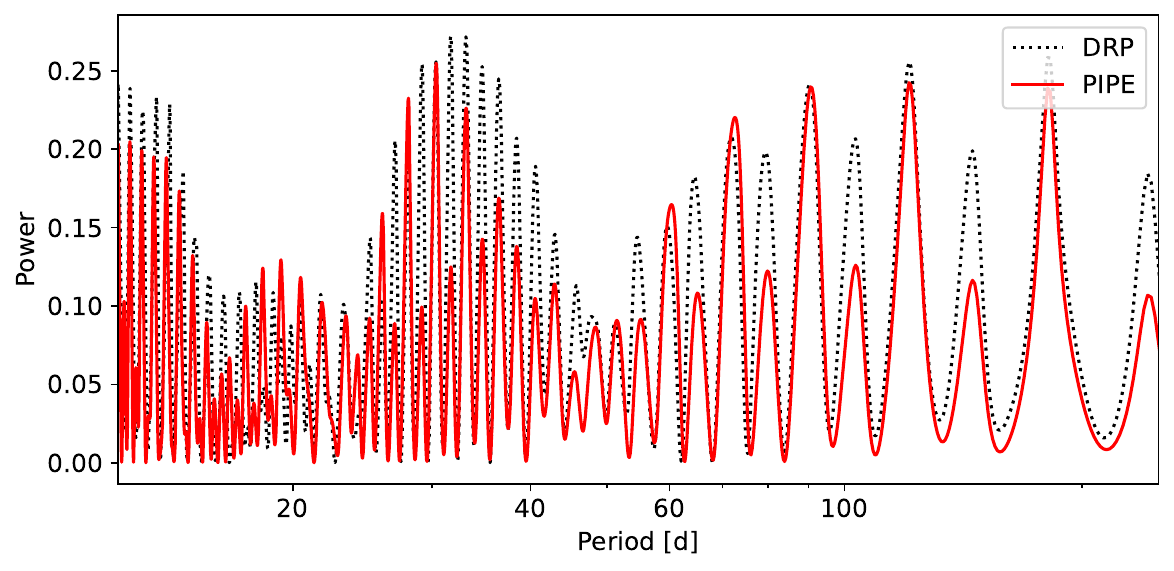}
    \caption{LS periodogram of the transit timing variations measured with CHEOPS, for the \texttt{DRP} and \texttt{PIPE} data reductions.}
    \label{fig:LS}
\end{figure}

Upon closer inspection of the derived CHEOPS mid-transit times, we noticed a sinusoidal variation that is only significant in these data due to the high precision of the CHEOPS photometry.
The first thing we did was to create a Lomb-Scargle (LS) periodogram \citep{2009A&A...496..577Z} of the transit timings to find possible periods, see Fig.~\ref{fig:LS}. For both, the \texttt{DRP} and \texttt{PIPE} reductions, we obtained multiple peaks of similar powers.

\begin{figure*}
    \centering
    \includegraphics[width=\linewidth]{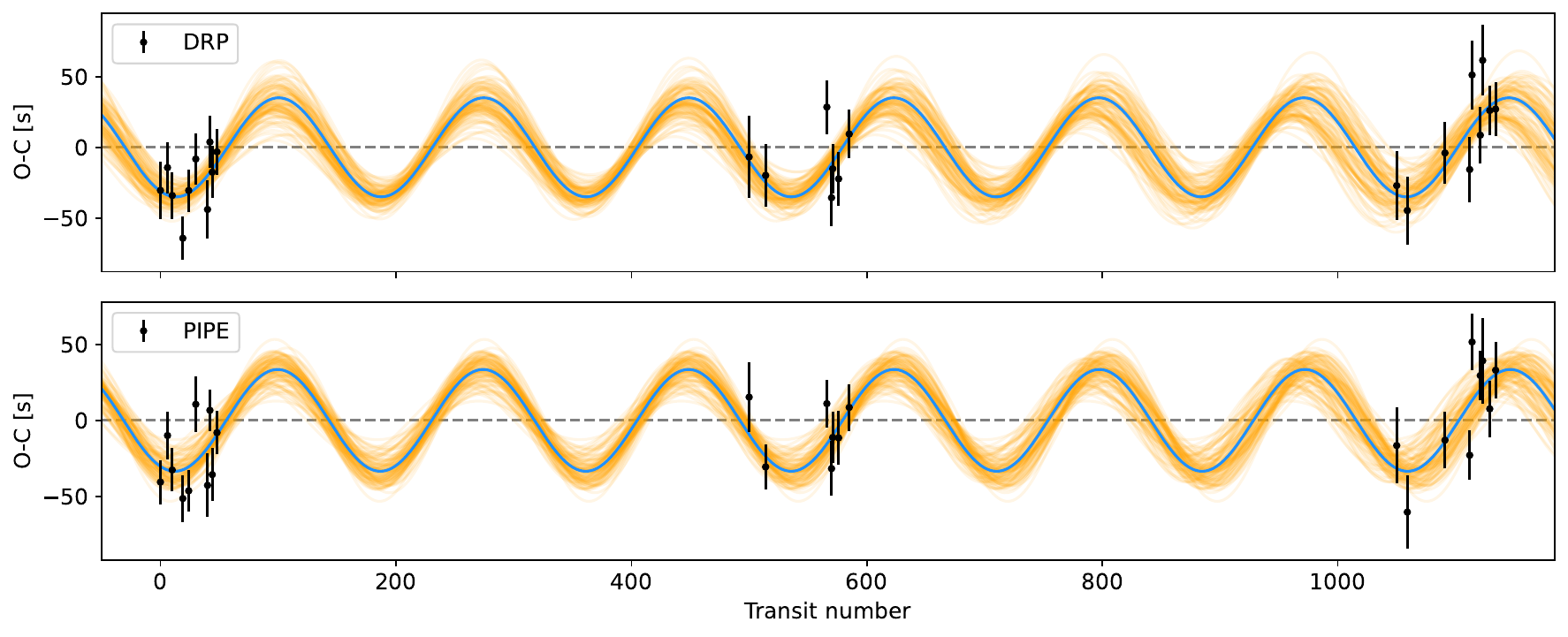}
    \caption{Timing variation plots of the \texttt{DRP} (top) and \texttt{PIPE} (bottom) reductions, showing the lowest BIC models for both data sets. The blue lines show the median solutions, with the thin orange lines showing random samples drawn from the posterior distribution. The dashed gray lines indicate the best linear models.}
    \label{fig:chain_plot}
\end{figure*}

The next step was fitting a simple sinusoidal model to the CHEOPS timings to obtain more information about this sinusoidal variation. We employed the following model (see e.g. \citealt{2012ApJ...761..122L, 2018ApJS..234....9O}) and optimized it using \texttt{emcee}:
\begin{equation}\label{eq:sine_model}
    t_\mathrm{tra}(N) = T_0 + N\,P + A\,\cos\left(\frac{2\pi}{P_\mathrm{sup}}\,(N\,P - T_\mathrm{sup})\right), 
\end{equation}
where $A$ is the amplitude of the TTVs, $P_\mathrm{sup}$ is the super-period and $T_\mathrm{sup}$ the super-epoch.
To differentiate between the different periods of similar power in the LS periodogram, and to find the best fitting one, we employed multiple MCMC optimisations of our simple TTV model, trying to find the period that leads to the highest log-probability. From this, we determined the parameters of each high-power peak, and sampled them individually for inter-comparison in a further MCMC optimisation (100 walkers, 10000 steps with 1000 steps of burn in).
We ensured convergence by checking that the number of steps was at least 50 times the autocorrelation time of all our parameters.
We find that $P_\mathrm{sup} \approx 117\,$d leads to the lowest BIC value, which is only slightly better ($\Delta\mathrm{BIC} = 0.16$) than $P_\mathrm{sup} \approx 88\,$d. Both are, however, better than a simple linear model for the mid-transit timings by $\Delta\mathrm{BIC} = 4.0$, giving us positive evidence for the apparent TTVs.
The final models are shown in Fig.~\ref{fig:chain_plot} and the respective corner plot in Fig.~\ref{fig:sine_corner} for both data sets, showing the good agreement between the two data reductions.
This, and the agreement of the orbital decay analysis between them prove the consistency of the derived mid-transit timings. Due to this fact, we use, going forward, only the retrieved \texttt{PIPE} timings because the roll-angle effect does not need to be accounted for, which should make the fits to the full data set and the individual timings more reliable.

Due to the relatively high mass of the HJ ($5.02\pm0.75\,$M$_\mathrm{J}$, see \citealt{2021AJ....162..256W}), and amplitude and period of the apparent TTVs, the companion, if present must be relatively close by \citep{2005MNRAS.359..567A}. Since we do not find additional transits in the photometric data that is available, we argue that planet c has to orbit outside of the orbit of the HJ. 
Making use of the Transit Least Squares algorithm \citep{2019A&A...623A..39H} for the search of additional transits in the residuals of our phase curve fits, we do not find any significant signals. We searched for orbital periods in the range between $1.0\,$d and $15.0\,$d, with the default settings for the search parameters, and the stellar parameters from \citet{2021AJ....162..256W}.
Moreover, we also searched for additional transit signals using the Détection Spécialisée de Transits (DST) algorithm \citep{2012A&A...548A..44C}, but we did not find any significant evidence of additional planets here either.

The relatively high impact parameter of planet b means that, if planet c is coplanar with planet b, and orbits further out, it will almost certainly not transit the star. In fact, the orbital distance to obtain $b = 1.0$, would be $a = 5.148\,$R$_\odot$ or $P = 1.125\,$d.
In theory, wide ranges of orbital periods and planet masses are possible. Especially the low precision of the RVs due to the fast stellar rotation is a major drawback in this regard, as it is almost impossible to gain any more information about the system, except for planet b, leaving only the observed TTVs as a source of information for the possible companion.

As a first estimate of planet mass and orbital period of the possible outer companion, we use Eq.~33 from \citet{2005MNRAS.359..567A}. This equation describes the relation between planet masses and TTV amplitude for two planets in MMR, in the case that the more massive planet is transiting. From this, we find masses in the range from 6 to 30 Earth masses for the outer companion, with orbital periods between 1 and 5 days (i.e. MMRs ranging from 3:2 to 7:1).


\subsection{Chaotic and regular orbits}
Due to the suspected close proximity of the planet candidate, we first explore regular regions in the mass-period space using \texttt{REBOUND} \citep{2012A&A...537A.128R} and its MEGNO (Mean Exponential Growth of Nearby Orbits, \citealt{2000A&AS..147..205C}) implementation. MEGNO is a chaos indicator for examining the dynamics of planetary systems. 

The setup that we have used for the chaos analysis is comprised of a 100 by 100 grid of orbital periods and planetary masses of the outer companion. For each of these grid points, we calculate the MEGNO value. For the setup of the system, we used the parameters given in \citet{2021AJ....162..256W}. Making use of the ``WHFast'' integrator \citep{2015MNRAS.452..376R}, we integrated the system for more than 270.000 orbits of the inner planet. Both, circular and eccentric systems were probed for completeness. The result for circular orbits is shown in Fig.~\ref{fig:megno_circ}, and for various eccentricities in Appendix~\ref{sec:app_megno}.

\begin{figure}
    \centering
    \includegraphics[width=\linewidth]{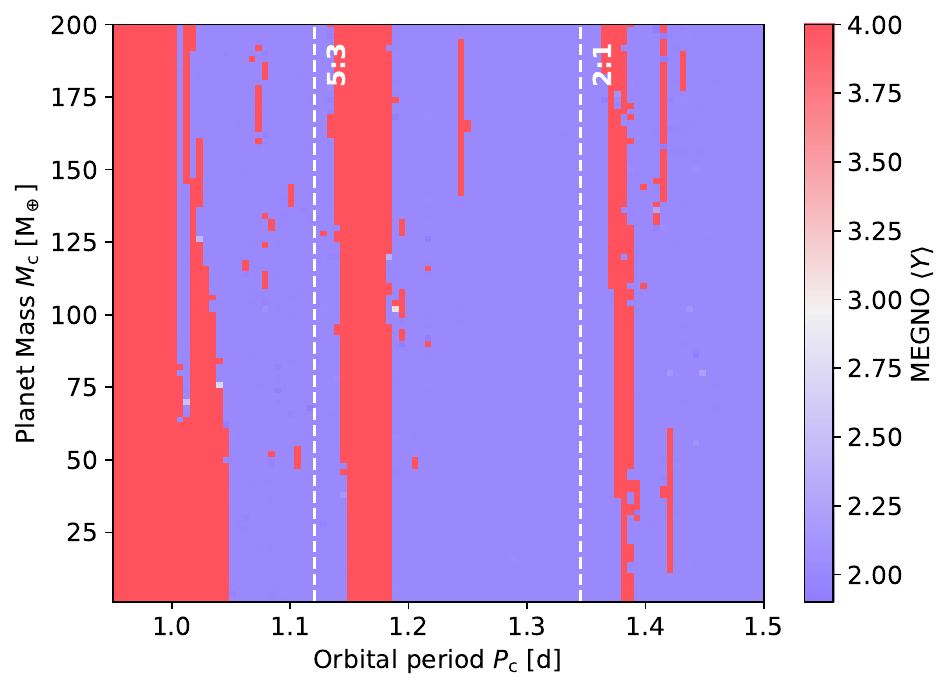}
    \caption{Mass-period MEGNO diagram showing regular and chaotic system configurations for circular orbits of both planets. MEGNO numbers around 2 indicating a regular system (blue), and numbers around 4 indicating a chaotic system (red). The specific mean-motion resonances that are within this period range are indicated as vertical dashed lines.}
    \label{fig:megno_circ}
\end{figure}

Probing orbital periods of the outer companion from 0.9 to 1.5 days, and planet masses ranging from 1 to 200 Earth masses, we find that there are a few chaotic areas, especially for very close orbits with $P_\mathrm{c} < 1.0\,$d. This chaotic area extends up to about $1.05\,$d for low $M_\mathrm{c}$ values. There is another chaotic region between about $1.15\,$d and $1.19\,$d for this system. A further, but narrow, chaotic area is present at $P_\mathrm{c} \approx 1.39\,$d, with a few more chaotic regions being between these areas. These are, however, mostly small and don't show the same structure over all examined planetary masses. Beyond orbital periods of $1.5\,$d, there are no significant chaotic areas in the $M_\mathrm{c}$ -- $P_\mathrm{c}$ space for circular orbits of the two planets.

In the case of eccentric orbits, we find that the innermost regular region gets pushed outwards due to the orbits coming closer together at one point. Depending on the eccentricities used in the simulations, the closest regular area might only be in a narrow area around $1.5\,$d or even farther out (see Fig.~\ref{fig:array_e0005} -- \ref{fig:array_e0325}). 

\subsection{\texttt{TRADES} TTV modelling\label{sec:TRADES_modelling}}

To model the apparent TTVs in more detail, we employed \texttt{TRADES}\footnote{\href{https://github.com/lucaborsato/trades}{https://github.com/lucaborsato/trades}} (TRAnsits and Dynamics of Exoplanetary Systems) \citep{2014A&A...571A..38B, 2019MNRAS.484.3233B}. \texttt{TRADES} is a publicly available N-body code to model the dynamics of exoplanet systems via TTVs and RVs. There are several approaches to obtain the final dynamical solution of the examined system, with us choosing the standard MCMC approach. In two cases, strictly circular or eccentric orbits, we fit for the masses of both planets ($M_\mathrm{b}$, $M_\mathrm{c}$), the orbital periods ($P_\mathrm{b}$, $P_\mathrm{c}$) and their mean anomalies ($MA_\mathrm{b}$, $MA_\mathrm{c}$). This gives us six free parameters for the circular case. In the case of eccentric orbits, we additionally fit for the eccentricities ($e_\mathrm{b}$, $e_\mathrm{c}$) and their arguments of pericenter ($\omega_\mathrm{b}$, $\omega_\mathrm{c}$) via $\sqrt{e_\mathrm{b,c}}\sin\omega_\mathrm{b,c}$ and $\sqrt{e_\mathrm{b,c}}\cos\omega_\mathrm{b,c}$, giving us a total of ten free parameters. The prior intervals and starting values are given in Table~\ref{tab:trades_priors}. The orbital inclination $i$ was fixed to the angle determined by \citet{2021AJ....162..256W} for both planets, assuming they share the same orbital plane. Due to the scarcity of information that we expect to be able to extract from the available data, we fixed the longitude of the ascending node to $\Omega = 180^\circ$ for both planets to reduce the size of the parameter space.

\begin{table}[]
    \centering
    \caption{Priors used in the setup files for the analysis with \texttt{TRADES}. The values give the starting value and the intervals give the allowed range of the parameters.}
    \begin{tabular}{l c c}\hline\hline
        Parameter & Planet b & Planet c \\\hline
        $M$ [M$_\mathrm{J}]$& 5.02 [4.0, 6.0] & 0.2 [0.001, 2.0] \\
        $P$ [d] & 0.672474 [0.668, 0.674] & 1.9 [0.8, 5.0] \\
        $e$\tablefootmark{a} & 0.04 [0.0, 0.15] & 0.05 [0.0, 0.35] \\
        $\omega$\tablefootmark{a} [$^\circ$] & 90.0 [0.0, 360.0] & 90.0 [0.0, 360.0] \\
        $MA$ [$^\circ$] & 45.0 [0.0, 360.0] & 0.0 [0.0, 360.0] \\
        \hline
    \end{tabular}
    \tablefoot{\tablefoottext{a}{In the case when eccentric orbits were allowed, otherwise they were fixed to $e=0$ and $\omega = 90^\circ$}.}
    \label{tab:trades_priors}
\end{table}

Making use of the pre-optimisation that \texttt{TRADES} offers via PyDE\footnote{\href{https://github.com/hpparvi/PyDE}{https://github.com/hpparvi/PyDE}} \citep{2016zndo.....45602P}, with a population size of 128, generation size of $60\,000$, and a difference amplification factor and cross-over probability of 0.5, we use this initial optimisation as input for \texttt{emcee}. The latter is run for at least $200\,000$ steps, using 128 walkers.

We explored circular and eccentric orbits. In every case, the RVs are also used in the optimisation.
Besides these, we explored a few possible MMRs, namely 2:1, 3:1, 5:2, 5:3, 7:2 and 7:3 (ratio of outer to inner period). In these cases, we only allowed a narrow window around the respective resonant orbital period of planet c. All were examined in the cases of circular and eccentric orbits, giving us a total of 12 additional strictly MMR cases. The 3:2 resonance was not probed due to the instability of this close orbital configuration ($P_\mathrm{c} \approx 1.0\,$d), as indicated in Fig.~\ref{fig:megno_circ}.

\begin{figure}
    \centering
    \includegraphics[width=0.9\linewidth]{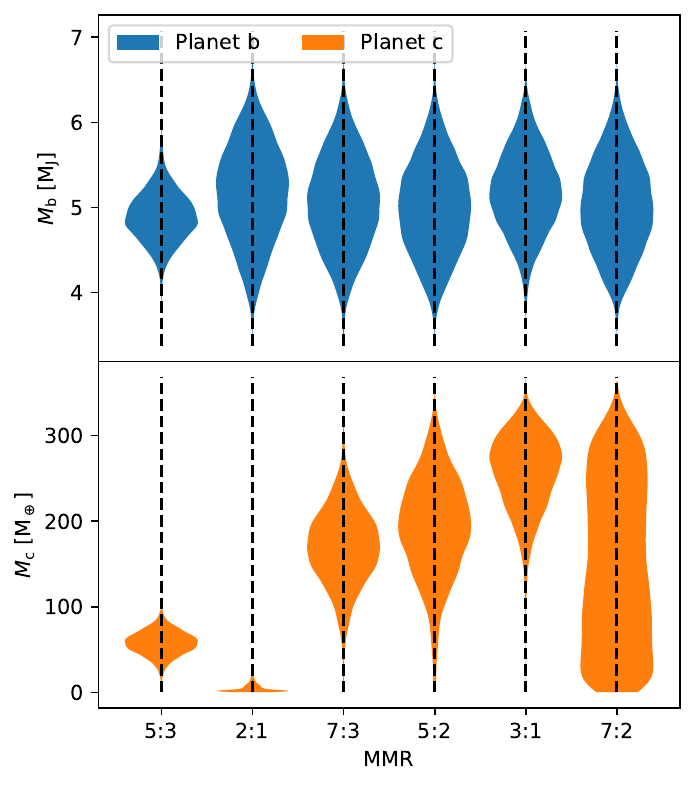}
    \caption{Violin plot of the posterior distributions of the masses of planets `b' (top, blue) and `c' (bottom, orange) derived from the \texttt{PIPE} data in the cases of the MMRs and circular orbits with \texttt{TRADES}.}
    \label{fig:violins_MMR_circ_PIPE}
\end{figure}

\begin{figure}
    \centering
    \includegraphics[width=0.9\linewidth]{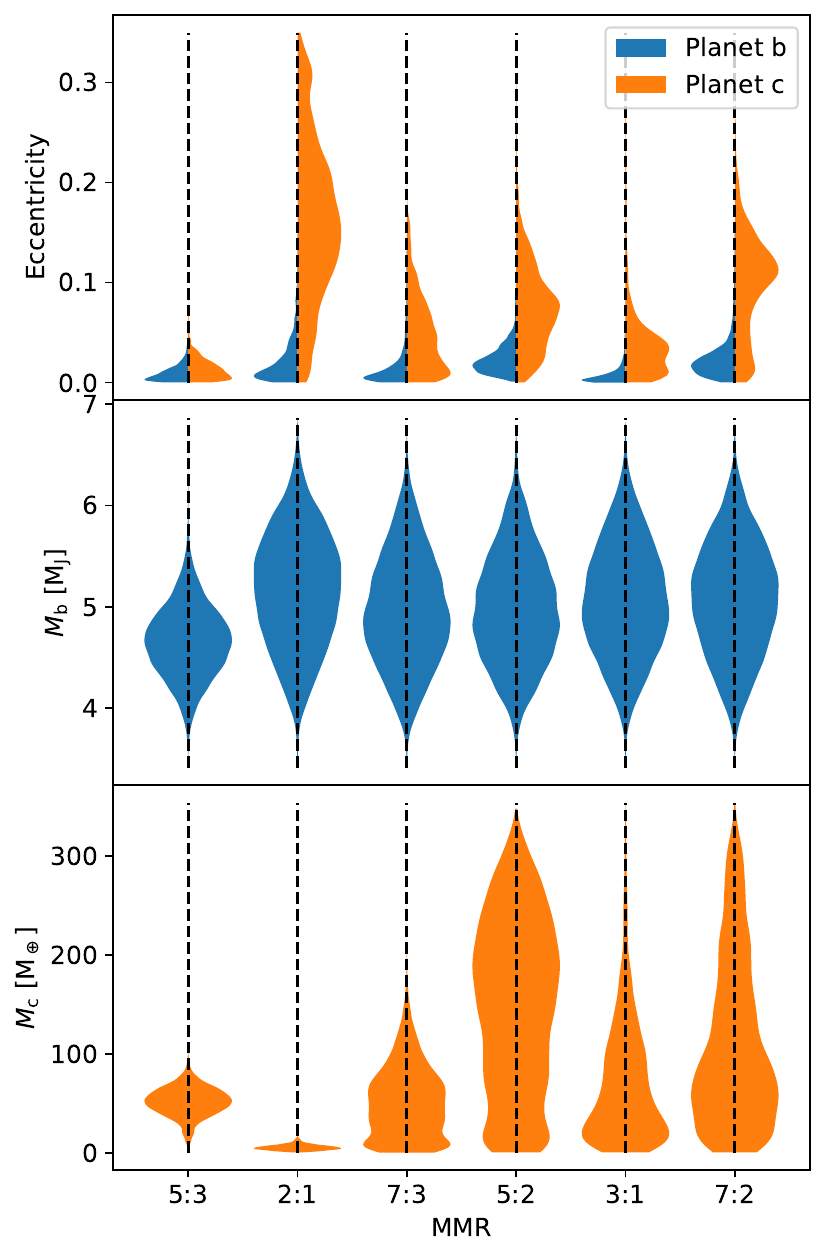}
    \caption{Violin plot of the posterior distributions of the masses and eccentricities from the \texttt{PIPE} data in the cases of the MMRs and eccentric orbits, derived with \texttt{TRADES}.}
    \label{fig:violins_MMR_ecc_PIPE}
\end{figure}

The posterior distributions of the \texttt{TRADES} runs featuring the various MMRs are shown in Fig.~\ref{fig:violins_MMR_circ_PIPE} and \ref{fig:violins_MMR_ecc_PIPE}, featuring circular and eccentric orbits, respectively.

\begin{figure}
    \centering
    \includegraphics[width=\linewidth]{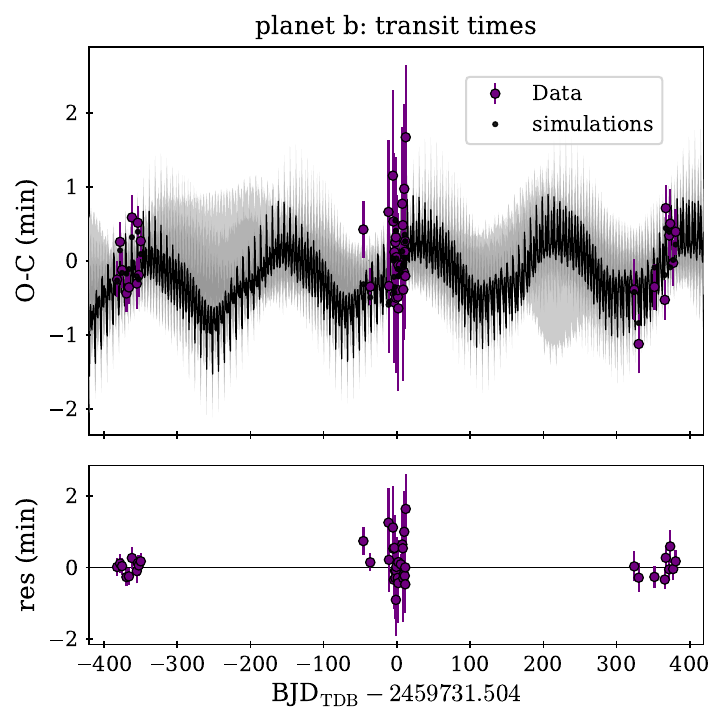}
    \caption{\texttt{TRADES} TTV plot of the best overall fit result, showing a 5:3 MMR and slightly eccentric orbits. Top: Fit to the transit timing measurements of TOI-2109\,b, including the final model and random samples from the MCMC analysis. Bottom: Residuals of the fit.}
    \label{fig:TRADES_53_ecc_pipe}
\end{figure}

Comparing the $\chi^2$ values of the final results using their maximum-a-posterior (MAP) values, we find the best fit for the 5:3 resonance and eccentric orbits with $\chi^2=68.4$, to the available transit timings from CHEOPS and TESS, and the RVs. Even though this specific setup delivers the best fit, the orbit of planet c with these parameters would likely lead to a chaotic system, with it being on the very edge between regular (``stable'') and chaotic (``unstable''). We note that while chaos is linked to instability, regular systems might become unstable after some time.
This is owed to the close orbits of the planets in this MMR. The orbital period of $P_\mathrm{c} = (1.1398\pm0.0017)\,$d (with $P_\mathrm{5:3} = 1.121\,$d being the exact resonance period), and eccentricities of $e_\mathrm{b} = 0.02$ and $e_\mathrm{c} = 0.03$, lead to the orbits coming close to each other, so that they may disrupt their orbits.
For the purpose of examining the stability of this configuration, we employed $N$-body simulations with \texttt{REBOUND}. First, we tested the stability of the MAP solution by simulating the system for $10^6$\,yr, which remained stable for this period. Secondly, we checked the stability of $1,000$ randomly drawn samples of the posterior for 100,000 orbits of the inner planet ($\sim 2,000$\,yr), and only found $3.5\%$ of them to become unstable within this time.
The resulting TTV plot of our fit is shown in Fig.~\ref{fig:TRADES_53_ecc_pipe}.
The other runs lead mostly to $\chi^2 = 73 - 88$, with the 7:2 resonance for circular orbits leading to the worst fits with $\chi^2\approx93$. Checking the stability of these solution for at least $10^6$\,yr, we find that the MAP values lead to stable orbits for each configuration. Randomly drawing 1,000 samples from each posterior, we find that only six of the resonant configurations have unstable solutions in this sample (including the already mentioned 5:3 MMR in the eccentric case). For the circular 2:1 MMR case, we find that 2 out of 1000 samples become unstable within a time period of 100,000 orbits of the inner planet, the eccentric 2:1 MMR case leads to 281 unstable configurations. For the eccentric cases, the 5:2 MMR leads to 15, the 7:2 MMR to 1, and the 7:3 MMR to 23 unstable solutions. The remaining configurations remain stable over this time frame.

\begin{figure*}
    \centering
    \includegraphics[width=0.7\textwidth]{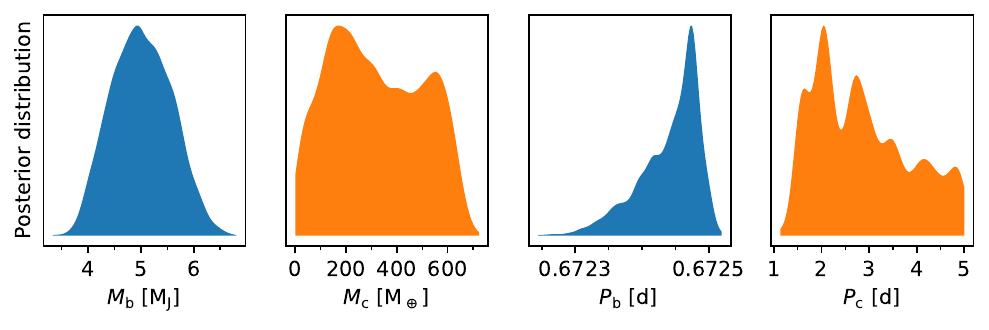}
    \caption{Posterior distributions of the masses and periods for the general run ($P_\mathrm{c}$ not constrained to be near a certain MMR) for circular orbits, derived with \texttt{TRADES}. The results for planet b are indicated in blue and for c in orange.}
    \label{fig:violins_general_circ}
\end{figure*}

\begin{figure*}
    \centering
    \includegraphics[width=\textwidth]{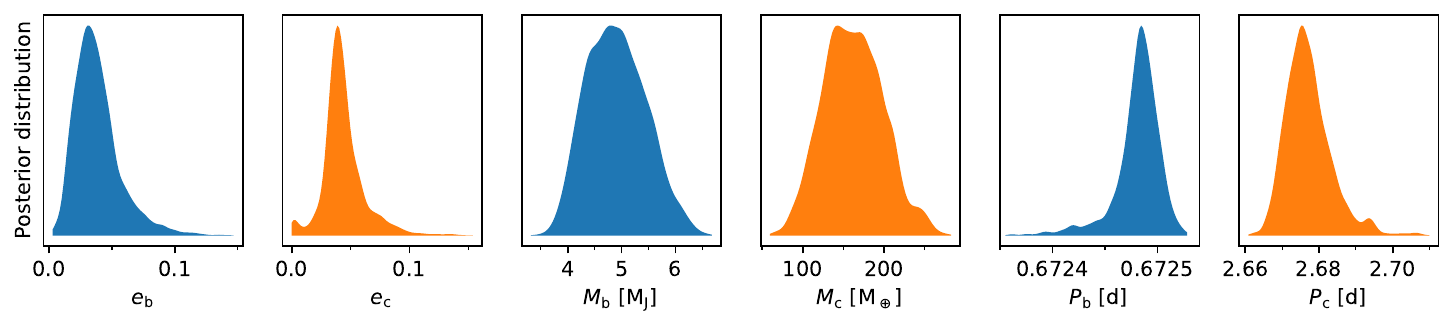}
    \caption{Posterior distributions of the eccentricities, masses and periods for the general run ($P_\mathrm{c}$ not constrained to be near a certain MMR) for eccentric orbits, derived with \texttt{TRADES}. The results for planet b are indicated in blue and for c in orange.}
    \label{fig:violins_general_ecc}
\end{figure*}

In the general runs (for the posteriors see Fig.~\ref{fig:violins_general_circ} and \ref{fig:violins_general_ecc}), without tight constraints on the orbital period of planet c, we obtain the best result in terms of $\chi^2$ value for eccentric orbits, resulting in $\chi^2=75.1$. The final orbital period of planet c in this case is actually very close to a 4:1 MMR with $P_c = 2.677\,$d, where the exact MMR period would be $P_\mathrm{4:1} = 2.688\,$d. The TTV plot is shown in Fig.~\ref{fig:TRADES_ecc}. It reproduces the super-period of about $\sim 117\,$d that we also found using the simple sinusoidal model (see Fig.~\ref{fig:chain_plot}).
The model allowing circular orbits fits worse by $\Delta\chi^2 \approx 3.5$, with a final period of the outer companion close to the 7:3 MMR. The respective TTV plot is shown in Fig.~\ref{fig:TRADES_general_best}.

\subsection{\pyttv analysis}
We carry out a set of photodynamical analyses modelling the CHEOPS \pipe transit timings, \tess photometry, and the TRES and FIES radial velocities jointly using \pyttv \citep{2023A&A...675A.115K}. The code models the three types of observables (transit photometry, radial velocity measurements, and transit centre time estimates) simultaneously using \rebound \citep{2012A&A...537A.128R, reboundias15, reboundx} for dynamical integration and \pytransit \citep{Parviainen2015,Parviainen2020a,Parviainen2020b} for transit modelling, and provides posterior densities for the model parameters estimated using MCMC sampling in a standard Bayesian parameter estimation framework. We employ \texttt{PyTTV} in addition to \texttt{TRADES} to make sure that our results are broadly consistent due to the weak constraints that we have of the system.

The model parameters and their priors are listed in Table~\ref{table:pyttv_parameters}. All the planet parameters except for the log$_{10}$ mass and radius ratio are defined at a reference time, $t_\mathrm{ref}=2459378.46$. We set uniform priors on the log masses of the two planets, with ranges designed to help optimisation but not to constrain the posteriors. We set a loosely informative prior on the inner planet's radius ratio based on the \citet{2021AJ....162..256W} estimate, and the outer planet's radius ratio is given a dummy normal prior since the data will not be able to constrain it.
We set a normal prior on the inner planet's transit center and a uniform prior on the outer planet's mean anomaly (the planets are parameterized differently since the hot Jupiter transits and the other does not). We do not constrain the mutual inclinations of the two planets. Instead, we set an informative prior on the inner planet's impact parameter based on the \citet{2021AJ....162..256W} estimate of $0.7481\pm0.0073$, and we set an uniform prior on the outer planet impact parameter. The lower limit of the outer planet's impact parameter prior is justified by the fact that the planet does not transit, and the upper limit was checked no to constrain the posterior after the MCMC sampling. Finally, we set additional zero-centered half-normal priors on the eccentricities, $\NP$(0.0, 0.023). These priors allow for eccentric orbits but bias against high eccentricities. 

The analysis begins with a global optimisation using the differential evolution method \citep{Storn1997a,Price2005} implemented in \pytransit \citep{Parviainen2015}. The optimiser starts with a population of parameter vectors drawn from the model prior, and clumps the population close to the global posterior mode. After the optimisation, we use the clumped parameter vector population to initialise the \emcee sampler \citep{2013PASP..125..306F}, which we then use for MCMC sampling to obtain a sample from the parameter posterior. 

We repeat the analysis for the 5:3, 2:1, 7:3, 5:2, 3:1, and 7:2 period commensurability scenarios. For each scenario, we set the outer planet period prior to $\NP(r \times 0.67247414, 0.1)$, where $r$ is the period ratio for the scenario.

\begin{figure}
    \centering
    \includegraphics[width=\linewidth]{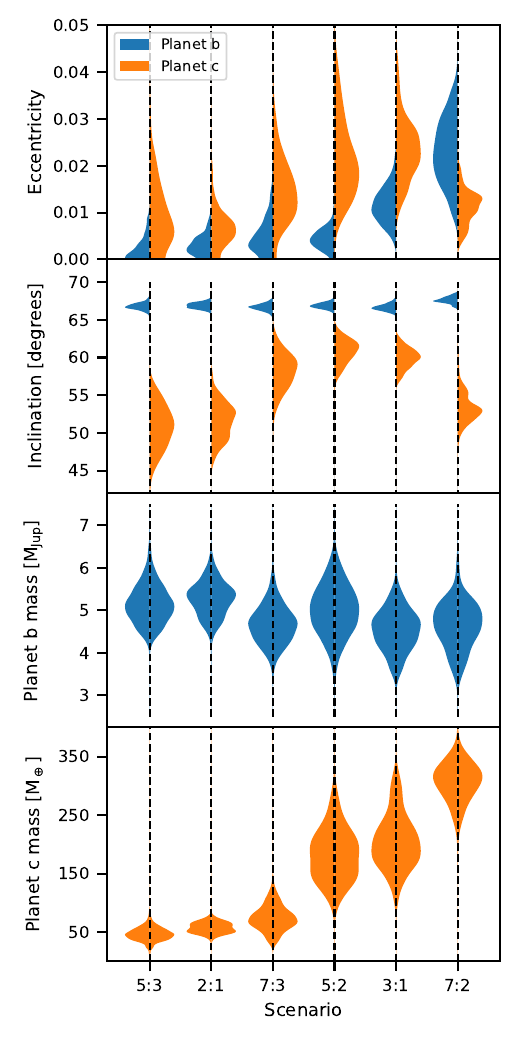}
    \caption{Posterior densities for the planet masses, orbital inclinations, and orbital eccentricities obtained from the \pyttv analysis for the six considered period commensurability scenarios.}
    \label{fig:pyttv_posteriors}
\end{figure}

We present the posterior distributions for the planet masses, orbital inclinations, and orbital eccentricities from the \pyttv analysis in Fig.~\ref{fig:pyttv_posteriors}. The Bayesian Information Criterion (BIC) favours the 3:1 period commensurability scenario, which also has a relatively small difference in the orbital inclinations. Looking at the individual likelihoods from the photometry, transit centres, and RVs, the 3:1 scenario is mainly favoured by the improved fit of the RV measurements. Indeed, the photometry and transit centre likelihood distributions corresponding to the posterior parameter distributions largely overlap, as do the ones for the RV data for all the other scenarios than 3:1. 

\begin{table}
\centering
\caption{\pyttv model parameters and priors.}
\label{table:pyttv_parameters}
\begin{tabular*}{\columnwidth}{@{\extracolsep{\fill}} lll}
\toprule\toprule
Description & Units & Prior \\
\midrule     
\multicolumn{3}{l}{\emph{Stellar parameters}} \\
\midrule
Stellar mass & [$M_\odot$] & $\NP(1.447, 0.077)$ \\
Stellar radius & [$R_\odot$] & $\NP(1.698, 0.060)$ \\
Limb darkening q$_1^a$ & [-] & $\UP(0, 1)$ \\
Limb darkening q$_2^a$ & [-] & $\UP(0, 1)$ \\
\\
\multicolumn{3}{l}{\emph{Planet b parameters}} \\
\midrule
log$_{10}$ mass  & [$\log_{10}\, M_\odot$] & $\UP(-2.8, -2.1)$ \\
Radius ratio  & [$R_\star$] & $\NP(0.08, 0.001)$ \\
Transit centre & [BJD] & $\NP(2459378.46, 0.02)$ \\
Orbital period & [d] & $\NP(0.67247414, 0.005)$\\
Impact parameter & [$R_\star$] & $\NP(0.75, 0.01)$ \\
$\sqrt{e} \cos\omega$ & [-]  & $\UP(-0.25, 0.25)$ \\
$\sqrt{e} \sin\omega$  & [-]  & $\UP(-0.25, 0.25)$ \\
$\Omega$ & [rad] & $\NP(\pi, 0.0001)$ \\
\\
\multicolumn{3}{l}{\emph{Planet c parameters}} \\
\midrule
log$_{10}$ mass & [$\log_{10}\, M_\odot$] & $\UP(-5.2, -3.0)$ \\
Radius ratio & [$R_\star$] & $\NP(0.05, 0.0001)$ \\
Mean anomaly at $T_\mathrm{ref}$  & [rad] & $\UP(0, 2\pi)$ \\
Orbital period & [d] & Depends on the scenario\\
Impact parameter & [$R_\star$] & $\UP(1.0, 3.0)$ \\
$\sqrt{e} \cos\omega$ & [-]  & $\UP(-0.25, 0.25)$ \\
$\sqrt{e} \sin\omega$ & [-]  & $\UP(-0.25, 0.25)$ \\
$\Omega$  & [rad] & $\NP(\pi, 0.0001)$ \\
\\
\multicolumn{3}{l}{\emph{RV parameters}} \\
\midrule
Linear trend  & [m/s/d] & $\NP(0, 1.0)$ \\
Systemic velocity 1 & [m/s] & $\NP(-25855, 200)$ \\
Systemic velocity 2 & [m/s] & $\NP(-25416, 200)$ \\
log$_{10}$ jitter 1 & [$\log_{10}$ m/s] & $\NP(-1, 0.1)$ \\
log$_{10}$ jitter 2 & [$\log_{10}$ m/s] & $\NP(-1, 0.1)$ \\
\\
\multicolumn{3}{l}{\emph{Additional priors}} \\
\midrule
Both eccentricities & &  $\NP(0.0, 0.023)$ \\
\bottomrule
\end{tabular*}
\tablefoot{
    All the planetary parameters except the radius ratio and log$_{10}$ mass are defined at a reference time, $t_\mathrm{ref}=2459378.46$. \\
    \tablefoottext{a}{We use the triangular parameterisation for quadratic limb darkening introduced by \citet{Kipping2013b}.}
    }
\end{table}

\subsection{Discussion of the TTV analysis}
Since we did not find additional transits in the available photometric data and due to the close proximity of the hot Jupiter to its host star, we establish that the companion has to be orbiting outside of the orbit of TOI-2109\,b, should it exist. 
Furthermore, if planet c's orbit is aligned with that of planet b, it would have to orbit with $P_\mathrm{c} < 1.125\,$d to be transiting (which would roughly correspond to a 5:3 MMR), indicating a longer orbital period than this lower limit, which all of our tested cases correspond to.
In theory, there are regular orbital solutions interior to the hot Jupiter, however the probability for the inner planet to not transit would be relatively small, requiring a large misalignment with a nearly polar orbit with the orbital plane having to be aligned almost perfectly perpendicular to our line of sight.

Unfortunately, the fast rotation of the host star hinders the precise measurement of RVs due to the inherent line broadening that this characteristic leads to. This means that it is unlikely that the close companion candidate will be confirmed using this method, even if many observations are taken with high-precision instruments. 
A major drawback is also the loose constraint on TOI-2109\,b's mass. The dynamical model would benefit from a reduced parameter space to better constrain the other fitted parameters.
A possible solution for the characterisation of this system could be to obtain at least two phase curve observations with JWST (see e.g., \citealt{2023ApJ...943L..17M, 2024NatAs...8..879B}). This would give us six precise timing measurements (four occultation timings and two transit timings) that, if spread out sufficiently over the phase of a sinusoid, have the potential to massively constrain the parameter space for the dynamical analysis of this system. 

The results of our \texttt{TRADES} TTV analysis reveal many possible orbital configurations that could, due to the loose contraints that we only have of the system, result in the apparent TTVs that have been observed. 
Some of the examined cases show poor convergence, which could not always be ensured owing to either the constraints that we enforced in the cases where we examined specific MMRs, or stochastic variations in the pre-optimisation procedure.
Checking the stability of the circular orbit MMR solutions, we find that all of them seem to lead to regular systems and are stable for at least 1\,Myr using the MAP solutions. 
However, the 5:3 MMR case leads to a final system configuration and orbital period ($P_\mathrm{c}=1.139\,$d) that is close to the chaotic area, but on the verge of the regular regime, as can be seen in Fig.~\ref{fig:megno_circ}. Nevertheless, we find the MAP solution to be stable for at least 1\,Myr using $N$-body simulations, all of the 1,000 randomly drawn samples of the posterior to be stable for 100,000 orbits of the hot Jupiter.
The overall best-fitting result was obtained from the 5:3 MMR case, where we allowed non-zero eccentricities. According to \texttt{REBOUND} simulations using the final MAP parameters from \texttt{TRADES}, we obtain stable orbits for at least 1\,Myr using $N$-body simulations. Even the grand majority of 1,000 randomly drawn samples of the posterior are stable for 100,000 orbits of the inner planet.
However, the regular area is relatively tight within the mean-longitude -- period space, using the MEGNO indicator.
Nevertheless, this was our overall best-fitting result.

The best-fitting result that we obtained from our general circular \texttt{TRADES} fit (see Fig.~\ref{fig:TRADES_general_best}) shows a relatively high-frequency modulation of the TTVs, with some of the chains reproducing the trend that is seen in Fig.~\ref{fig:TRADES_53_ecc_pipe}.
Our result allowing eccentric orbits in the general case (see Fig.~\ref{fig:TRADES_ecc}) is close to a 4:1 MMR and also reproduces the TTV super-period of $\sim 117\,$d that we found to be the best fit using the simple sinusoidal model (see. Fig.~\ref{fig:chain_plot} and Fig.~\ref{fig:sine_corner}).
Even if we cannot solve the system with the data that is available, this provides a good agreement between the two approaches and is evidence that the apparent sinusoidal TTVs are real. 
In theory, TTVs of this amplitude could be caused by starspot crossing events \citep{2016A&A...585A..72I}, but this seems quite unlikely in this case due to the star being an F-type star, and the very low probability of this effect to occur consistently over 3 observing seasons. In addition, we did not find signs of starspot crossings in the transits that we captured with CHEOPS.

Investigating the stability of the remaining cases that we examined, we find that the individual MAP solutions are all stable for 1\,Myr, as tested with \texttt{REBOUND}, and as has been seen for the circular orbits. Randomly drawing samples from the posterior and testing their stability also finds that most solutions are stable for 100,000 orbits of the hot Jupiter (see Section~\ref{sec:TRADES_modelling}.
For the general runs we observe that in both cases, circular and eccentric orbits, all tested solutions lead to stable systems over the examined time frames.

Regarding the \texttt{PyTTV} analysis, we find the 3:1 MMR to lead to the best fit, mainly due to the RV data. This means that the RV data may provide some evidence of planet c, though this evidence remains tentative at best, due to the large RV uncertainties, the TTVs have the highest impact in the fitting process. Before comparison of these results from those of \texttt{TRADES}, we first of all have to note the different approaches that were chosen for the two codes. 
While we are using the same data sets in both cases, we fit for more parameters using \texttt{PyTTV} due to its faster and more flexible approach, where especially the inclusion of the orbital inclination might have an impact on the results. Here, we also only focused on the specific MMRs in cases where we allow non-zero eccentricities.
Comparing the results (Fig.~\ref{fig:violins_MMR_ecc_PIPE} and Fig.~\ref{fig:pyttv_posteriors}), we can see the effect that \texttt{PyTTVs} photodynamical model has on the final fits by inspecting the final eccentricities.
\texttt{PyTTVs} achieves better convergence due to it being able to better break the mass-eccentricity degeneracies that can be seen for the TRADES equivalents of the runs.
Lower eccentricities should lead to more stable systems, especially for such tight systems, while in the case of \texttt{TRADES}, we find greater possible eccentricities for planet c, especially in the 2:1, 5:2 and 7:2 MMRs cases. Which can explain the differences in the posterior mass ranges. Other than this, the posteriors mostly agree with each other and show similar trends, remaining difference can be explained by the inclinations of planet c in the \texttt{PyTTV} runs.


\section{Conclusions\label{sec:conclusion}}

By examining newly obtained CHEOPS photometry of the TOI-2109 system, we discovered sinusoidal transit-timing variations (TTVs) while searching for signs of tidal orbital decay, 
the latter of which we find tentative evidence for when correcting the mid-transit timings using a simple sinusoidal model that describes the TTVs best.
Being unable to discover additional transits in the available photometry, we conclude that the candidate planet c must orbit outside of TOI-2109\,b's orbit with a period greater than $P_\mathrm{c} \approx 1.13\,$d owing to the high impact parameter and the orbital inclination of planet b, assuming co-planar orbits. 
Probing the apparent TTVs using a simple sine model, we found that two super-periods ($P_\mathrm{sup} \approx 88\,$d and $P_\mathrm{sup} \approx 117\,$d) fit the observations best, when compared to a simple linear model, giving us positive evidence ($\Delta\mathrm{BIC} = 4$) for the authenticity of the TTVs. 

Applying the N-body code \texttt{TRADES} \citep{2014A&A...571A..38B, 2019MNRAS.484.3233B} to our data set (for other examples see e.g., \citealt{2017AJ....153..224M, 2017A&A...601A.128N, 2019MNRAS.484.3233B, 2023A&A...673A..42N, 2024A&A...689A..52B}), we find that the best-fitting solution corresponds to a 5:3 mean-motion resonance (MMR) with $P_\mathrm{c} = (1.1398\pm0.0017)\,$d and slight eccentricities. This solution is stable for at least 1\,Myr, according to our $N$-body simulations using \texttt{REBOUND}.
From general, non-strictly MMR fits, we find orbital periods of the outer candidate companion of about $P_\mathrm{c} \approx 1.57\,$d (close to a 7:3 MMR) in the circular case and $P_\mathrm{c} \approx 2.67\,$d (close to a 4:1 MMR) in the eccentric case.
The free eccentric fit solutions are compatible with the best-fitting super-period that we found during our investigation of the TTVs with the simple sinusoidal model. And even some of the chains of the circular general fit reproduce our best-fitting model of the 5:3 MMR with eccentricities.
In addition to TTV analysis with \texttt{TRADES}, we also made use of \texttt{PyTTV} \citep{2023A&A...675A.115K}.
While we do not find the same preference of the 5:3 MMR, we find a slight preference of the 3:1 MMR and a similar posterior distributions between the codes in general. Differences can between the two can be explained by the different approach that was chosen in the modelling, with more parameters being included in \texttt{PyTTV} that were fixed in the \texttt{TRADES} analysis to achieve faster convergence.
Owing to the low amplitude of the TTVs and the quality of the radial velocity measurements caused by the fast rotation of the stellar host, we cannot fully confirm the new planetary candidate yet, or which of the models is preferred overall.
More precise timing measurements are necessary in this regard. Continued monitoring with CHEOPS would be a possible solution and would also support the search of orbital decay, since the TESS observations have not been precise enough in terms of the precision of the retrieved mid-transit timings.

Besides CHEOPS, HST or JWST observations of this system would not only allow atmospheric characterisation of this short period, ultra-hot Jupiter, but also give us the opportunity to obtain very precise timing measurements, even if only a few were obtained. Spread out over the phase of the possible sinusoids, six precise timings from two phase curve observations (four occultation timings and two transit timings) could massively constrain the parameter space, and should enable us to fully confirm this planet and better determine its characteristics. Due to the close orbit, these observations wouldn't be very expensive in terms of observation time.

Confirming the existence of the planet candidate would add TOI-2109 to the rare class of hot Jupiters with close companions and would make it only the second of these systems to host an outer companion, like it is present in the WASP-47 system \citep{2015ApJ...812L..18B, 2016A&A...595L...5A, 2016A&A...586A..93N, 2017AJ....154..237V, 2023A&A...673A..42N}.

\begin{acknowledgements}
The authors thank the anonymous referee for their helpful comments, which helped to improve the clarity of the manuscript.
CHEOPS is an ESA mission in partnership with Switzerland with important contributions to the payload and the ground segment from Austria, Belgium, France, Germany, Hungary, Italy, Portugal, Spain, Sweden, and the United Kingdom. The CHEOPS Consortium would like to gratefully acknowledge the support received by all the agencies, offices, universities, and industries involved. Their flexibility and willingness to explore new approaches were essential to the success of this mission. CHEOPS data analysed in this article will be made available in the CHEOPS mission archive (\url{https://cheops.unige.ch/archive_browser/}). 
JVH is funded by the DFG priority programme SPP 1992 ``Exploring the
Diversity of Extrasolar Planets (SM 486/2-1)''.
S.C.C.B. acknowledges support from FCT through FCT contracts nr. IF/01312/2014/CP1215/CT0004. 
LBo, GBr, VNa, IPa, GPi, RRa, GSc, VSi, and TZi acknowledge support from CHEOPS ASI-INAF agreement n. 2019-29-HH.0. 
ABr was supported by the SNSA. 
ACC acknowledges support from STFC consolidated grant number ST/V000861/1, and UKSA grant number ST/X002217/1. 
ML acknowledges support of the Swiss National Science Foundation under grant number PCEFP2\_194576. 
TWi acknowledges support from the UKSA and the University of Warwick. 
ACMC acknowledges support from the FCT, Portugal, through the CFisUC projects UIDB/04564/2020 and UIDP/04564/2020, with DOI identifiers 10.54499/UIDB/04564/2020 and 10.54499/UIDP/04564/2020, respectively.
YAl acknowledges support from the Swiss National Science Foundation (SNSF) under grant 200020\_192038. 
We acknowledge financial support from the Agencia Estatal de Investigación of the Ministerio de Ciencia e Innovación MCIN/AEI/10.13039/501100011033 and the ERDF “A way of making Europe” through projects PID2019-107061GB-C61, PID2019-107061GB-C66, PID2021-125627OB-C31, and PID2021-125627OB-C32, from the Centre of Excellence “Severo Ochoa” award to the Instituto de Astrofísica de Canarias (CEX2019-000920-S), from the Centre of Excellence “María de Maeztu” award to the Institut de Ciències de l’Espai (CEX2020-001058-M), and from the Generalitat de Catalunya/CERCA programme. 
We acknowledge financial support from the Agencia Estatal de Investigación of the Ministerio de Ciencia e Innovación MCIN/AEI/10.13039/501100011033 and the ERDF “A way of making Europe” through projects PID2019-107061GB-C61, PID2019-107061GB-C66, PID2021-125627OB-C31, and PID2021-125627OB-C32, from the Centre of Excellence “Severo Ochoa'' award to the Instituto de Astrofísica de Canarias (CEX2019-000920-S), from the Centre of Excellence “María de Maeztu” award to the Institut de Ciències de l’Espai (CEX2020-001058-M), and from the Generalitat de Catalunya/CERCA programme. 
C.B. acknowledges support from the Swiss Space Office through the ESA PRODEX program. 
This work has been carried out within the framework of the NCCR PlanetS supported by the Swiss National Science Foundation under grants 51NF40\_182901 and 51NF40\_205606. 
P.E.C. is funded by the Austrian Science Fund (FWF) Erwin Schroedinger Fellowship, program J4595-N. 
This project was supported by the CNES. 
The Belgian participation to CHEOPS has been supported by the Belgian Federal Science Policy Office (BELSPO) in the framework of the PRODEX Program, and by the University of Liège through an ARC grant for Concerted Research Actions financed by the Wallonia-Brussels Federation. 
L.D. thanks the Belgian Federal Science Policy Office (BELSPO) for the provision of financial support in the framework of the PRODEX Programme of the European Space Agency (ESA) under contract number 4000142531. 
This work was supported by FCT - Funda\c{c}\~{a}o para a Ci\^{e}ncia e a Tecnologia through national funds and by FEDER through COMPETE2020 through the research grants UIDB/04434/2020, UIDP/04434/2020, 2022.06962.PTDC. 
O.D.S.D. is supported in the form of work contract (DL 57/2016/CP1364/CT0004) funded by national funds through FCT. 
B.-O. D. acknowledges support from the Swiss State Secretariat for Education, Research and Innovation (SERI) under contract number MB22.00046. 
This project has received funding from the Swiss National Science Foundation for project 200021\_200726. It has also been carried out within the framework of the National Centre of Competence in Research PlanetS supported by the Swiss National Science Foundation under grant 51NF40\_205606. The authors acknowledge the financial support of the SNSF. 
MF and CMP gratefully acknowledge the support of the Swedish National Space Agency (DNR 65/19, 174/18). 
DG gratefully acknowledges financial support from the CRT foundation under Grant No. 2018.2323 “Gaseousor rocky? Unveiling the nature of small worlds”. 
M.G. is an F.R.S.-FNRS Senior Research Associate. 
MNG is the ESA CHEOPS Project Scientist and Mission Representative, and as such also responsible for the Guest Observers (GO) Programme. MNG does not relay proprietary information between the GO and Guaranteed Time Observation (GTO) Programmes, and does not decide on the definition and target selection of the GTO Programme. 
CHe acknowledges support from the European Union H2020-MSCA-ITN-2019 under Grant Agreement no. 860470 (CHAMELEON). 
KGI is the ESA CHEOPS Project Scientist and is responsible for the ESA CHEOPS Guest Observers Programme. She does not participate in, or contribute to, the definition of the Guaranteed Time Programme of the CHEOPS mission through which observations described in this paper have been taken, nor to any aspect of target selection for the programme. 
K.W.F.L. was supported by Deutsche Forschungsgemeinschaft grants RA714/14-1 within the DFG Schwerpunkt SPP 1992, Exploring the Diversity of Extrasolar Planets. 
This work was granted access to the HPC resources of MesoPSL financed by the Region Ile de France and the project Equip@Meso (reference ANR-10-EQPX-29-01) of the programme Investissements d'Avenir supervised by the Agence Nationale pour la Recherche. 
PM acknowledges support from STFC research grant number ST/R000638/1. 
This work was also partially supported by a grant from the Simons Foundation (PI Queloz, grant number 327127). 
NCSa acknowledges funding by the European Union (ERC, FIERCE, 101052347). Views and opinions expressed are however those of the author(s) only and do not necessarily reflect those of the European Union or the European Research Council. Neither the European Union nor the granting authority can be held responsible for them. 
A. S. acknowledges support from the Swiss Space Office through the ESA PRODEX program. 
S.G.S. acknowledge support from FCT through FCT contract nr. CEECIND/00826/2018 and POPH/FSE (EC). 
The Portuguese team thanks the Portuguese Space Agency for the provision of financial support in the framework of the PRODEX Programme of the European Space Agency (ESA) under contract number 4000142255. 
GyMSz acknowledges the support of the Hungarian National Research, Development and Innovation Office (NKFIH) grant K-125015, a a PRODEX Experiment Agreement No. 4000137122, the Lendulet LP2018-7/2021 grant of the Hungarian Academy of Science and the support of the city of Szombathely. 
V.V.G. is an F.R.S-FNRS Research Associate. 
JV acknowledges support from the Swiss National Science Foundation (SNSF) under grant PZ00P2\_208945. 
NAW acknowledges UKSA grant ST/R004838/1. 
This work made use of the \texttt{Python} libraries \texttt{NumPy} \citep{harris2020array}, \texttt{Matplotlib} \citep{Hunter:2007}, \texttt{corner} \citep{corner}, and \texttt{Astropy} \citep{2022ApJ...935..167A}.
\end{acknowledgements}

\bibliographystyle{aa}
\bibliography{aanda}

\begin{thebibliography}{121}
\expandafter\ifx\csname natexlab\endcsname\relax\def\natexlab#1{#1}\fi

\bibitem[{{Agol} {et~al.}(2005){Agol}, {Steffen}, {Sari}, \& {Clarkson}}]{2005MNRAS.359..567A}
{Agol}, E., {Steffen}, J., {Sari}, R., \& {Clarkson}, W. 2005, \mnras, 359, 567

\bibitem[{{Albrecht} {et~al.}(2022){Albrecht}, {Dawson}, \& {Winn}}]{2022PASP..134h2001A}
{Albrecht}, S.~H., {Dawson}, R.~I., \& {Winn}, J.~N. 2022, \pasp, 134, 082001

\bibitem[{{Albrecht} {et~al.}(2021){Albrecht}, {Marcussen}, {Winn}, {Dawson}, \& {Knudstrup}}]{2021ApJ...916L...1A}
{Albrecht}, S.~H., {Marcussen}, M.~L., {Winn}, J.~N., {Dawson}, R.~I., \& {Knudstrup}, E. 2021, \apjl, 916, L1

\bibitem[{{Almenara} {et~al.}(2016){Almenara}, {D{\'\i}az}, {Bonfils}, \& {Udry}}]{2016A&A...595L...5A}
{Almenara}, J.~M., {D{\'\i}az}, R.~F., {Bonfils}, X., \& {Udry}, S. 2016, \aap, 595, L5

\bibitem[{{Anderson} {et~al.}(2014){Anderson}, {Collier Cameron}, {Delrez}, {Doyle}, {Faedi}, {Fumel}, {Gillon}, {G{\'o}mez Maqueo Chew}, {Hellier}, {Jehin}, {Lendl}, {Maxted}, {Pepe}, {Pollacco}, {Queloz}, {S{\'e}gransan}, {Skillen}, {Smalley}, {Smith}, {Southworth}, {Triaud}, {Turner}, {Udry}, \& {West}}]{2014MNRAS.445.1114A}
{Anderson}, D.~R., {Collier Cameron}, A., {Delrez}, L., {et~al.} 2014, \mnras, 445, 1114

\bibitem[{{Astropy Collaboration} {et~al.}(2022){Astropy Collaboration}, {Price-Whelan}, {Lim}, {Earl}, {Starkman}, {Bradley}, {Shupe}, {Patil}, {Corrales}, {Brasseur}, {N{\"o}the}, {Donath}, {Tollerud}, {Morris}, {Ginsburg}, {Vaher}, {Weaver}, {Tocknell}, {Jamieson}, {van Kerkwijk}, {Robitaille}, {Merry}, {Bachetti}, {G{\"u}nther}, {Aldcroft}, {Alvarado-Montes}, {Archibald}, {B{\'o}di}, {Bapat}, {Barentsen}, {Baz{\'a}n}, {Biswas}, {Boquien}, {Burke}, {Cara}, {Cara}, {Conroy}, {Conseil}, {Craig}, {Cross}, {Cruz}, {D'Eugenio}, {Dencheva}, {Devillepoix}, {Dietrich}, {Eigenbrot}, {Erben}, {Ferreira}, {Foreman-Mackey}, {Fox}, {Freij}, {Garg}, {Geda}, {Glattly}, {Gondhalekar}, {Gordon}, {Grant}, {Greenfield}, {Groener}, {Guest}, {Gurovich}, {Handberg}, {Hart}, {Hatfield-Dodds}, {Homeier}, {Hosseinzadeh}, {Jenness}, {Jones}, {Joseph}, {Kalmbach}, {Karamehmetoglu}, {Ka{\l}uszy{\'n}ski}, {Kelley}, {Kern}, {Kerzendorf}, {Koch}, {Kulumani}, {Lee}, {Ly}, {Ma}, {MacBride}, {Maljaars}, {Muna}, {Murphy}, {Norman},
  {O'Steen}, {Oman}, {Pacifici}, {Pascual}, {Pascual-Granado}, {Patil}, {Perren}, {Pickering}, {Rastogi}, {Roulston}, {Ryan}, {Rykoff}, {Sabater}, {Sakurikar}, {Salgado}, {Sanghi}, {Saunders}, {Savchenko}, {Schwardt}, {Seifert-Eckert}, {Shih}, {Jain}, {Shukla}, {Sick}, {Simpson}, {Singanamalla}, {Singer}, {Singhal}, {Sinha}, {Sip{\H{o}}cz}, {Spitler}, {Stansby}, {Streicher}, {{\v{S}}umak}, {Swinbank}, {Taranu}, {Tewary}, {Tremblay}, {de Val-Borro}, {Van Kooten}, {Vasovi{\'c}}, {Verma}, {de Miranda Cardoso}, {Williams}, {Wilson}, {Winkel}, {Wood-Vasey}, {Xue}, {Yoachim}, {Zhang}, {Zonca}, \& {Astropy Project Contributors}}]{2022ApJ...935..167A}
{Astropy Collaboration}, {Price-Whelan}, A.~M., {Lim}, P.~L., {et~al.} 2022, \apj, 935, 167

\bibitem[{{Barros} {et~al.}(2022){Barros}, {Akinsanmi}, {Bou{\'e}}, {Smith}, {Laskar}, {Ulmer-Moll}, {Lillo-Box}, {Queloz}, {Cameron}, {Sousa}, {Ehrenreich}, {Hooton}, {Bruno}, {Demory}, {Correia}, {Demangeon}, {Wilson}, {Bonfanti}, {Hoyer}, {Alibert}, {Alonso}, {Escud{\'e}}, {Barbato}, {B{\'a}rczy}, {Barrado}, {Baumjohann}, {Beck}, {Beck}, {Benz}, {Bergomi}, {Billot}, {Bonfils}, {Bouchy}, {Brandeker}, {Broeg}, {Cabrera}, {Cessa}, {Charnoz}, {Damme}, {Davies}, {Deleuil}, {Deline}, {Delrez}, {Erikson}, {Fortier}, {Fossati}, {Fridlund}, {Gandolfi}, {Mu{\~n}oz}, {Gillon}, {G{\"u}del}, {Isaak}, {Heng}, {Kiss}, {des Etangs}, {Lendl}, {Lovis}, {Magrin}, {Nascimbeni}, {Maxted}, {Olofsson}, {Ottensamer}, {Pagano}, {Pall{\'e}}, {Parviainen}, {Peter}, {Piotto}, {Pollacco}, {Ragazzoni}, {Rando}, {Rauer}, {Ribas}, {Santos}, {Scandariato}, {S{\'e}gransan}, {Simon}, {Steller}, {Szab{\'o}}, {Thomas}, {Udry}, {Ulmer}, {Van Grootel}, \& {Walton}}]{2022A&A...657A..52B}
{Barros}, S.~C.~C., {Akinsanmi}, B., {Bou{\'e}}, G., {et~al.} 2022, \aap, 657, A52

\bibitem[{{Becker} {et~al.}(2015){Becker}, {Vanderburg}, {Adams}, {Rappaport}, \& {Schwengeler}}]{2015ApJ...812L..18B}
{Becker}, J.~C., {Vanderburg}, A., {Adams}, F.~C., {Rappaport}, S.~A., \& {Schwengeler}, H.~M. 2015, \apjl, 812, L18

\bibitem[{{Bell} {et~al.}(2024){Bell}, {Crouzet}, {Cubillos}, {Kreidberg}, {Piette}, {Roman}, {Barstow}, {Blecic}, {Carone}, {Coulombe}, {Ducrot}, {Hammond}, {Mendon{\c{c}}a}, {Moses}, {Parmentier}, {Stevenson}, {Teinturier}, {Zhang}, {Batalha}, {Bean}, {Benneke}, {Charnay}, {Chubb}, {Demory}, {Gao}, {Lee}, {L{\'o}pez-Morales}, {Morello}, {Rauscher}, {Sing}, {Tan}, {Venot}, {Wakeford}, {Aggarwal}, {Ahrer}, {Alam}, {Baeyens}, {Barrado}, {Caceres}, {Carter}, {Casewell}, {Challener}, {Crossfield}, {Decin}, {D{\'e}sert}, {Dobbs-Dixon}, {Dyrek}, {Espinoza}, {Feinstein}, {Gibson}, {Harrington}, {Helling}, {Hu}, {Iro}, {Kempton}, {Kendrew}, {Komacek}, {Krick}, {Lagage}, {Leconte}, {Lendl}, {Lewis}, {Lothringer}, {Malsky}, {Mancini}, {Mansfield}, {Mayne}, {Evans-Soma}, {Molaverdikhani}, {Nikolov}, {Nixon}, {Palle}, {Petit dit de la Roche}, {Piaulet}, {Powell}, {Rackham}, {Schneider}, {Steinrueck}, {Taylor}, {Welbanks}, {Yurchenko}, {Zhang}, \& {Zieba}}]{2024NatAs...8..879B}
{Bell}, T.~J., {Crouzet}, N., {Cubillos}, P.~E., {et~al.} 2024, Nature Astronomy, 8, 879

\bibitem[{{Benz} {et~al.}(2021){Benz}, {Broeg}, {Fortier}, {Rando}, {Beck}, {Beck}, {Queloz}, {Ehrenreich}, {Maxted}, {Isaak}, {Billot}, {Alibert}, {Alonso}, {Ant{\'o}nio}, {Asquier}, {Bandy}, {B{\'a}rczy}, {Barrado}, {Barros}, {Baumjohann}, {Bekkelien}, {Bergomi}, {Biondi}, {Bonfils}, {Borsato}, {Brandeker}, {Busch}, {Cabrera}, {Cessa}, {Charnoz}, {Chazelas}, {Collier Cameron}, {Corral Van Damme}, {Cortes}, {Davies}, {Deleuil}, {Deline}, {Delrez}, {Demangeon}, {Demory}, {Erikson}, {Farinato}, {Fossati}, {Fridlund}, {Futyan}, {Gandolfi}, {Garcia Munoz}, {Gillon}, {Guterman}, {Gutierrez}, {Hasiba}, {Heng}, {Hernandez}, {Hoyer}, {Kiss}, {Kovacs}, {Kuntzer}, {Laskar}, {Lecavelier des Etangs}, {Lendl}, {L{\'o}pez}, {Lora}, {Lovis}, {L{\"u}ftinger}, {Magrin}, {Malvasio}, {Marafatto}, {Michaelis}, {de Miguel}, {Modrego}, {Munari}, {Nascimbeni}, {Olofsson}, {Ottacher}, {Ottensamer}, {Pagano}, {Palacios}, {Pall{\'e}}, {Peter}, {Piazza}, {Piotto}, {Pizarro}, {Pollaco}, {Ragazzoni}, {Ratti}, {Rauer}, {Ribas}, {Rieder},
  {Rohlfs}, {Safa}, {Salatti}, {Santos}, {Scandariato}, {S{\'e}gransan}, {Simon}, {Smith}, {Sordet}, {Sousa}, {Steller}, {Szab{\'o}}, {Szoke}, {Thomas}, {Tschentscher}, {Udry}, {Van Grootel}, {Viotto}, {Walter}, {Walton}, {Wildi}, \& {Wolter}}]{2021ExA....51..109B}
{Benz}, W., {Broeg}, C., {Fortier}, A., {et~al.} 2021, Experimental Astronomy, 51, 109

\bibitem[{{Borsato} {et~al.}(2024){Borsato}, {Degen}, {Leleu}, {Hooton}, {Egger}, {Bekkelien}, {Brandeker}, {Collier Cameron}, {G{\"u}nther}, {Nascimbeni}, {Persson}, {Bonfanti}, {Wilson}, {Correia}, {Zingales}, {Guillot}, {Triaud}, {Piotto}, {Gandolfi}, {Abe}, {Alibert}, {Alonso}, {B{\'a}rczy}, {Navascues}, {Barros}, {Baumjohann}, {Beck}, {Bendjoya}, {Benz}, {Billot}, {Broeg}, {Busch}, {Csizmadia}, {Cubillos}, {Davies}, {Deleuil}, {Deline}, {Delrez}, {Demangeon}, {Demory}, {Derekas}, {Edwards}, {Ehrenreich}, {Erikson}, {Fortier}, {Fossati}, {Fridlund}, {Gazeas}, {Gillon}, {G{\"u}del}, {Heitzmann}, {Helling}, {Hoyer}, {Isaak}, {Kiss}, {Korth}, {Lam}, {Laskar}, {Lecavelier des Etangs}, {Lendl}, {Magrin}, {Marafatto}, {Maxted}, {Mecina}, {M{\'e}karnia}, {Mordasini}, {Mura}, {Olofsson}, {Ottensamer}, {Pagano}, {Pall{\'e}}, {Peter}, {Pollacco}, {Queloz}, {Ragazzoni}, {Rando}, {Ratti}, {Rauer}, {Ribas}, {Salmon}, {Santos}, {Scandariato}, {S{\'e}gransan}, {Simon}, {Smith}, {Sousa}, {Stalport}, {Suarez}, {Sulis},
  {Szab{\'o}}, {Udry}, {Van Grootel}, {Venturini}, {Villaver}, {Walton}, \& {Wolter}}]{2024A&A...689A..52B}
{Borsato}, L., {Degen}, D., {Leleu}, A., {et~al.} 2024, \aap, 689, A52

\bibitem[{{Borsato} {et~al.}(2019){Borsato}, {Malavolta}, {Piotto}, {Buchhave}, {Mortier}, {Rice}, {Collier Cameron}, {Coffinet}, {Sozzetti}, {Charbonneau}, {Cosentino}, {Dumusque}, {Figueira}, {Latham}, {Lopez-Morales}, {Mayor}, {Micela}, {Molinari}, {Pepe}, {Phillips}, {Poretti}, {Udry}, \& {Watson}}]{2019MNRAS.484.3233B}
{Borsato}, L., {Malavolta}, L., {Piotto}, G., {et~al.} 2019, \mnras, 484, 3233

\bibitem[{{Borsato} {et~al.}(2014){Borsato}, {Marzari}, {Nascimbeni}, {Piotto}, {Granata}, {Bedin}, \& {Malavolta}}]{2014A&A...571A..38B}
{Borsato}, L., {Marzari}, F., {Nascimbeni}, V., {et~al.} 2014, \aap, 571, A38

\bibitem[{{Bou{\'e}} \& {Efroimsky}(2019)}]{2019CeMDA.131...30B}
{Bou{\'e}}, G. \& {Efroimsky}, M. 2019, Celestial Mechanics and Dynamical Astronomy, 131, 30

\bibitem[{{Brandeker} {et~al.}(2022){Brandeker}, {Heng}, {Lendl}, {Patel}, {Morris}, {Broeg}, {Guterman}, {Beck}, {Maxted}, {Demangeon}, {Delrez}, {Demory}, {Kitzmann}, {Santos}, {Singh}, {Alibert}, {Alonso}, {Anglada}, {B{\'a}rczy}, {Barrado y Navascues}, {Barros}, {Baumjohann}, {Beck}, {Benz}, {Billot}, {Bonfils}, {Bruno}, {Cabrera}, {Charnoz}, {Collier Cameron}, {Corral van Damme}, {Csizmadia}, {Davies}, {Deleuil}, {Deline}, {Ehrenreich}, {Erikson}, {Farinato}, {Fortier}, {Fossati}, {Fridlund}, {Gandolfi}, {Gillon}, {G{\"u}del}, {Hoyer}, {Isaak}, {Kiss}, {Laskar}, {Lecavelier des Etangs}, {Lovis}, {Luntzer}, {Magrin}, {Nascimbeni}, {Olofsson}, {Ottensamer}, {Pagano}, {Pall{\'e}}, {Peter}, {Piotto}, {Pollacco}, {Queloz}, {Ragazzoni}, {Rando}, {Rauer}, {Ribas}, {Scandariato}, {S{\'e}gransan}, {Simon}, {Smith}, {Sousa}, {Steller}, {Szab{\'o}}, {Thomas}, {Udry}, {Van Grootel}, {Walton}, \& {Wolter}}]{2022A&A...659L...4B}
{Brandeker}, A., {Heng}, K., {Lendl}, M., {et~al.} 2022, \aap, 659, L4

\bibitem[{{Ca{\~n}as} {et~al.}(2019){Ca{\~n}as}, {Wang}, {Mahadevan}, {Bender}, {De Lee}, {Fleming}, {Garc{\'\i}a-Hern{\'a}ndez}, {Hearty}, {Majewski}, {Roman-Lopes}, {Schneider}, \& {Stassun}}]{2019ApJ...870L..17C}
{Ca{\~n}as}, C.~I., {Wang}, S., {Mahadevan}, S., {et~al.} 2019, \apjl, 870, L17

\bibitem[{{Cabrera} {et~al.}(2012){Cabrera}, {Csizmadia}, {Erikson}, {Rauer}, \& {Kirste}}]{2012A&A...548A..44C}
{Cabrera}, J., {Csizmadia}, S., {Erikson}, A., {Rauer}, H., \& {Kirste}, S. 2012, \aap, 548, A44

\bibitem[{{Cincotta} \& {Sim{\'o}}(2000)}]{2000A&AS..147..205C}
{Cincotta}, P.~M. \& {Sim{\'o}}, C. 2000, \aaps, 147, 205

\bibitem[{{Claret}(2017)}]{2017A&A...600A..30C}
{Claret}, A. 2017, \aap, 600, A30

\bibitem[{{Claret}(2021)}]{2021RNAAS...5...13C}
{Claret}, A. 2021, Research Notes of the American Astronomical Society, 5, 13

\bibitem[{{Collier Cameron} \& {Jardine}(2018)}]{2018MNRAS.476.2542C}
{Collier Cameron}, A. \& {Jardine}, M. 2018, \mnras, 476, 2542

\bibitem[{{Counselman}(1973)}]{1973ApJ...180..307C}
{Counselman}, Charles~C., I. 1973, \apj, 180, 307

\bibitem[{{Csizmadia}(2020)}]{2020MNRAS.496.4442C}
{Csizmadia}, S. 2020, \mnras, 496, 4442

\bibitem[{{Dawson} \& {Johnson}(2018)}]{2018ARA&A..56..175D}
{Dawson}, R.~I. \& {Johnson}, J.~A. 2018, \araa, 56, 175

\bibitem[{{Deline} {et~al.}(2022){Deline}, {Hooton}, {Lendl}, {Morris}, {Salmon}, {Olofsson}, {Broeg}, {Ehrenreich}, {Beck}, {Brandeker}, {Hoyer}, {Sulis}, {Van Grootel}, {Bourrier}, {Demangeon}, {Demory}, {Heng}, {Parviainen}, {Serrano}, {Singh}, {Bonfanti}, {Fossati}, {Kitzmann}, {Sousa}, {Wilson}, {Alibert}, {Alonso}, {Anglada}, {B{\'a}rczy}, {Barrado Navascues}, {Barros}, {Baumjohann}, {Beck}, {Bekkelien}, {Benz}, {Billot}, {Bonfils}, {Cabrera}, {Charnoz}, {Collier Cameron}, {Corral van Damme}, {Csizmadia}, {Davies}, {Deleuil}, {Delrez}, {de Roche}, {Erikson}, {Fortier}, {Fridlund}, {Futyan}, {Gandolfi}, {Gillon}, {G{\"u}del}, {Gutermann}, {Hasiba}, {Isaak}, {Kiss}, {Laskar}, {Lecavelier des Etangs}, {Lovis}, {Magrin}, {Maxted}, {Munari}, {Nascimbeni}, {Ottensamer}, {Pagano}, {Pall{\'e}}, {Peter}, {Piotto}, {Pollacco}, {Queloz}, {Ragazzoni}, {Rando}, {Rauer}, {Ribas}, {Santos}, {Scandariato}, {S{\'e}gransan}, {Simon}, {Smith}, {Steller}, {Szab{\'o}}, {Thomas}, {Udry}, {Walter}, \&
  {Walton}}]{2022A&A...659A..74D}
{Deline}, A., {Hooton}, M.~J., {Lendl}, M., {et~al.} 2022, \aap, 659, A74

\bibitem[{{Dos Santos} {et~al.}(2023){Dos Santos}, {Alam}, {Espinoza}, \& {Vissapragada}}]{2023AJ....165..244D}
{Dos Santos}, L.~A., {Alam}, M.~K., {Espinoza}, N., \& {Vissapragada}, S. 2023, \aj, 165, 244

\bibitem[{Foreman-Mackey(2016)}]{corner}
Foreman-Mackey, D. 2016, The Journal of Open Source Software, 1, 24

\bibitem[{{Foreman-Mackey} {et~al.}(2013){Foreman-Mackey}, {Hogg}, {Lang}, \& {Goodman}}]{2013PASP..125..306F}
{Foreman-Mackey}, D., {Hogg}, D.~W., {Lang}, D., \& {Goodman}, J. 2013, \pasp, 125, 306

\bibitem[{{Gaia Collaboration} {et~al.}(2016){Gaia Collaboration}, {Prusti}, {de Bruijne}, {Brown}, {Vallenari}, {Babusiaux}, {Bailer-Jones}, {Bastian}, {Biermann}, {Evans}, {Eyer}, {Jansen}, {Jordi}, {Klioner}, {Lammers}, {Lindegren}, {Luri}, {Mignard}, {Milligan}, {Panem}, {Poinsignon}, {Pourbaix}, {Randich}, {Sarri}, {Sartoretti}, {Siddiqui}, {Soubiran}, {Valette}, {van Leeuwen}, {Walton}, {Aerts}, {Arenou}, {Cropper}, {Drimmel}, {H{\o}g}, {Katz}, {Lattanzi}, {O'Mullane}, {Grebel}, {Holland}, {Huc}, {Passot}, {Bramante}, {Cacciari}, {Casta{\~n}eda}, {Chaoul}, {Cheek}, {De Angeli}, {Fabricius}, {Guerra}, {Hern{\'a}ndez}, {Jean-Antoine-Piccolo}, {Masana}, {Messineo}, {Mowlavi}, {Nienartowicz}, {Ord{\'o}{\~n}ez-Blanco}, {Panuzzo}, {Portell}, {Richards}, {Riello}, {Seabroke}, {Tanga}, {Th{\'e}venin}, {Torra}, {Els}, {Gracia-Abril}, {Comoretto}, {Garcia-Reinaldos}, {Lock}, {Mercier}, {Altmann}, {Andrae}, {Astraatmadja}, {Bellas-Velidis}, {Benson}, {Berthier}, {Blomme}, {Busso}, {Carry}, {Cellino}, {Clementini},
  {Cowell}, {Creevey}, {Cuypers}, {Davidson}, {De Ridder}, {de Torres}, {Delchambre}, {Dell'Oro}, {Ducourant}, {Fr{\'e}mat}, {Garc{\'\i}a-Torres}, {Gosset}, {Halbwachs}, {Hambly}, {Harrison}, {Hauser}, {Hestroffer}, {Hodgkin}, {Huckle}, {Hutton}, {Jasniewicz}, {Jordan}, {Kontizas}, {Korn}, {Lanzafame}, {Manteiga}, {Moitinho}, {Muinonen}, {Osinde}, {Pancino}, {Pauwels}, {Petit}, {Recio-Blanco}, {Robin}, {Sarro}, {Siopis}, {Smith}, {Smith}, {Sozzetti}, {Thuillot}, {van Reeven}, {Viala}, {Abbas}, {Abreu Aramburu}, {Accart}, {Aguado}, {Allan}, {Allasia}, {Altavilla}, {{\'A}lvarez}, {Alves}, {Anderson}, {Andrei}, {Anglada Varela}, {Antiche}, {Antoja}, {Ant{\'o}n}, {Arcay}, {Atzei}, {Ayache}, {Bach}, {Baker}, {Balaguer-N{\'u}{\~n}ez}, {Barache}, {Barata}, {Barbier}, {Barblan}, {Baroni}, {Barrado y Navascu{\'e}s}, {Barros}, {Barstow}, {Becciani}, {Bellazzini}, {Bellei}, {Bello Garc{\'\i}a}, {Belokurov}, {Bendjoya}, {Berihuete}, {Bianchi}, {Bienaym{\'e}}, {Billebaud}, {Blagorodnova}, {Blanco-Cuaresma}, {Boch},
  {Bombrun}, {Borrachero}, {Bouquillon}, {Bourda}, {Bouy}, {Bragaglia}, {Breddels}, {Brouillet}, {Br{\"u}semeister}, {Bucciarelli}, {Budnik}, {Burgess}, {Burgon}, {Burlacu}, {Busonero}, {Buzzi}, {Caffau}, {Cambras}, {Campbell}, {Cancelliere}, {Cantat-Gaudin}, {Carlucci}, {Carrasco}, {Castellani}, {Charlot}, {Charnas}, {Charvet}, {Chassat}, {Chiavassa}, {Clotet}, {Cocozza}, {Collins}, {Collins}, {Costigan}, {Crifo}, {Cross}, {Crosta}, {Crowley}, {Dafonte}, {Damerdji}, {Dapergolas}, {David}, {David}, {De Cat}, {de Felice}, {de Laverny}, {De Luise}, {De March}, {de Martino}, {de Souza}, {Debosscher}, {del Pozo}, {Delbo}, {Delgado}, {Delgado}, {di Marco}, {Di Matteo}, {Diakite}, {Distefano}, {Dolding}, {Dos Anjos}, {Drazinos}, {Dur{\'a}n}, {Dzigan}, {Ecale}, {Edvardsson}, {Enke}, {Erdmann}, {Escolar}, {Espina}, {Evans}, {Eynard Bontemps}, {Fabre}, {Fabrizio}, {Faigler}, {Falc{\~a}o}, {Farr{\`a}s Casas}, {Faye}, {Federici}, {Fedorets}, {Fern{\'a}ndez-Hern{\'a}ndez}, {Fernique}, {Fienga}, {Figueras}, {Filippi},
  {Findeisen}, {Fonti}, {Fouesneau}, {Fraile}, {Fraser}, {Fuchs}, {Furnell}, {Gai}, {Galleti}, {Galluccio}, {Garabato}, {Garc{\'\i}a-Sedano}, {Gar{\'e}}, {Garofalo}, {Garralda}, {Gavras}, {Gerssen}, {Geyer}, {Gilmore}, {Girona}, {Giuffrida}, {Gomes}, {Gonz{\'a}lez-Marcos}, {Gonz{\'a}lez-N{\'u}{\~n}ez}, {Gonz{\'a}lez-Vidal}, {Granvik}, {Guerrier}, {Guillout}, {Guiraud}, {G{\'u}rpide}, {Guti{\'e}rrez-S{\'a}nchez}, {Guy}, {Haigron}, {Hatzidimitriou}, {Haywood}, {Heiter}, {Helmi}, {Hobbs}, {Hofmann}, {Holl}, {Holland}, {Hunt}, {Hypki}, {Icardi}, {Irwin}, {Jevardat de Fombelle}, {Jofr{\'e}}, {Jonker}, {Jorissen}, {Julbe}, {Karampelas}, {Kochoska}, {Kohley}, {Kolenberg}, {Kontizas}, {Koposov}, {Kordopatis}, {Koubsky}, {Kowalczyk}, {Krone-Martins}, {Kudryashova}, {Kull}, {Bachchan}, {Lacoste-Seris}, {Lanza}, {Lavigne}, {Le Poncin-Lafitte}, {Lebreton}, {Lebzelter}, {Leccia}, {Leclerc}, {Lecoeur-Taibi}, {Lemaitre}, {Lenhardt}, {Leroux}, {Liao}, {Licata}, {Lindstr{\o}m}, {Lister}, {Livanou}, {Lobel}, {L{\"o}ffler},
  {L{\'o}pez}, {Lopez-Lozano}, {Lorenz}, {Loureiro}, {MacDonald}, {Magalh{\~a}es Fernandes}, {Managau}, {Mann}, {Mantelet}, {Marchal}, {Marchant}, {Marconi}, {Marie}, {Marinoni}, {Marrese}, {Marschalk{\'o}}, {Marshall}, {Mart{\'\i}n-Fleitas}, {Martino}, {Mary}, {Matijevi{\v{c}}}, {Mazeh}, {McMillan}, {Messina}, {Mestre}, {Michalik}, {Millar}, {Miranda}, {Molina}, {Molinaro}, {Molinaro}, {Moln{\'a}r}, {Moniez}, {Montegriffo}, {Monteiro}, {Mor}, {Mora}, {Morbidelli}, {Morel}, {Morgenthaler}, {Morley}, {Morris}, {Mulone}, {Muraveva}, {Musella}, {Narbonne}, {Nelemans}, {Nicastro}, {Noval}, {Ord{\'e}novic}, {Ordieres-Mer{\'e}}, {Osborne}, {Pagani}, {Pagano}, {Pailler}, {Palacin}, {Palaversa}, {Parsons}, {Paulsen}, {Pecoraro}, {Pedrosa}, {Pentik{\"a}inen}, {Pereira}, {Pichon}, {Piersimoni}, {Pineau}, {Plachy}, {Plum}, {Poujoulet}, {Pr{\v{s}}a}, {Pulone}, {Ragaini}, {Rago}, {Rambaux}, {Ramos-Lerate}, {Ranalli}, {Rauw}, {Read}, {Regibo}, {Renk}, {Reyl{\'e}}, {Ribeiro}, {Rimoldini}, {Ripepi}, {Riva}, {Rixon},
  {Roelens}, {Romero-G{\'o}mez}, {Rowell}, {Royer}, {Rudolph}, {Ruiz-Dern}, {Sadowski}, {Sagrist{\`a} Sell{\'e}s}, {Sahlmann}, {Salgado}, {Salguero}, {Sarasso}, {Savietto}, {Schnorhk}, {Schultheis}, {Sciacca}, {Segol}, {Segovia}, {Segransan}, {Serpell}, {Shih}, {Smareglia}, {Smart}, {Smith}, {Solano}, {Solitro}, {Sordo}, {Soria Nieto}, {Souchay}, {Spagna}, {Spoto}, {Stampa}, {Steele}, {Steidelm{\"u}ller}, {Stephenson}, {Stoev}, {Suess}, {S{\"u}veges}, {Surdej}, {Szabados}, {Szegedi-Elek}, {Tapiador}, {Taris}, {Tauran}, {Taylor}, {Teixeira}, {Terrett}, {Tingley}, {Trager}, {Turon}, {Ulla}, {Utrilla}, {Valentini}, {van Elteren}, {Van Hemelryck}, {van Leeuwen}, {Varadi}, {Vecchiato}, {Veljanoski}, {Via}, {Vicente}, {Vogt}, {Voss}, {Votruba}, {Voutsinas}, {Walmsley}, {Weiler}, {Weingrill}, {Werner}, {Wevers}, {Whitehead}, {Wyrzykowski}, {Yoldas}, {{\v{Z}}erjal}, {Zucker}, {Zurbach}, {Zwitter}, {Alecu}, {Allen}, {Allende Prieto}, {Amorim}, {Anglada-Escud{\'e}}, {Arsenijevic}, {Azaz}, {Balm}, {Beck}, {Bernstein},
  {Bigot}, {Bijaoui}, {Blasco}, {Bonfigli}, {Bono}, {Boudreault}, {Bressan}, {Brown}, {Brunet}, {Bunclark}, {Buonanno}, {Butkevich}, {Carret}, {Carrion}, {Chemin}, {Ch{\'e}reau}, {Corcione}, {Darmigny}, {de Boer}, {de Teodoro}, {de Zeeuw}, {Delle Luche}, {Domingues}, {Dubath}, {Fodor}, {Fr{\'e}zouls}, {Fries}, {Fustes}, {Fyfe}, {Gallardo}, {Gallegos}, {Gardiol}, {Gebran}, {Gomboc}, {G{\'o}mez}, {Grux}, {Gueguen}, {Heyrovsky}, {Hoar}, {Iannicola}, {Isasi Parache}, {Janotto}, {Joliet}, {Jonckheere}, {Keil}, {Kim}, {Klagyivik}, {Klar}, {Knude}, {Kochukhov}, {Kolka}, {Kos}, {Kutka}, {Lainey}, {LeBouquin}, {Liu}, {Loreggia}, {Makarov}, {Marseille}, {Martayan}, {Martinez-Rubi}, {Massart}, {Meynadier}, {Mignot}, {Munari}, {Nguyen}, {Nordlander}, {Ocvirk}, {O'Flaherty}, {Olias Sanz}, {Ortiz}, {Osorio}, {Oszkiewicz}, {Ouzounis}, {Palmer}, {Park}, {Pasquato}, {Peltzer}, {Peralta}, {P{\'e}turaud}, {Pieniluoma}, {Pigozzi}, {Poels}, {Prat}, {Prod'homme}, {Raison}, {Rebordao}, {Risquez}, {Rocca-Volmerange}, {Rosen},
  {Ruiz-Fuertes}, {Russo}, {Sembay}, {Serraller Vizcaino}, {Short}, {Siebert}, {Silva}, {Sinachopoulos}, {Slezak}, {Soffel}, {Sosnowska}, {Strai{\v{z}}ys}, {ter Linden}, {Terrell}, {Theil}, {Tiede}, {Troisi}, {Tsalmantza}, {Tur}, {Vaccari}, {Vachier}, {Valles}, {Van Hamme}, {Veltz}, {Virtanen}, {Wallut}, {Wichmann}, {Wilkinson}, {Ziaeepour}, \& {Zschocke}}]{2016A&A...595A...1G}
{Gaia Collaboration}, {Prusti}, T., {de Bruijne}, J.~H.~J., {et~al.} 2016, \aap, 595, A1

\bibitem[{{Galazutdinov} {et~al.}(2023){Galazutdinov}, {Baluev}, {Valyavin}, {Aitov}, {Gadelshin}, {Valeev}, {Sendzikas}, {Sokov}, {Mitiani}, {Burlakova}, {Yakunin}, {Antonyuk}, {Vlasyuk}, {Romanyuk}, {Rzaev}, {Yushkin}, {Ivanova}, {Tavrov}, \& {Korablev}}]{2023MNRAS.526L.111G}
{Galazutdinov}, G.~A., {Baluev}, R.~V., {Valyavin}, G., {et~al.} 2023, \mnras, 526, L111

\bibitem[{{Gardner} {et~al.}(2006){Gardner}, {Mather}, {Clampin}, {Doyon}, {Greenhouse}, {Hammel}, {Hutchings}, {Jakobsen}, {Lilly}, {Long}, {Lunine}, {McCaughrean}, {Mountain}, {Nella}, {Rieke}, {Rieke}, {Rix}, {Smith}, {Sonneborn}, {Stiavelli}, {Stockman}, {Windhorst}, \& {Wright}}]{2006SSRv..123..485G}
{Gardner}, J.~P., {Mather}, J.~C., {Clampin}, M., {et~al.} 2006, \ssr, 123, 485

\bibitem[{{Gaudi} {et~al.}(2017){Gaudi}, {Stassun}, {Collins}, {Beatty}, {Zhou}, {Latham}, {Bieryla}, {Eastman}, {Siverd}, {Crepp}, {Gonzales}, {Stevens}, {Buchhave}, {Pepper}, {Johnson}, {Colon}, {Jensen}, {Rodriguez}, {Bozza}, {Novati}, {D'Ago}, {Dumont}, {Ellis}, {Gaillard}, {Jang-Condell}, {Kasper}, {Fukui}, {Gregorio}, {Ito}, {Kielkopf}, {Manner}, {Matt}, {Narita}, {Oberst}, {Reed}, {Scarpetta}, {Stephens}, {Yeigh}, {Zambelli}, {Fulton}, {Howard}, {James}, {Penny}, {Bayliss}, {Curtis}, {Depoy}, {Esquerdo}, {Gould}, {Joner}, {Kuhn}, {Labadie-Bartz}, {Lund}, {Marshall}, {McLeod}, {Pogge}, {Relles}, {Stockdale}, {Tan}, {Trueblood}, \& {Trueblood}}]{2017Natur.546..514G}
{Gaudi}, B.~S., {Stassun}, K.~G., {Collins}, K.~A., {et~al.} 2017, \nat, 546, 514

\bibitem[{{Gillon} {et~al.}(2017){Gillon}, {Triaud}, {Demory}, {Jehin}, {Agol}, {Deck}, {Lederer}, {de Wit}, {Burdanov}, {Ingalls}, {Bolmont}, {Leconte}, {Raymond}, {Selsis}, {Turbet}, {Barkaoui}, {Burgasser}, {Burleigh}, {Carey}, {Chaushev}, {Copperwheat}, {Delrez}, {Fernandes}, {Holdsworth}, {Kotze}, {Van Grootel}, {Almleaky}, {Benkhaldoun}, {Magain}, \& {Queloz}}]{2017Natur.542..456G}
{Gillon}, M., {Triaud}, A. H.~M.~J., {Demory}, B.-O., {et~al.} 2017, \nat, 542, 456

\bibitem[{{Gim{\'e}nez} \& {Bastero}(1995)}]{1995Ap&SS.226...99G}
{Gim{\'e}nez}, A. \& {Bastero}, M. 1995, \apss, 226, 99

\bibitem[{{Goldreich} \& {Soter}(1966)}]{1966Icar....5..375G}
{Goldreich}, P. \& {Soter}, S. 1966, \icarus, 5, 375

\bibitem[{{Grimm} {et~al.}(2018){Grimm}, {Demory}, {Gillon}, {Dorn}, {Agol}, {Burdanov}, {Delrez}, {Sestovic}, {Triaud}, {Turbet}, {Bolmont}, {Caldas}, {de Wit}, {Jehin}, {Leconte}, {Raymond}, {Van Grootel}, {Burgasser}, {Carey}, {Fabrycky}, {Heng}, {Hernandez}, {Ingalls}, {Lederer}, {Selsis}, \& {Queloz}}]{2018A&A...613A..68G}
{Grimm}, S.~L., {Demory}, B.-O., {Gillon}, M., {et~al.} 2018, \aap, 613, A68

\bibitem[{{Grunblatt} {et~al.}(2022){Grunblatt}, {Saunders}, {Sun}, {Chontos}, {Soares-Furtado}, {Eisner}, {Pereira}, {Komacek}, {Huber}, {Collins}, {Wang}, {Stockdale}, {Quinn}, {Tronsgaard}, {Zhou}, {Nowak}, {Deeg}, {Ciardi}, {Boyle}, {Rice}, {Dai}, {Blunt}, {Van Zandt}, {Beard}, {Akana Murphy}, {Dalba}, {Lubin}, {Polanski}, {Brinkman}, {Howard}, {Buchhave}, {Angus}, {Ricker}, {Jenkins}, {Wohler}, {Goeke}, {Levine}, {Colon}, {Huang}, {Kunimoto}, {Shporer}, {Latham}, {Seager}, {Vanderspek}, \& {Winn}}]{2022AJ....163..120G}
{Grunblatt}, S.~K., {Saunders}, N., {Sun}, M., {et~al.} 2022, \aj, 163, 120

\bibitem[{{Hadden} \& {Lithwick}(2014)}]{2014ApJ...787...80H}
{Hadden}, S. \& {Lithwick}, Y. 2014, \apj, 787, 80

\bibitem[{{Hadden} \& {Lithwick}(2017)}]{2017AJ....154....5H}
{Hadden}, S. \& {Lithwick}, Y. 2017, \aj, 154, 5

\bibitem[{{Harre} \& {Smith}(2023)}]{2023Univ....9..506H}
{Harre}, J.-V. \& {Smith}, A. M.~S. 2023, Universe, 9, 506

\bibitem[{{Harre} {et~al.}(2023){Harre}, {Smith}, {Barros}, {Bou{\'e}}, {Csizmadia}, {Ehrenreich}, {Flor{\'e}n}, {Fortier}, {Maxted}, {Hooton}, {Akinsanmi}, {Serrano}, {Ros{\'a}rio}, {Demory}, {Jones}, {Laskar}, {Adibekyan}, {Alibert}, {Alonso}, {Anderson}, {Anglada}, {Asquier}, {B{\'a}rczy}, {Barrado y Navascues}, {Baumjohann}, {Beck}, {Beck}, {Benz}, {Billot}, {Biondi}, {Bonfanti}, {Bonfils}, {Brandeker}, {Broeg}, {Cabrera}, {Cessa}, {Charnoz}, {Collier Cameron}, {Davies}, {Deleuil}, {Delrez}, {Demangeon}, {Erikson}, {Fossati}, {Fridlund}, {Gandolfi}, {Gillon}, {G{\"u}del}, {Hellier}, {Heng}, {Hoyer}, {Isaak}, {Kiss}, {Lecavelier des Etangs}, {Lendl}, {Lovis}, {Luntzer}, {Magrin}, {Nascimbeni}, {Olofsson}, {Ottensamer}, {Pagano}, {Pall{\'e}}, {Persson}, {Peter}, {Piotto}, {Pollacco}, {Queloz}, {Ragazzoni}, {Rando}, {Rauer}, {Ribas}, {Ricker}, {Salmon}, {Santos}, {Scandariato}, {Seager}, {S{\'e}gransan}, {Simon}, {Sousa}, {Steller}, {Szab{\'o}}, {Thomas}, {Udry}, {Ulmer}, {Van Grootel}, {Walton}, {Wilson},
  {Winn}, \& {Wohler}}]{2023A&A...669A.124H}
{Harre}, J.~V., {Smith}, A.~M.~S., {Barros}, S.~C.~C., {et~al.} 2023, \aap, 669, A124

\bibitem[{Harris {et~al.}(2020)Harris, Millman, van~der Walt, Gommers, Virtanen, Cournapeau, Wieser, Taylor, Berg, Smith, Kern, Picus, Hoyer, van Kerkwijk, Brett, Haldane, del R{\'{i}}o, Wiebe, Peterson, G{\'{e}}rard-Marchant, Sheppard, Reddy, Weckesser, Abbasi, Gohlke, \& Oliphant}]{harris2020array}
Harris, C.~R., Millman, K.~J., van~der Walt, S.~J., {et~al.} 2020, Nature, 585, 357

\bibitem[{{H{\'e}brard} {et~al.}(2020){H{\'e}brard}, {D{\'\i}az}, {Correia}, {Collier Cameron}, {Laskar}, {Pollacco}, {Almenara}, {Anderson}, {Barros}, {Boisse}, {Bonomo}, {Bouchy}, {Bou{\'e}}, {Boumis}, {Brown}, {Dalal}, {Deleuil}, {Demangeon}, {Doyle}, {Haswell}, {Hellier}, {Osborn}, {Kiefer}, {Kolb}, {Lam}, {Lecavelier des {\'E}tangs}, {Lopez}, {Martin-Lagarde}, {Maxted}, {McCormac}, {Nielsen}, {Pall{\'e}}, {Prieto-Arranz}, {Queloz}, {Santerne}, {Smalley}, {Turner}, {Udry}, {Verilhac}, {West}, {Wheatley}, \& {Wilson}}]{2020A&A...640A..32H}
{H{\'e}brard}, G., {D{\'\i}az}, R.~F., {Correia}, A.~C.~M., {et~al.} 2020, \aap, 640, A32

\bibitem[{{Hellier} {et~al.}(2017){Hellier}, {Anderson}, {Collier Cameron}, {Delrez}, {Gillon}, {Jehin}, {Lendl}, {Maxted}, {Neveu-VanMalle}, {Pepe}, {Pollacco}, {Queloz}, {S{\'e}gransan}, {Smalley}, {Southworth}, {Triaud}, {Udry}, {Wagg}, \& {West}}]{2017MNRAS.465.3693H}
{Hellier}, C., {Anderson}, D.~R., {Collier Cameron}, A., {et~al.} 2017, \mnras, 465, 3693

\bibitem[{{Hellier} {et~al.}(2012){Hellier}, {Anderson}, {Collier Cameron}, {Doyle}, {Fumel}, {Gillon}, {Jehin}, {Lendl}, {Maxted}, {Pepe}, {Pollacco}, {Queloz}, {S{\'e}gransan}, {Smalley}, {Smith}, {Southworth}, {Triaud}, {Udry}, \& {West}}]{2012MNRAS.426..739H}
{Hellier}, C., {Anderson}, D.~R., {Collier Cameron}, A., {et~al.} 2012, \mnras, 426, 739

\bibitem[{{Hippke} \& {Heller}(2019)}]{2019A&A...623A..39H}
{Hippke}, M. \& {Heller}, R. 2019, \aap, 623, A39

\bibitem[{{Holman} {et~al.}(2010){Holman}, {Fabrycky}, {Ragozzine}, {Ford}, {Steffen}, {Welsh}, {Lissauer}, {Latham}, {Marcy}, {Walkowicz}, {Batalha}, {Jenkins}, {Rowe}, {Cochran}, {Fressin}, {Torres}, {Buchhave}, {Sasselov}, {Borucki}, {Koch}, {Basri}, {Brown}, {Caldwell}, {Charbonneau}, {Dunham}, {Gautier}, {Geary}, {Gilliland}, {Haas}, {Howell}, {Ciardi}, {Endl}, {Fischer}, {F{\"u}r{\'e}sz}, {Hartman}, {Isaacson}, {Johnson}, {MacQueen}, {Moorhead}, {Morehead}, \& {Orosz}}]{2010Sci...330...51H}
{Holman}, M.~J., {Fabrycky}, D.~C., {Ragozzine}, D., {et~al.} 2010, Science, 330, 51

\bibitem[{{Hord} {et~al.}(2022){Hord}, {Col{\'o}n}, {Berger}, {Kostov}, {Silverstein}, {Stassun}, {Lissauer}, {Collins}, {Schwarz}, {Sefako}, {Ziegler}, {Brice{\~n}o}, {Law}, {Mann}, {Ricker}, {Latham}, {Seager}, {Winn}, {Jenkins}, {Bouma}, {Falk}, {Torres}, {Twicken}, \& {Vanderburg}}]{2022AJ....164...13H}
{Hord}, B.~J., {Col{\'o}n}, K.~D., {Berger}, T.~A., {et~al.} 2022, \aj, 164, 13

\bibitem[{{Hoyer} {et~al.}(2020){Hoyer}, {Guterman}, {Demangeon}, {Sousa}, {Deleuil}, {Meunier}, \& {Benz}}]{2020A&A...635A..24H}
{Hoyer}, S., {Guterman}, P., {Demangeon}, O., {et~al.} 2020, \aap, 635, A24

\bibitem[{{Huang} {et~al.}(2020){Huang}, {Quinn}, {Vanderburg}, {Becker}, {Rodriguez}, {Pozuelos}, {Gandolfi}, {Zhou}, {Mann}, {Collins}, {Crossfield}, {Barkaoui}, {Collins}, {Fridlund}, {Gillon}, {Gonzales}, {G{\"u}nther}, {Henry}, {Howell}, {James}, {Jao}, {Jehin}, {Jensen}, {Kane}, {Lissauer}, {Matthews}, {Matson}, {Paredes}, {Schlieder}, {Stassun}, {Shporer}, {Sha}, {Tan}, {Georgieva}, {Mathur}, {Palle}, {Persson}, {Van Eylen}, {Ricker}, {Vanderspek}, {Latham}, {Winn}, {Seager}, {Jenkins}, {Burke}, {Goeke}, {Rinehart}, {Rose}, {Ting}, {Torres}, \& {Wong}}]{2020ApJ...892L...7H}
{Huang}, C.~X., {Quinn}, S.~N., {Vanderburg}, A., {et~al.} 2020, \apjl, 892, L7

\bibitem[{Hunter(2007)}]{Hunter:2007}
Hunter, J.~D. 2007, Computing in Science \& Engineering, 9, 90

\bibitem[{{Husnoo} {et~al.}(2012){Husnoo}, {Pont}, {Mazeh}, {Fabrycky}, {H{\'e}brard}, {Bouchy}, \& {Shporer}}]{2012MNRAS.422.3151H}
{Husnoo}, N., {Pont}, F., {Mazeh}, T., {et~al.} 2012, \mnras, 422, 3151

\bibitem[{{Husser} {et~al.}(2013){Husser}, {Wende-von Berg}, {Dreizler}, {Homeier}, {Reiners}, {Barman}, \& {Hauschildt}}]{2013A&A...553A...6H}
{Husser}, T.~O., {Wende-von Berg}, S., {Dreizler}, S., {et~al.} 2013, \aap, 553, A6

\bibitem[{{Hut}(1981)}]{1981A&A....99..126H}
{Hut}, P. 1981, \aap, 99, 126

\bibitem[{{Ioannidis} {et~al.}(2016){Ioannidis}, {Huber}, \& {Schmitt}}]{2016A&A...585A..72I}
{Ioannidis}, P., {Huber}, K.~F., \& {Schmitt}, J.~H.~M.~M. 2016, \aap, 585, A72

\bibitem[{{Ivshina} \& {Winn}(2022)}]{2022ApJS..259...62I}
{Ivshina}, E.~S. \& {Winn}, J.~N. 2022, \apjs, 259, 62

\bibitem[{{Jackson} {et~al.}(2008){Jackson}, {Greenberg}, \& {Barnes}}]{2008ApJ...678.1396J}
{Jackson}, B., {Greenberg}, R., \& {Barnes}, R. 2008, \apj, 678, 1396

\bibitem[{{Jones} {et~al.}(2022){Jones}, {Morris}, {Demory}, {Heng}, {Hooton}, {Billot}, {Ehrenreich}, {Hoyer}, {Simon}, {Lendl}, {Demangeon}, {Sousa}, {Bonfanti}, {Wilson}, {Salmon}, {Csizmadia}, {Parviainen}, {Bruno}, {Alibert}, {Alonso}, {Anglada}, {B{\'a}rczy}, {Barrado}, {Barros}, {Baumjohann}, {Beck}, {Beck}, {Benz}, {Bonfils}, {Brandeker}, {Broeg}, {Cabrera}, {Charnoz}, {Collier Cameron}, {Davies}, {Deleuil}, {Deline}, {Delrez}, {Erikson}, {Fortier}, {Fossati}, {Fridlund}, {Gandolfi}, {Gillon}, {G{\"u}del}, {Isaak}, {Kiss}, {Laskar}, {Lecavelier des Etangs}, {Lovis}, {Magrin}, {Maxted}, {Nascimbeni}, {Olofsson}, {Ottensamer}, {Pagano}, {Pall{\'e}}, {Peter}, {Piotto}, {Pollacco}, {Queloz}, {Ragazzoni}, {Rando}, {Ratti}, {Rauer}, {Reimers}, {Ribas}, {Santos}, {Scandariato}, {S{\'e}gransan}, {Smith}, {Steller}, {Szab{\'o}}, {Thomas}, {Udry}, {Van Grootel}, {Walter}, {Walton}, \& {Wang Jungo}}]{2022A&A...666A.118J}
{Jones}, K., {Morris}, B.~M., {Demory}, B.~O., {et~al.} 2022, \aap, 666, A118

\bibitem[{{K{\'a}lm{\'a}n} {et~al.}(2024){K{\'a}lm{\'a}n}, {Derekas}, {Csizmadia}, {P{\'a}l}, {Szab{\'o}}, {Smith}, {Nagy}, {Heged{\H{u}}s}, {Mitnyan}, {Szigeti}, \& {Szab{\'o}}}]{2024arXiv240319468K}
{K{\'a}lm{\'a}n}, S., {Derekas}, A., {Csizmadia}, S., {et~al.} 2024, arXiv e-prints, arXiv:2403.19468

\bibitem[{Kipping(2013)}]{Kipping2013b}
Kipping, D.~M. 2013, Monthly Notices of the Royal Astronomical Society, 435, 2152

\bibitem[{{Korth} {et~al.}(2024){Korth}, {Chaturvedi}, {Parviainen}, {Carleo}, {Endl}, {Guenther}, {Nowak}, {Persson}, {MacQueen}, {Mustill}, {Cabrera}, {Cochran}, {Lillo-Box}, {Hobbs}, {Murgas}, {Greklek-McKeon}, {Kellermann}, {H{\'e}brard}, {Fukui}, {Pall{\'e}}, {Jenkins}, {Twicken}, {Collins}, {Quinn}, {{\v{S}}ubjak}, {Beck}, {Gandolfi}, {Mathur}, {Deeg}, {Latham}, {Albrecht}, {Barrado}, {Boisse}, {Bouy}, {Delfosse}, {Demangeon}, {Garc{\'\i}a}, {Hatzes}, {Heidari}, {Ikuta}, {Kab{\'a}th}, {Knutson}, {Livingston}, {Martioli}, {Morales-Calder{\'o}n}, {Morello}, {Narita}, {Orell-Miquel}, {Osborne}, {Palakkatharappil}, {Pinter}, {Redfield}, {Relles}, {Schwarz}, {Seager}, {Shporer}, {Skarka}, {Srdoc}, {Stangret}, {Thomas}, {Van Eylen}, {Watanabe}, \& {Winn}}]{2024ApJ...971L..28K}
{Korth}, J., {Chaturvedi}, P., {Parviainen}, H., {et~al.} 2024, \apjl, 971, L28

\bibitem[{{Korth} {et~al.}(2023){Korth}, {Gandolfi}, {{\v{S}}ubjak}, {Howard}, {Ataiee}, {Collins}, {Quinn}, {Mustill}, {Guillot}, {Lodieu}, {Smith}, {Esposito}, {Rodler}, {Muresan}, {Abe}, {Albrecht}, {Alqasim}, {Barkaoui}, {Beck}, {Burke}, {Butler}, {Conti}, {Collins}, {Crane}, {Dai}, {Deeg}, {Evans}, {Grziwa}, {Hatzes}, {Hirano}, {Horne}, {Huang}, {Jenkins}, {Kab{\'a}th}, {Kielkopf}, {Knudstrup}, {Latham}, {Livingston}, {Luque}, {Mathur}, {Murgas}, {Osborne}, {Palle}, {Persson}, {Rodriguez}, {Rose}, {Rowden}, {Schwarz}, {Seager}, {Serrano}, {Sha}, {Shectman}, {Shporer}, {Srdoc}, {Stockdale}, {Tan}, {Teske}, {Van Eylen}, {Vanderburg}, {Vanderspek}, {Wang}, \& {Winn}}]{2023A&A...675A.115K}
{Korth}, J., {Gandolfi}, D., {{\v{S}}ubjak}, J., {et~al.} 2023, \aap, 675, A115

\bibitem[{{Kraft}(1967)}]{1967ApJ...150..551K}
{Kraft}, R.~P. 1967, \apj, 150, 551

\bibitem[{{Kurucz}(1970)}]{1970SAOSR.309.....K}
{Kurucz}, R.~L. 1970, SAO Special Report, 309

\bibitem[{{Lanza} {et~al.}(2011){Lanza}, {Damiani}, \& {Gandolfi}}]{2011A&A...529A..50L}
{Lanza}, A.~F., {Damiani}, C., \& {Gandolfi}, D. 2011, \aap, 529, A50

\bibitem[{{Laskar} {et~al.}(2012){Laskar}, {Bou{\'e}}, \& {Correia}}]{2012A&A...538A.105L}
{Laskar}, J., {Bou{\'e}}, G., \& {Correia}, A.~C.~M. 2012, \aap, 538, A105

\bibitem[{{Lee} \& {Peale}(2003)}]{2003ApJ...592.1201L}
{Lee}, M.~H. \& {Peale}, S.~J. 2003, \apj, 592, 1201

\bibitem[{{Leleu} {et~al.}(2021){Leleu}, {Alibert}, {Hara}, {Hooton}, {Wilson}, {Robutel}, {Delisle}, {Laskar}, {Hoyer}, {Lovis}, {Bryant}, {Ducrot}, {Cabrera}, {Delrez}, {Acton}, {Adibekyan}, {Allart}, {Allende Prieto}, {Alonso}, {Alves}, {Anderson}, {Angerhausen}, {Anglada Escud{\'e}}, {Asquier}, {Barrado}, {Barros}, {Baumjohann}, {Bayliss}, {Beck}, {Beck}, {Bekkelien}, {Benz}, {Billot}, {Bonfanti}, {Bonfils}, {Bouchy}, {Bourrier}, {Bou{\'e}}, {Brandeker}, {Broeg}, {Buder}, {Burdanov}, {Burleigh}, {B{\'a}rczy}, {Cameron}, {Chamberlain}, {Charnoz}, {Cooke}, {Corral Van Damme}, {Correia}, {Cristiani}, {Damasso}, {Davies}, {Deleuil}, {Demangeon}, {Demory}, {Di Marcantonio}, {Di Persio}, {Dumusque}, {Ehrenreich}, {Erikson}, {Figueira}, {Fortier}, {Fossati}, {Fridlund}, {Futyan}, {Gandolfi}, {Garc{\'\i}a Mu{\~n}oz}, {Garcia}, {Gill}, {Gillen}, {Gillon}, {Goad}, {Gonz{\'a}lez Hern{\'a}ndez}, {Guedel}, {G{\"u}nther}, {Haldemann}, {Henderson}, {Heng}, {Hogan}, {Isaak}, {Jehin}, {Jenkins}, {Jord{\'a}n}, {Kiss},
  {Kristiansen}, {Lam}, {Lavie}, {Lecavelier des Etangs}, {Lendl}, {Lillo-Box}, {Lo Curto}, {Magrin}, {Martins}, {Maxted}, {McCormac}, {Mehner}, {Micela}, {Molaro}, {Moyano}, {Murray}, {Nascimbeni}, {Nunes}, {Olofsson}, {Osborn}, {Oshagh}, {Ottensamer}, {Pagano}, {Pall{\'e}}, {Pedersen}, {Pepe}, {Persson}, {Peter}, {Piotto}, {Polenta}, {Pollacco}, {Poretti}, {Pozuelos}, {Queloz}, {Ragazzoni}, {Rando}, {Ratti}, {Rauer}, {Raynard}, {Rebolo}, {Reimers}, {Ribas}, {Santos}, {Scandariato}, {Schneider}, {Sebastian}, {Sestovic}, {Simon}, {Smith}, {Sousa}, {Sozzetti}, {Steller}, {Su{\'a}rez Mascare{\~n}o}, {Szab{\'o}}, {S{\'e}gransan}, {Thomas}, {Thompson}, {Tilbrook}, {Triaud}, {Turner}, {Udry}, {Van Grootel}, {Venus}, {Verrecchia}, {Vines}, {Walton}, {West}, {Wheatley}, {Wolter}, \& {Zapatero Osorio}}]{2021A&A...649A..26L}
{Leleu}, A., {Alibert}, Y., {Hara}, N.~C., {et~al.} 2021, \aap, 649, A26

\bibitem[{{Lendl} {et~al.}(2020){Lendl}, {Csizmadia}, {Deline}, {Fossati}, {Kitzmann}, {Heng}, {Hoyer}, {Salmon}, {Benz}, {Broeg}, {Ehrenreich}, {Fortier}, {Queloz}, {Bonfanti}, {Brandeker}, {Collier Cameron}, {Delrez}, {Garcia Mu{\~n}oz}, {Hooton}, {Maxted}, {Morris}, {Van Grootel}, {Wilson}, {Alibert}, {Alonso}, {Asquier}, {Bandy}, {B{\'a}rczy}, {Barrado}, {Barros}, {Baumjohann}, {Beck}, {Beck}, {Bekkelien}, {Bergomi}, {Billot}, {Biondi}, {Bonfils}, {Bourrier}, {Busch}, {Cabrera}, {Cessa}, {Charnoz}, {Chazelas}, {Corral Van Damme}, {Davies}, {Deleuil}, {Demangeon}, {Demory}, {Erikson}, {Farinato}, {Fridlund}, {Futyan}, {Gandolfi}, {Gillon}, {Guterman}, {Hasiba}, {Hernandez}, {Isaak}, {Kiss}, {Kuntzer}, {Lecavelier des Etangs}, {L{\"u}ftinger}, {Laskar}, {Lovis}, {Magrin}, {Malvasio}, {Marafatto}, {Michaelis}, {Munari}, {Nascimbeni}, {Olofsson}, {Ottacher}, {Ottensamer}, {Pagano}, {Pall{\'e}}, {Peter}, {Piazza}, {Piotto}, {Pollacco}, {Ratti}, {Rauer}, {Ragazzoni}, {Rando}, {Ribas}, {Rieder}, {Rohlfs},
  {Safa}, {Santos}, {Scandariato}, {S{\'e}gransan}, {Simon}, {Singh}, {Smith}, {Sordet}, {Sousa}, {Steller}, {Szab{\'o}}, {Thomas}, {Tschentscher}, {Udry}, {Viotto}, {Walter}, {Walton}, {Wildi}, \& {Wolter}}]{2020A&A...643A..94L}
{Lendl}, M., {Csizmadia}, S., {Deline}, A., {et~al.} 2020, \aap, 643, A94

\bibitem[{{Lithwick} {et~al.}(2012){Lithwick}, {Xie}, \& {Wu}}]{2012ApJ...761..122L}
{Lithwick}, Y., {Xie}, J., \& {Wu}, Y. 2012, \apj, 761, 122

\bibitem[{{MacDonald} {et~al.}(2016){MacDonald}, {Ragozzine}, {Fabrycky}, {Ford}, {Holman}, {Isaacson}, {Lissauer}, {Lopez}, {Mazeh}, {Rogers}, {Rowe}, {Steffen}, \& {Torres}}]{2016AJ....152..105M}
{MacDonald}, M.~G., {Ragozzine}, D., {Fabrycky}, D.~C., {et~al.} 2016, \aj, 152, 105

\bibitem[{{Maciejewski} {et~al.}(2023){Maciejewski}, {Golonka}, {{\L}oboda}, {Ohlert}, {Fern{\'a}ndez}, \& {Aceituno}}]{2023MNRAS.525L..43M}
{Maciejewski}, G., {Golonka}, J., {{\L}oboda}, W., {et~al.} 2023, \mnras, 525, L43

\bibitem[{{Maciejewski} {et~al.}(2020){Maciejewski}, {Knutson}, {Howard}, {Isaacson}, {Fern{\'a}ndez-Laj{\'u}s}, {DiSisto}, \& {Migaszewski}}]{2020AcA....70....1M}
{Maciejewski}, G., {Knutson}, H.~A., {Howard}, A.~W., {et~al.} 2020, \actaa, 70, 1

\bibitem[{{Malavolta} {et~al.}(2017){Malavolta}, {Borsato}, {Granata}, {Piotto}, {Lopez}, {Vanderburg}, {Figueira}, {Mortier}, {Nascimbeni}, {Affer}, {Bonomo}, {Bouchy}, {Buchhave}, {Charbonneau}, {Collier Cameron}, {Cosentino}, {Dressing}, {Dumusque}, {Fiorenzano}, {Harutyunyan}, {Haywood}, {Johnson}, {Latham}, {Lopez-Morales}, {Lovis}, {Mayor}, {Micela}, {Molinari}, {Motalebi}, {Pepe}, {Phillips}, {Pollacco}, {Queloz}, {Rice}, {Sasselov}, {S{\'e}gransan}, {Sozzetti}, {Udry}, \& {Watson}}]{2017AJ....153..224M}
{Malavolta}, L., {Borsato}, L., {Granata}, V., {et~al.} 2017, \aj, 153, 224

\bibitem[{{Mantovan} {et~al.}(2024){Mantovan}, {Malavolta}, {Desidera}, {Zingales}, {Borsato}, {Piotto}, {Maggio}, {Locci}, {Polychroni}, {Turrini}, {Baratella}, {Biazzo}, {Nardiello}, {Stassun}, {Nascimbeni}, {Benatti}, {Anna John}, {Watkins}, {Bieryla}, {Lissauer}, {Twicken}, {Lanza}, {Winn}, {Messina}, {Montalto}, {Sozzetti}, {Boffin}, {Cheryasov}, {Strakhov}, {Murgas}, {D'Arpa}, {Barkaoui}, {Benni}, {Bignamini}, {Bonomo}, {Borsa}, {Cabona}, {Cameron}, {Claudi}, {Cochran}, {Collins}, {Damasso}, {Dong}, {Endl}, {Fukui}, {F{\H{u}}r{\'e}sz}, {Gandolfi}, {Ghedina}, {Jenkins}, {Kab{\'a}th}, {Latham}, {Lorenzi}, {Luque}, {Maldonado}, {McLeod}, {Molinaro}, {Narita}, {Nowak}, {Orell-Miquel}, {Pall{\'e}}, {Parviainen}, {Pedani}, {Quinn}, {Relles}, {Rowden}, {Scandariato}, {Schwarz}, {Seager}, {Shporer}, {Vanderburg}, \& {Wilson}}]{2024A&A...682A.129M}
{Mantovan}, G., {Malavolta}, L., {Desidera}, S., {et~al.} 2024, \aap, 682, A129

\bibitem[{{Mardling}(2007)}]{2007MNRAS.382.1768M}
{Mardling}, R.~A. 2007, \mnras, 382, 1768

\bibitem[{{Mardling}(2010)}]{2010MNRAS.407.1048M}
{Mardling}, R.~A. 2010, \mnras, 407, 1048

\bibitem[{{Maxted} {et~al.}(2022){Maxted}, {Ehrenreich}, {Wilson}, {Alibert}, {Cameron}, {Hoyer}, {Sousa}, {Olofsson}, {Bekkelien}, {Deline}, {Delrez}, {Bonfanti}, {Borsato}, {Alonso}, {Anglada Escud{\'e}}, {Barrado}, {Barros}, {Baumjohann}, {Beck}, {Beck}, {Benz}, {Billot}, {Biondi}, {Bonfils}, {Brandeker}, {Broeg}, {B{\'a}rczy}, {Cabrera}, {Charnoz}, {Corral Van Damme}, {Csizmadia}, {Davies}, {Deleuil}, {Demangeon}, {Demory}, {Erikson}, {Flor{\'e}n}, {Fortier}, {Fossati}, {Fridlund}, {Futyan}, {Gandolfi}, {Gillon}, {Guedel}, {Guterman}, {Heng}, {Isaak}, {Kiss}, {Laskar}, {Lecavelier des Etangs}, {Lendl}, {Lovis}, {Magrin}, {Nascimbeni}, {Ottensamer}, {Pagano}, {Pall{\'e}}, {Peter}, {Piotto}, {Pollacco}, {Pozuelos}, {Queloz}, {Ragazzoni}, {Rando}, {Rauer}, {Reimers}, {Ribas}, {Salmon}, {Santos}, {Scandariato}, {Simon}, {Smith}, {Steller}, {Swayne}, {Szab{\'o}}, {S{\'e}gransan}, {Thomas}, {Udry}, {Van Grootel}, \& {Walton}}]{2022MNRAS.514...77M}
{Maxted}, P.~F.~L., {Ehrenreich}, D., {Wilson}, T.~G., {et~al.} 2022, \mnras, 514, 77

\bibitem[{{Meibom} \& {Mathieu}(2005)}]{2005ApJ...620..970M}
{Meibom}, S. \& {Mathieu}, R.~D. 2005, \apj, 620, 970

\bibitem[{{Mikal-Evans} {et~al.}(2023){Mikal-Evans}, {Sing}, {Dong}, {Foreman-Mackey}, {Kataria}, {Barstow}, {Goyal}, {Lewis}, {Lothringer}, {Mayne}, {Wakeford}, {Christie}, \& {Rustamkulov}}]{2023ApJ...943L..17M}
{Mikal-Evans}, T., {Sing}, D.~K., {Dong}, J., {et~al.} 2023, \apjl, 943, L17

\bibitem[{{Morris} {et~al.}(2021{\natexlab{a}}){Morris}, {Delrez}, {Brandeker}, {Cameron}, {Simon}, {Futyan}, {Olofsson}, {Hoyer}, {Fortier}, {Demory}, {Lendl}, {Wilson}, {Oshagh}, {Heng}, {Ehrenreich}, {Sulis}, {Alibert}, {Alonso}, {Anglada Escud{\'e}}, {Barrado}, {Barros}, {Baumjohann}, {Beck}, {Beck}, {Bekkelien}, {Benz}, {Bergomi}, {Billot}, {Bonfils}, {Bourrier}, {Broeg}, {B{\'a}rczy}, {Cabrera}, {Charnoz}, {Davies}, {De Miguel Ferreras}, {Deleuil}, {Deline}, {Demangeon}, {Erikson}, {Floren}, {Fossati}, {Fridlund}, {Gandolfi}, {Garc{\'\i}a Mu{\~n}oz}, {Gillon}, {Guedel}, {Guterman}, {Isaak}, {Kiss}, {Laskar}, {Lecavelier des Etangs}, {Lieder}, {Lovis}, {Magrin}, {Maxted}, {Nascimbeni}, {Ottensamer}, {Pagano}, {Pall{\'e}}, {Peter}, {Piotto}, {Pizarro Rubio}, {Pollacco}, {Pozuelos}, {Queloz}, {Ragazzoni}, {Rando}, {Rauer}, {Ribas}, {Santos}, {Scandariato}, {Smith}, {Sousa}, {Steller}, {Szab{\'o}}, {S{\'e}gransan}, {Thomas}, {Udry}, {Ulmer}, {Van Grootel}, \& {Walton}}]{2021A&A...653A.173M}
{Morris}, B.~M., {Delrez}, L., {Brandeker}, A., {et~al.} 2021{\natexlab{a}}, \aap, 653, A173

\bibitem[{{Morris} {et~al.}(2021{\natexlab{b}}){Morris}, {Heng}, {Brandeker}, {Swan}, \& {Lendl}}]{2021A&A...651L..12M}
{Morris}, B.~M., {Heng}, K., {Brandeker}, A., {Swan}, A., \& {Lendl}, M. 2021{\natexlab{b}}, \aap, 651, L12

\bibitem[{{Nascimbeni} {et~al.}(2023){Nascimbeni}, {Borsato}, {Zingales}, {Piotto}, {Pagano}, {Beck}, {Broeg}, {Ehrenreich}, {Hoyer}, {Majidi}, {Granata}, {Sousa}, {Wilson}, {Van Grootel}, {Bonfanti}, {Salmon}, {Mustill}, {Delrez}, {Alibert}, {Alonso}, {Anglada}, {B{\'a}rczy}, {Barrado}, {Barros}, {Baumjohann}, {Beck}, {Benz}, {Bergomi}, {Billot}, {Bonfils}, {Brandeker}, {Cabrera}, {Charnoz}, {Collier Cameron}, {Csizmadia}, {Cubillos}, {Davies}, {Deleuil}, {Deline}, {Demangeon}, {Demory}, {Erikson}, {Fortier}, {Fossati}, {Fridlund}, {Gandolfi}, {Gillon}, {G{\"u}del}, {Isaak}, {Kiss}, {Laskar}, {Lecavelier des Etangs}, {Lendl}, {Lovis}, {Luque}, {Magrin}, {Maxted}, {Mordasini}, {Olofsson}, {Ottensamer}, {Pall{\'e}}, {Peter}, {Piazza}, {Pollacco}, {Queloz}, {Ragazzoni}, {Rando}, {Ratti}, {Rauer}, {Ribas}, {Santos}, {Scandariato}, {S{\'e}gransan}, {Simon}, {Smith}, {Steinberger}, {Steller}, {Szab{\'o}}, {Thomas}, {Udry}, {Venturini}, {Walton}, \& {Wolter}}]{2023A&A...673A..42N}
{Nascimbeni}, V., {Borsato}, L., {Zingales}, T., {et~al.} 2023, \aap, 673, A42

\bibitem[{{Nespral} {et~al.}(2017){Nespral}, {Gandolfi}, {Deeg}, {Borsato}, {Fridlund}, {Barrag{\'a}n}, {Alonso}, {Grziwa}, {Korth}, {Albrecht}, {Cabrera}, {Csizmadia}, {Nowak}, {Kuutma}, {Saario}, {Eigm{\"u}ller}, {Erikson}, {Guenther}, {Hatzes}, {Monta{\~n}{\'e}s Rodr{\'\i}guez}, {Palle}, {P{\"a}tzold}, {Prieto-Arranz}, {Rauer}, \& {Sebastian}}]{2017A&A...601A.128N}
{Nespral}, D., {Gandolfi}, D., {Deeg}, H.~J., {et~al.} 2017, \aap, 601, A128

\bibitem[{{Neveu-VanMalle} {et~al.}(2016){Neveu-VanMalle}, {Queloz}, {Anderson}, {Brown}, {Collier Cameron}, {Delrez}, {D{\'\i}az}, {Gillon}, {Hellier}, {Jehin}, {Lister}, {Pepe}, {Rojo}, {S{\'e}gransan}, {Triaud}, {Turner}, \& {Udry}}]{2016A&A...586A..93N}
{Neveu-VanMalle}, M., {Queloz}, D., {Anderson}, D.~R., {et~al.} 2016, \aap, 586, A93

\bibitem[{{Ofir} {et~al.}(2018){Ofir}, {Xie}, {Jiang}, {Sari}, \& {Aharonson}}]{2018ApJS..234....9O}
{Ofir}, A., {Xie}, J.-W., {Jiang}, C.-F., {Sari}, R., \& {Aharonson}, O. 2018, \apjs, 234, 9

\bibitem[{{Ogilvie}(2014)}]{2014ARA&A..52..171O}
{Ogilvie}, G.~I. 2014, \araa, 52, 171

\bibitem[{{Ogilvie} \& {Lin}(2007)}]{2007ApJ...661.1180O}
{Ogilvie}, G.~I. \& {Lin}, D.~N.~C. 2007, \apj, 661, 1180

\bibitem[{Parviainen(2015)}]{Parviainen2015}
Parviainen, H. 2015, Monthly Notices of the Royal Astronomical Society, 450, 3233

\bibitem[{{Parviainen}(2016)}]{2016zndo.....45602P}
{Parviainen}, H. 2016, {PyDE: v1.5}

\bibitem[{Parviainen(2020)}]{Parviainen2020b}
Parviainen, H. 2020, Monthly Notices of the Royal Astronomical Society, 499, 1633

\bibitem[{Parviainen \& Korth(2020)}]{Parviainen2020a}
Parviainen, H. \& Korth, J. 2020, Monthly Notices of the Royal Astronomical Society [\eprint[arxiv]{2009.09965}]

\bibitem[{{Patra} {et~al.}(2020){Patra}, {Winn}, {Holman}, {Gillon}, {Burdanov}, {Jehin}, {Delrez}, {Pozuelos}, {Barkaoui}, {Benkhaldoun}, {Narita}, {Fukui}, {Kusakabe}, {Kawauchi}, {Terada}, {Bouma}, {Weinberg}, \& {Broome}}]{2020AJ....159..150P}
{Patra}, K.~C., {Winn}, J.~N., {Holman}, M.~J., {et~al.} 2020, \aj, 159, 150

\bibitem[{{Patra} {et~al.}(2017){Patra}, {Winn}, {Holman}, {Yu}, {Deming}, \& {Dai}}]{2017AJ....154....4P}
{Patra}, K.~C., {Winn}, J.~N., {Holman}, M.~J., {et~al.} 2017, \aj, 154, 4

\bibitem[{{Penev} {et~al.}(2018){Penev}, {Bouma}, {Winn}, \& {Hartman}}]{2018AJ....155..165P}
{Penev}, K., {Bouma}, L.~G., {Winn}, J.~N., \& {Hartman}, J.~D. 2018, \aj, 155, 165

\bibitem[{{Penev} {et~al.}(2012){Penev}, {Jackson}, {Spada}, \& {Thom}}]{2012ApJ...751...96P}
{Penev}, K., {Jackson}, B., {Spada}, F., \& {Thom}, N. 2012, \apj, 751, 96

\bibitem[{{Penev} \& {Sasselov}(2011)}]{2011ApJ...731...67P}
{Penev}, K. \& {Sasselov}, D. 2011, \apj, 731, 67

\bibitem[{{Pollacco} {et~al.}(2006){Pollacco}, {Skillen}, {Collier Cameron}, {Christian}, {Hellier}, {Irwin}, {Lister}, {Street}, {West}, {Anderson}, {Clarkson}, {Deeg}, {Enoch}, {Evans}, {Fitzsimmons}, {Haswell}, {Hodgkin}, {Horne}, {Kane}, {Keenan}, {Maxted}, {Norton}, {Osborne}, {Parley}, {Ryans}, {Smalley}, {Wheatley}, \& {Wilson}}]{2006PASP..118.1407P}
{Pollacco}, D.~L., {Skillen}, I., {Collier Cameron}, A., {et~al.} 2006, \pasp, 118, 1407

\bibitem[{{Poon} {et~al.}(2021){Poon}, {Nelson}, \& {Coleman}}]{2021MNRAS.505.2500P}
{Poon}, S. T.~S., {Nelson}, R.~P., \& {Coleman}, G. A.~L. 2021, \mnras, 505, 2500

\bibitem[{Price {et~al.}(2005)Price, Storn, \& Lampinen}]{Price2005}
Price, K., Storn, R., \& Lampinen, J. 2005, Differential {{Evolution}} (Berlin: Springer)

\bibitem[{{Rasio} {et~al.}(1996){Rasio}, {Tout}, {Lubow}, \& {Livio}}]{1996ApJ...470.1187R}
{Rasio}, F.~A., {Tout}, C.~A., {Lubow}, S.~H., \& {Livio}, M. 1996, \apj, 470, 1187

\bibitem[{{Rein} \& {Liu}(2012)}]{2012A&A...537A.128R}
{Rein}, H. \& {Liu}, S.~F. 2012, \aap, 537, A128

\bibitem[{{Rein} \& {Spiegel}(2015)}]{reboundias15}
{Rein}, H. \& {Spiegel}, D.~S. 2015, \mnras, 446, 1424

\bibitem[{{Rein} \& {Tamayo}(2015)}]{2015MNRAS.452..376R}
{Rein}, H. \& {Tamayo}, D. 2015, \mnras, 452, 376

\bibitem[{{Rice} {et~al.}(2022){Rice}, {Wang}, \& {Laughlin}}]{2022ApJ...926L..17R}
{Rice}, M., {Wang}, S., \& {Laughlin}, G. 2022, \apjl, 926, L17

\bibitem[{{Ros{\'a}rio} {et~al.}(2022){Ros{\'a}rio}, {Barros}, {Demangeon}, \& {Santos}}]{2022A&A...668A.114R}
{Ros{\'a}rio}, N.~M., {Barros}, S.~C.~C., {Demangeon}, O.~D.~S., \& {Santos}, N.~C. 2022, \aap, 668, A114

\bibitem[{{Sha} {et~al.}(2023){Sha}, {Vanderburg}, {Huang}, {Armstrong}, {Brahm}, {Giacalone}, {Wood}, {Collins}, {Nielsen}, {Hobson}, {Ziegler}, {Howell}, {Torres-Miranda}, {Mann}, {Zhou}, {Delgado-Mena}, {Rojas}, {Abe}, {Trifonov}, {Adibekyan}, {Sousa}, {Fajardo-Acosta}, {Guillot}, {Howard}, {Littlefield}, {Hawthorn}, {Schmider}, {Eberhardt}, {Tan}, {Osborn}, {Schwarz}, {Str{\o}m}, {Jord{\'a}n}, {Wang}, {Henning}, {Massey}, {Law}, {Stockdale}, {Furlan}, {Srdoc}, {Wheatley}, {Barrado Navascu{\'e}s}, {Lissauer}, {Stassun}, {Ricker}, {Vanderspek}, {Latham}, {Winn}, {Seager}, {Jenkins}, {Barclay}, {Bouma}, {Christiansen}, {Guerrero}, \& {Rose}}]{2023MNRAS.524.1113S}
{Sha}, L., {Vanderburg}, A.~M., {Huang}, C.~X., {et~al.} 2023, \mnras, 524, 1113

\bibitem[{Storn \& Price(1997)}]{Storn1997a}
Storn, R. \& Price, K. 1997, Journal of Global Optimization, 11, 341

\bibitem[{{Szab{\'o}} {et~al.}(2021){Szab{\'o}}, {Gandolfi}, {Brandeker}, {Csizmadia}, {Garai}, {Billot}, {Broeg}, {Ehrenreich}, {Fortier}, {Fossati}, {Hoyer}, {Kiss}, {Lecavelier des Etangs}, {Maxted}, {Ribas}, {Alibert}, {Alonso}, {Anglada Escud{\'e}}, {B{\'a}rczy}, {Barros}, {Barrado}, {Baumjohann}, {Beck}, {Beck}, {Bekkelien}, {Bonfils}, {Benz}, {Borsato}, {Busch}, {Cabrera}, {Charnoz}, {Collier Cameron}, {Van Damme}, {Davies}, {Delrez}, {Deleuil}, {Demangeon}, {Demory}, {Erikson}, {Fridlund}, {Futyan}, {Garc{\'\i}a Mu{\~n}oz}, {Gillon}, {Guedel}, {Guterman}, {Heng}, {Isaak}, {Lacedelli}, {Laskar}, {Lendl}, {Lovis}, {Luntzer}, {Magrin}, {Nascimbeni}, {Olofsson}, {Osborn}, {Ottensamer}, {Pagano}, {Pall{\'e}}, {Peter}, {Piazza}, {Piotto}, {Pollacco}, {Queloz}, {Ragazzoni}, {Rando}, {Rauer}, {Santos}, {Scandariato}, {S{\'e}gransan}, {Serrano}, {Sicilia}, {Simon}, {Smith}, {Sousa}, {Steller}, {Thomas}, {Udry}, {Van Grootel}, {Walton}, \& {Wilson}}]{2021A&A...654A.159S}
{Szab{\'o}}, G.~M., {Gandolfi}, D., {Brandeker}, A., {et~al.} 2021, \aap, 654, A159

\bibitem[{{Tamayo} {et~al.}(2020){Tamayo}, {Rein}, {Shi}, \& {Hernandez}}]{reboundx}
{Tamayo}, D., {Rein}, H., {Shi}, P., \& {Hernandez}, D.~M. 2020, \mnras, 491, 2885

\bibitem[{{Thompson} {et~al.}(2018){Thompson}, {Coughlin}, {Hoffman}, {Mullally}, {Christiansen}, {Burke}, {Bryson}, {Batalha}, {Haas}, {Catanzarite}, {Rowe}, {Barentsen}, {Caldwell}, {Clarke}, {Jenkins}, {Li}, {Latham}, {Lissauer}, {Mathur}, {Morris}, {Seader}, {Smith}, {Klaus}, {Twicken}, {Van Cleve}, {Wohler}, {Akeson}, {Ciardi}, {Cochran}, {Henze}, {Howell}, {Huber}, {Pr{\v{s}}a}, {Ram{\'\i}rez}, {Morton}, {Barclay}, {Campbell}, {Chaplin}, {Charbonneau}, {Christensen-Dalsgaard}, {Dotson}, {Doyle}, {Dunham}, {Dupree}, {Ford}, {Geary}, {Girouard}, {Isaacson}, {Kjeldsen}, {Quintana}, {Ragozzine}, {Shabram}, {Shporer}, {Silva Aguirre}, {Steffen}, {Still}, {Tenenbaum}, {Welsh}, {Wolfgang}, {Zamudio}, {Koch}, \& {Borucki}}]{2018ApJS..235...38T}
{Thompson}, S.~E., {Coughlin}, J.~L., {Hoffman}, K., {et~al.} 2018, \apjs, 235, 38

\bibitem[{{Vanderburg} {et~al.}(2017){Vanderburg}, {Becker}, {Buchhave}, {Mortier}, {Lopez}, {Malavolta}, {Haywood}, {Latham}, {Charbonneau}, {L{\'o}pez-Morales}, {Adams}, {Bonomo}, {Bouchy}, {Collier Cameron}, {Cosentino}, {Di Fabrizio}, {Dumusque}, {Fiorenzano}, {Harutyunyan}, {Johnson}, {Lorenzi}, {Lovis}, {Mayor}, {Micela}, {Molinari}, {Pedani}, {Pepe}, {Piotto}, {Phillips}, {Rice}, {Sasselov}, {S{\'e}gransan}, {Sozzetti}, {Udry}, \& {Watson}}]{2017AJ....154..237V}
{Vanderburg}, A., {Becker}, J.~C., {Buchhave}, L.~A., {et~al.} 2017, \aj, 154, 237

\bibitem[{{Vissapragada} {et~al.}(2022){Vissapragada}, {Chontos}, {Greklek-McKeon}, {Knutson}, {Dai}, {P{\'e}rez Gonz{\'a}lez}, {Grunblatt}, {Huber}, \& {Saunders}}]{2022ApJ...941L..31V}
{Vissapragada}, S., {Chontos}, A., {Greklek-McKeon}, M., {et~al.} 2022, \apjl, 941, L31

\bibitem[{{Wang} {et~al.}(2024){Wang}, {Zhang}, {Chen}, {Wang}, {Yu}, \& {Ma}}]{2024ApJS..270...14W}
{Wang}, W., {Zhang}, Z., {Chen}, Z., {et~al.} 2024, \apjs, 270, 14

\bibitem[{{Wong} {et~al.}(2021){Wong}, {Shporer}, {Zhou}, {Kitzmann}, {Komacek}, {Tan}, {Tronsgaard}, {Buchhave}, {Vissapragada}, {Greklek-McKeon}, {Rodriguez}, {Ahlers}, {Quinn}, {Furlan}, {Howell}, {Bieryla}, {Heng}, {Knutson}, {Collins}, {McLeod}, {Berlind}, {Brown}, {Calkins}, {de Leon}, {Esparza-Borges}, {Esquerdo}, {Fukui}, {Gan}, {Girardin}, {Gnilka}, {Ikoma}, {Jensen}, {Kielkopf}, {Kodama}, {Kurita}, {Lester}, {Lewin}, {Marino}, {Murgas}, {Narita}, {Pall{\'e}}, {Schwarz}, {Stassun}, {Tamura}, {Watanabe}, {Benneke}, {Ricker}, {Latham}, {Vanderspek}, {Seager}, {Winn}, {Jenkins}, {Caldwell}, {Fong}, {Huang}, {Mireles}, {Schlieder}, {Shiao}, \& {Noel Villase{\~n}or}}]{2021AJ....162..256W}
{Wong}, I., {Shporer}, A., {Zhou}, G., {et~al.} 2021, \aj, 162, 256

\bibitem[{{Wu} {et~al.}(2023){Wu}, {Rice}, \& {Wang}}]{2023AJ....165..171W}
{Wu}, D.-H., {Rice}, M., \& {Wang}, S. 2023, \aj, 165, 171

\bibitem[{{Wu} \& {Lithwick}(2013)}]{2013ApJ...772...74W}
{Wu}, Y. \& {Lithwick}, Y. 2013, \apj, 772, 74

\bibitem[{{Wyttenbach} {et~al.}(2020){Wyttenbach}, {Molli{\`e}re}, {Ehrenreich}, {Cegla}, {Bourrier}, {Lovis}, {Pino}, {Allart}, {Seidel}, {Hoeijmakers}, {Nielsen}, {Lavie}, {Pepe}, {Bonfils}, \& {Snellen}}]{2020A&A...638A..87W}
{Wyttenbach}, A., {Molli{\`e}re}, P., {Ehrenreich}, D., {et~al.} 2020, \aap, 638, A87

\bibitem[{{Yee} {et~al.}(2020){Yee}, {Winn}, {Knutson}, {Patra}, {Vissapragada}, {Zhang}, {Holman}, {Shporer}, \& {Wright}}]{2020ApJ...888L...5Y}
{Yee}, S.~W., {Winn}, J.~N., {Knutson}, H.~A., {et~al.} 2020, \apjl, 888, L5

\bibitem[{{Zechmeister} \& {K{\"u}rster}(2009)}]{2009A&A...496..577Z}
{Zechmeister}, M. \& {K{\"u}rster}, M. 2009, \aap, 496, 577

\bibitem[{{Zhu} {et~al.}(2018){Zhu}, {Dai}, \& {Masuda}}]{2018RNAAS...2..160Z}
{Zhu}, W., {Dai}, F., \& {Masuda}, K. 2018, Research Notes of the American Astronomical Society, 2, 160

\end{thebibliography}

\begin{appendix}
\onecolumn
\section{Individual transit plots}

\begin{figure}[h]
    \centering
    \includegraphics[height=.8\textheight]{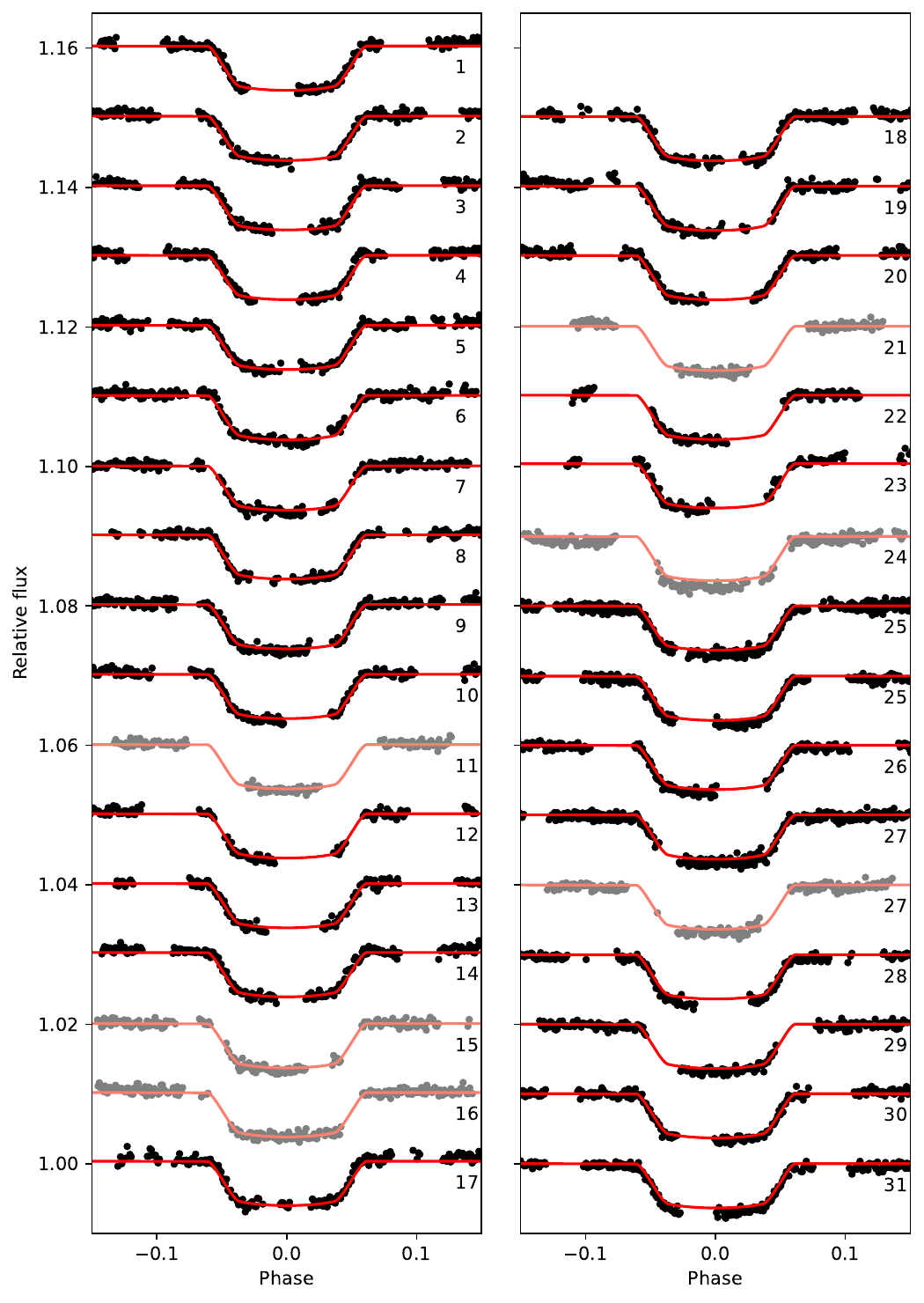}
    \caption{All individual transits from the \texttt{PIPE} data (black) that were fitted using \texttt{TLCM} with the respective transit models (red). The numbers below each transit light curve indicate the number of the corresponding visit, see Table~\ref{table:obs_log}. Duplications in visit number refer to the first and second transit in a phase curve observation. If the resulting transit timings were not used in the TTV analysis, they are greyed out. 
    }
    \label{fig:transits}
\end{figure}

\FloatBarrier
\pagebreak

\section{Retrieved mid-transit timings}
\vspace{-3cm}
\begin{table*}[h]
    \centering
    \caption{Mid-transit times of TOI-2109\,b, retrieved from the CHEOPS \texttt{DRP} and \texttt{PIPE} data reductions. Potentially biased timings were filtered out. Transit epochs given relative to our timing from the phase curve fits. Times given in BJD$_\mathrm{TDB} - 2450000$.}
    \begin{tabular}{c c c c c}\hline\hline
        Epoch & Time$_\mathrm{DRP}$ & Error$_\mathrm{DRP}$ [d] & Time$_\mathrm{PIPE}$ & Error$_\mathrm{PIPE}$ [d] \\\hline
        -1041 & 9349.542546 & 0.000232 & 9349.542445 & 0.000169 \\
        -1035 & 9353.577576 & 0.000211 & 9353.577643 & 0.000182 \\
        -1031 & 9356.267243 & 0.000191 & 9356.267276 & 0.000165 \\
        -1022 & 9362.319158 & 0.000177 & 9362.319324 & 0.000177 \\
        -1017 & 9365.681918 & 0.000173 & 9365.681752 & 0.000159 \\
        -1011 & 9369.717019 & 0.000213 & 9369.717254 & 0.000212 \\
        -1001 & 9376.441345 & 0.000240 & 9376.441376 & 0.000241 \\
        -999 & 9377.786845 & 0.000212 & 9377.786895 & 0.000156 \\
        -997 & 9379.131546 & 0.000213 & 9379.131353 & 0.000202 \\
        -993 & 9381.821608 & 0.000187 & 9381.821568 & 0.000166 \\
        -541 & 9685.779757 & 0.000335 & 9685.780035 & 0.000269 \\
        -527 & 9695.194240 & 0.000254 & 9695.194138 & 0.000172 \\
        -475 & 9730.163441 & 0.000220 & 9730.163262 & 0.000181 \\
        -471 & 9732.852594 & 0.000230 & 9732.852663 & 0.000208 \\
        -470 & 9733.525305 & 0.000203 & 9733.525374 & 0.000194 \\
        -465 & 9736.887592 & 0.000222 & 9736.887740 & 0.000206 \\
        -456 & 9742.940224 & 0.000199 & 9742.940237 & 0.000178 \\
        9 & 10055.640153 & 0.000280 & 10055.640304 & 0.000289 \\
        18 & 10061.692214 & 0.000278 & 10061.692062 & 0.000277 \\
        50 & 10083.211849 & 0.000252 & 10083.211774 & 0.000216 \\
        71 & 10097.333665 & 0.000266 & 10097.333611 & 0.000187 \\
        73 & 10098.679387 & 0.000284 & 10098.679421 & 0.000216 \\
        80 & 10103.386211 & 0.000235 & 10103.386483 & 0.000191 \\
        82 & 10104.731772 & 0.000289 & 10104.731543 & 0.000327 \\
        88 & 10108.766204 & 0.000203 & 10108.766020 & 0.000215 \\
        93 & 10112.128585 & 0.000224 & 10112.128684 & 0.000217 \\
        \hline
    \end{tabular}
    \label{tab:tts}
\end{table*}

\vspace{-5cm}
\begin{table*}[h]
    \centering
    \caption{Mid-transit times of TOI-2109\,b, retrieved from the ground-based data published in \citet{2021AJ....162..256W}, WASP and TESS. Transit epochs given relative to our timing from the phase curve fits. Times given in BJD$_\mathrm{TDB} - 2450000$. Only a portion of the table is shown to indicate its form and contents. The full table can be retrieved from the CDS.}
    \begin{tabular}{c c c c c}\hline\hline
        Epoch & Time & Error [d] & Source \\\hline
        -9166 & 3885.688618 & 0.001079 & WASP \\
        -1584 & 8984.387648 & 0.002520 & TESS \\
        -1583 & 8985.063895 & 0.002520 & TESS \\
        -1582 & 8985.733954 & 0.001407 & TESS \\
        \dots & \dots & \dots & \dots \\
        \hline
    \end{tabular}
    \label{tab:tts_rest}
\end{table*}

\FloatBarrier
\pagebreak
\section{Sine fit corner plot}

\begin{figure}[h]
    \centering
    \includegraphics[width=\linewidth]{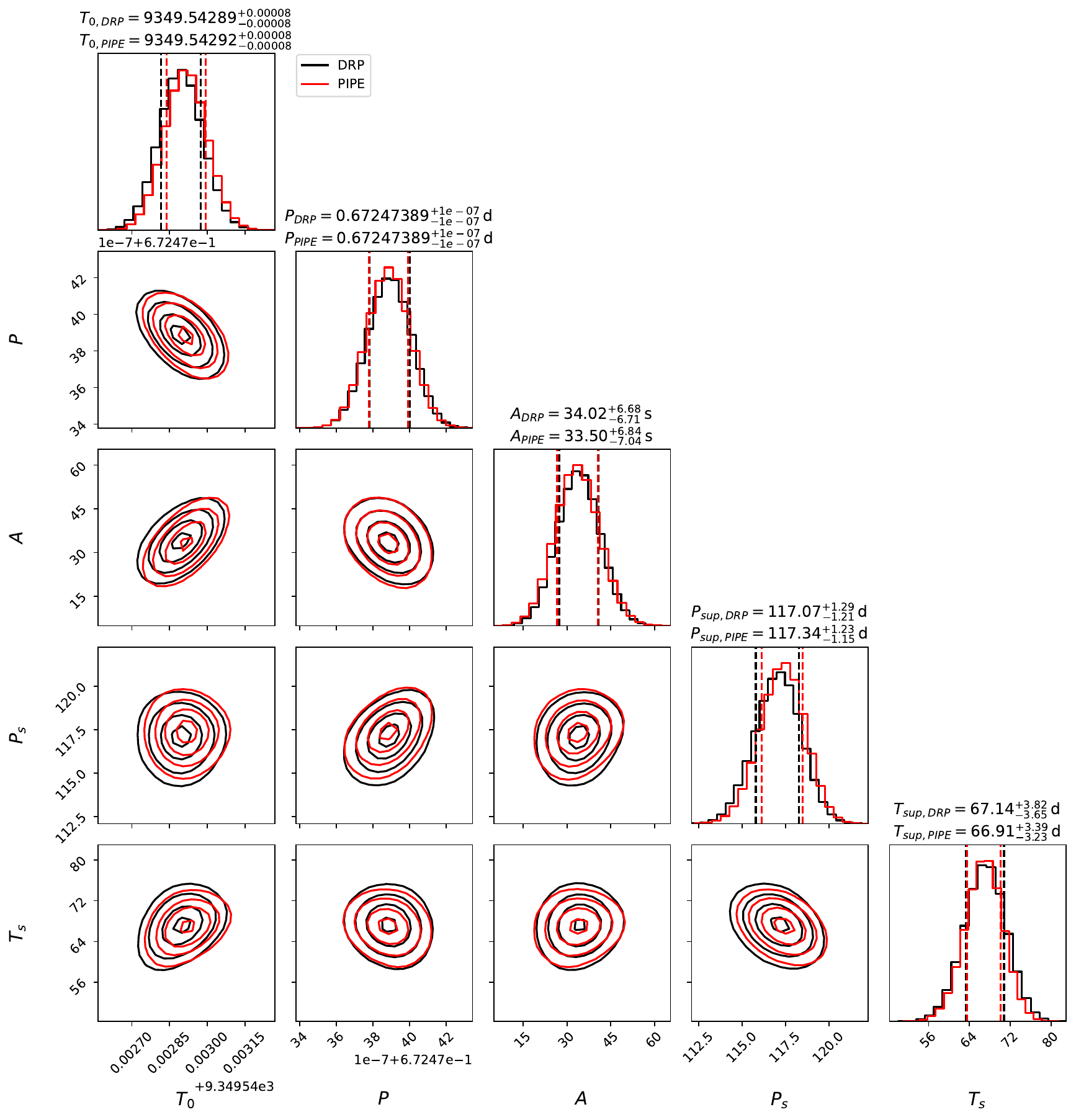}
    \caption{Final corner plot of the fit of the simple sine model (Eq.~\ref{eq:sine_model}) to the mid-transit timings, measured with CHEOPS using the \texttt{DRP} and \texttt{PIPE} reductions. The dashed lines indicate the $1\sigma$ errors around the median value. The contours show the $0.5\sigma$, $1\sigma$, $1.5\sigma$ and $2\sigma$ levels. $T_0$ is given in BJD$_\mathrm{TDB} - 2450000$.
    }
    \label{fig:sine_corner}
\end{figure}

\pagebreak

\section{Additional MEGNO eccentricity plots\label{sec:app_megno}}
\vspace{-0.2cm}
\begin{figure*}[h]
    \centering
    \begin{minipage}[b]{0.43\linewidth}
        \centering
        \includegraphics[width=\linewidth]{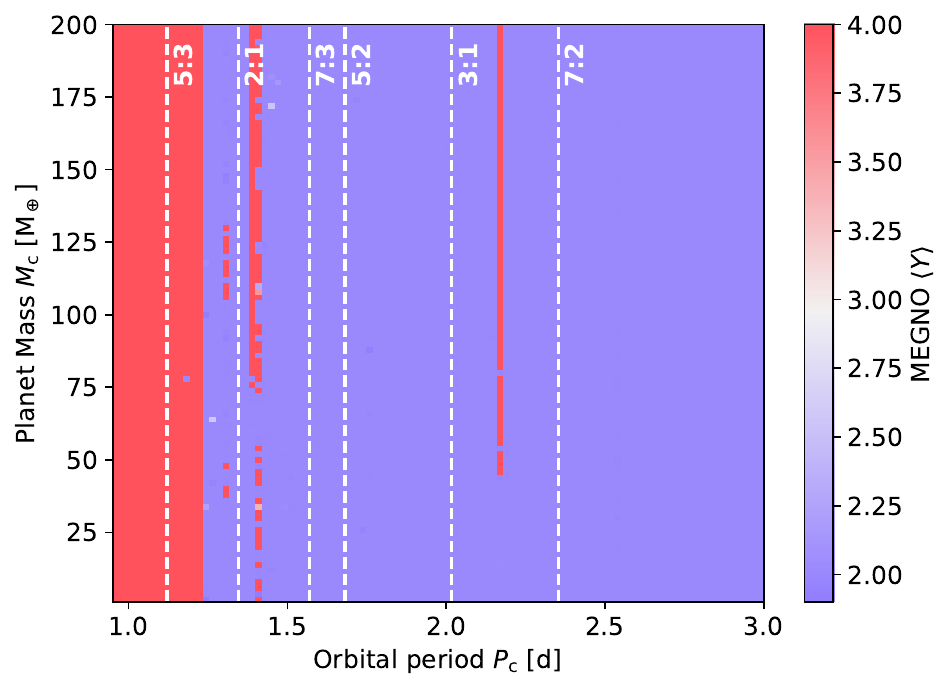}
        \caption{MEGNO plot of the TOI-2109 system for a circular orbit of the hot Jupiter and a slightly eccentric orbit of planet c ($e_\mathrm{c}=0.05$), for $270\,000$ orbits of the inner planet.}
        \label{fig:array_e0005}
    \end{minipage}
    \hfill
    \begin{minipage}[b]{0.43\linewidth}
        \centering
        \includegraphics[width=\linewidth]{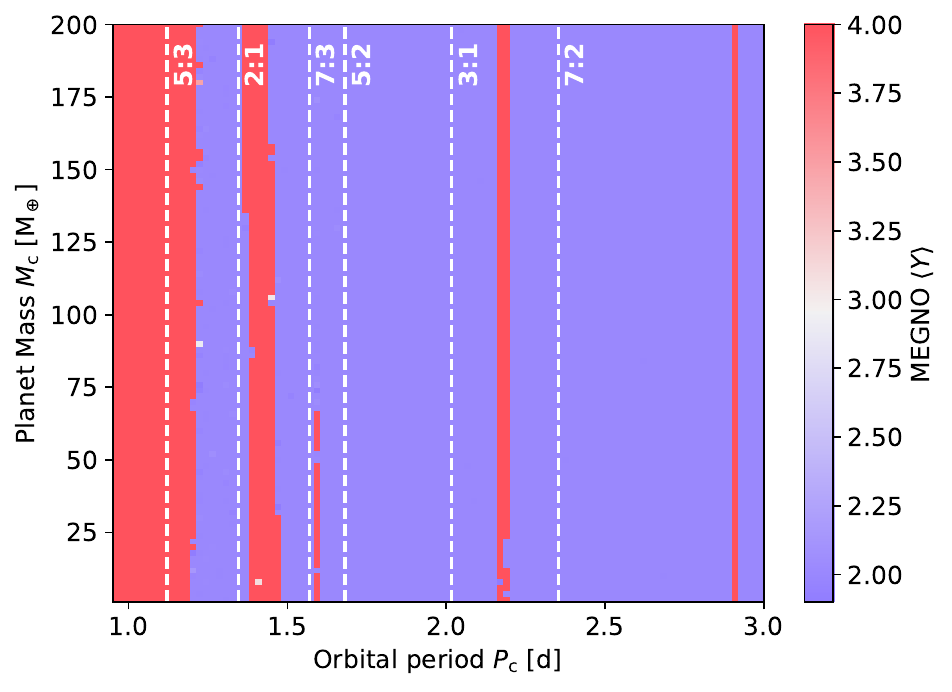}
        \caption{MEGNO plot of TOI-2109 for eccentric orbits of both planets with $e_\mathrm{b} = 0.03$ and $e_\mathrm{c} = 0.05$, computed for $270\,000$ orbits of the inner planet.}
        \label{fig:array_e0305}
    \end{minipage}
\end{figure*}
\begin{figure*}[h]
    \centering
    \begin{minipage}[b]{0.43\linewidth}
        \centering
        \includegraphics[width=\linewidth]{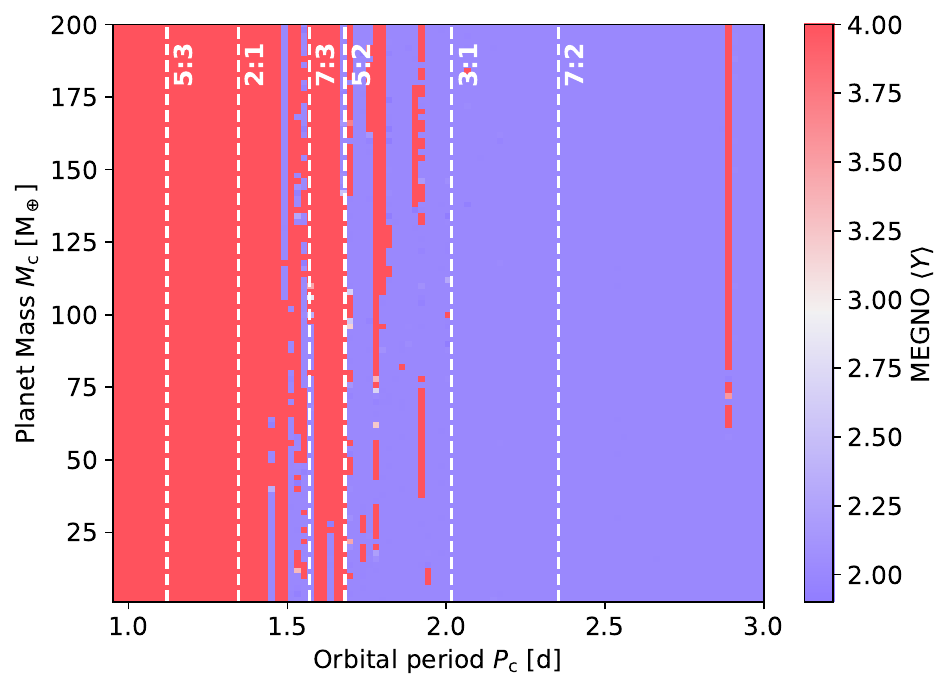}
        \caption{MEGNO plot of the TOI-2109 system for a circular orbit of the hot Jupiter and a moderately eccentric orbit of planet c ($e_\mathrm{c}=0.15$), for $270\,000$ orbits of the inner planet.}
        \label{fig:array_e0015}
    \end{minipage}
    \hfill
    \begin{minipage}[b]{0.43\linewidth}
        \centering
        \includegraphics[width=\linewidth]{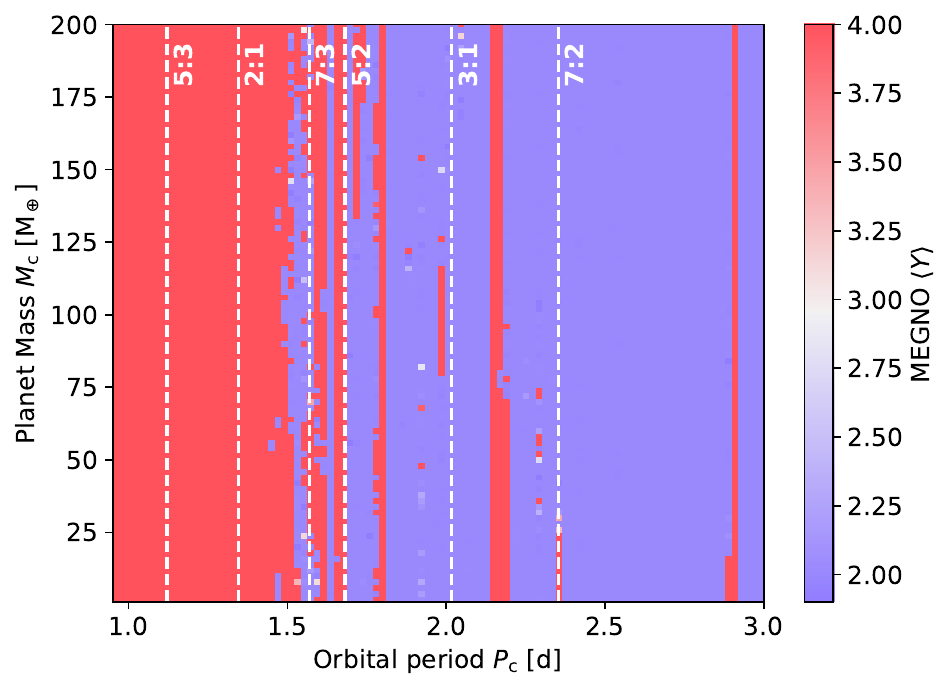}
        \caption{MEGNO plot of TOI-2109 for eccentric orbits of both planets with $e_\mathrm{b} = 0.03$ and $e_\mathrm{c} = 0.15$, computed for $270\,000$ orbits of the inner planet.}
        \label{fig:array_e0315}
    \end{minipage}
\end{figure*}
\begin{figure*}[h]
    \centering
    \begin{minipage}[b]{0.43\linewidth}
        \centering
        \includegraphics[width=\linewidth]{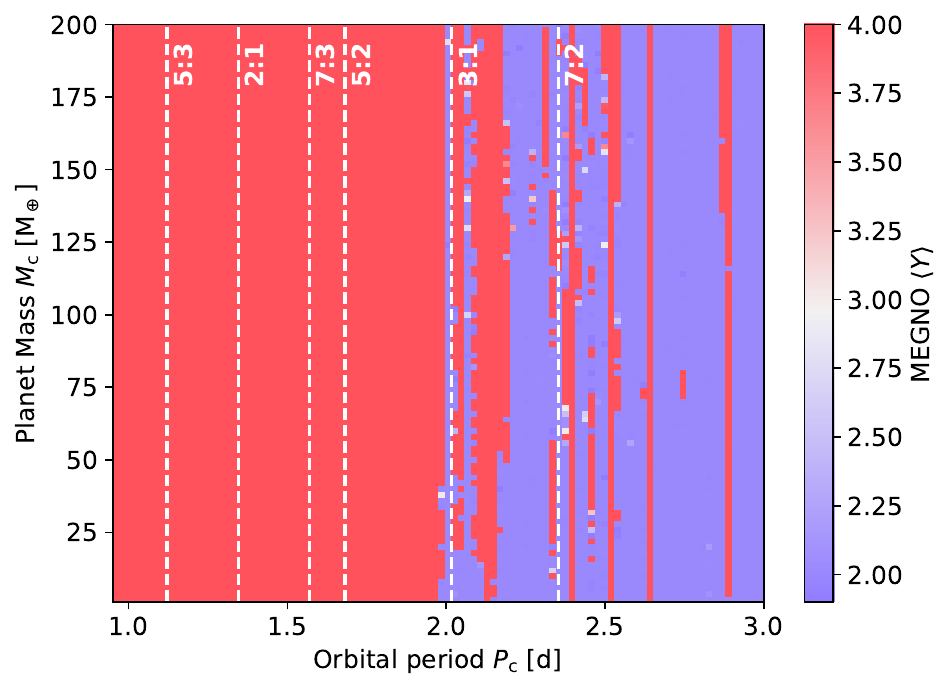}
        \caption{MEGNO plot of the TOI-2109 system for a circular orbit of the hot Jupiter and an eccentric orbit of planet c ($e_\mathrm{c}=0.25$), for $270\,000$ orbits of the inner planet.}
        \label{fig:array_e0025}
    \end{minipage}
    \hfill
    \begin{minipage}[b]{0.43\linewidth}
        \centering
        \includegraphics[width=\linewidth]{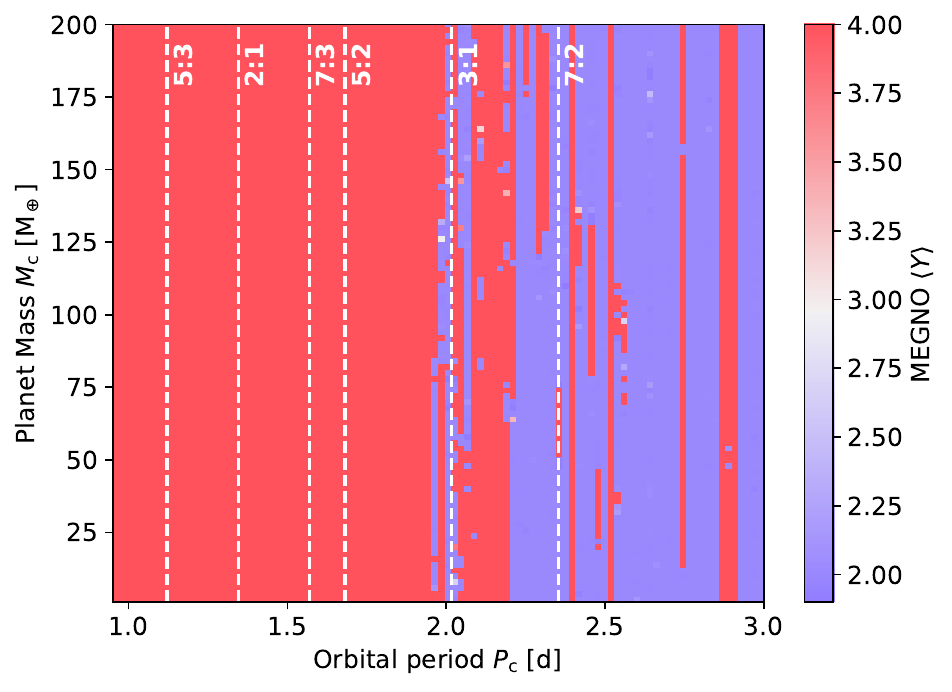}
        \caption{MEGNO plot of TOI-2109 for eccentric orbits of both planets with $e_\mathrm{b} = 0.03$ and $e_\mathrm{c} = 0.25$, computed for $270\,000$ orbits of the inner planet.}
        \label{fig:array_e0325}
    \end{minipage}
\end{figure*}

\FloatBarrier
\section{Additional \texttt{TRADES} TTV plots}
\begin{figure}[h]
    \centering
    \includegraphics[width=0.5\linewidth]{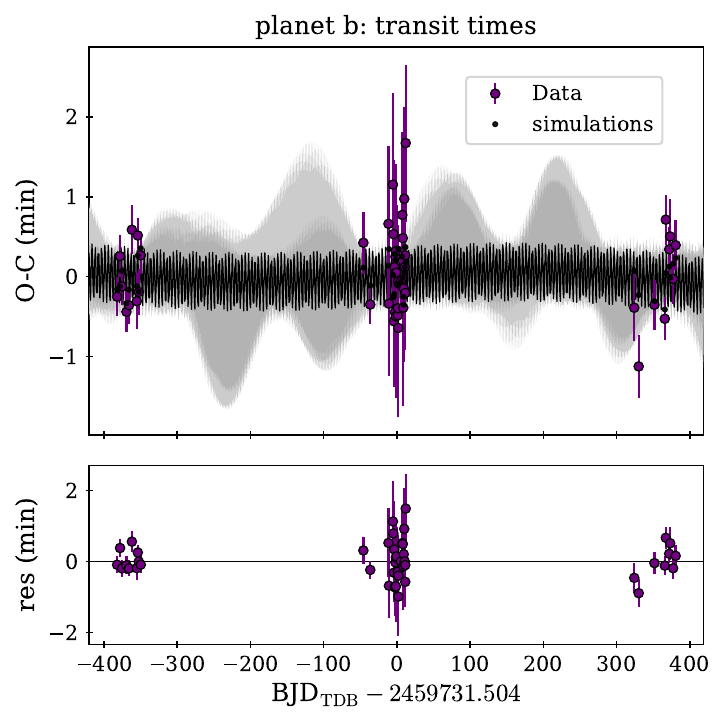}
    \caption{\texttt{TRADES} TTV plot of the general circular orbit fit. Top: Fit to the transit timing measurements of TOI-2109\,b, including the final MAP model and random samples from the MCMC analysis. Bottom: Residuals of the fit.}
    \label{fig:TRADES_general_best}
\end{figure}

\begin{figure}[h]
    \centering
    \includegraphics[width=.5\linewidth]{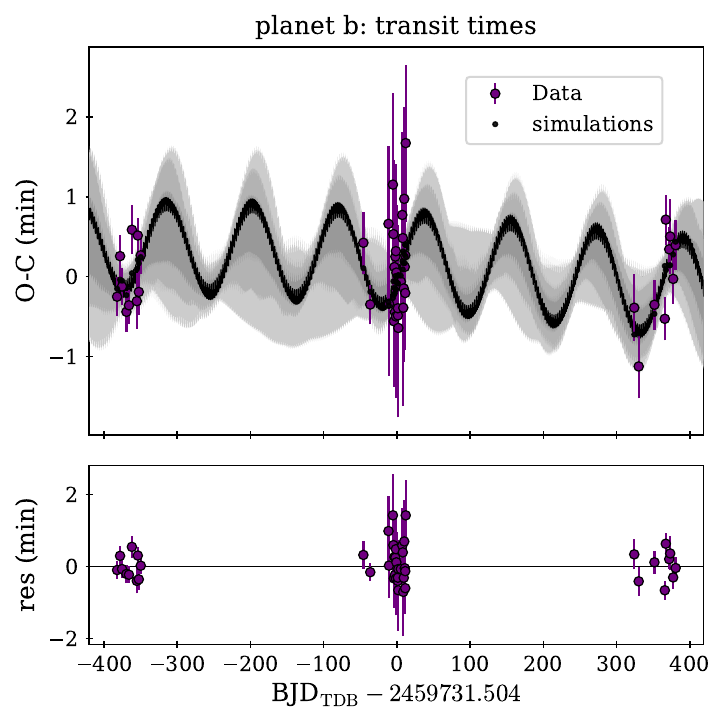}
    \caption{\texttt{TRADES} TTV plot of the general eccentric orbit fit. Top: Fit to the transit timing measurements of TOI-2109\,b, including the final MAP model and random samples from the MCMC analysis. Bottom: Residuals of the fit.}
    \label{fig:TRADES_ecc}
\end{figure}




\end{appendix}

\end{document}